\documentclass[iop]{emulateapj}

\newcommand{\noprint}[1]{}
\newcommand{\figsetstart}{{\bf Fig. Set} }
\newcommand{\figsetend}{}
\newcommand{\figsetgrpstart}{}
\newcommand{\figsetgrpend}{}
\newcommand{\figsetnum}[1]{{\bf #1.}}
\newcommand{\figsettitle}[1]{ {\bf #1} }
\newcommand{\figsetgrpnum}[1]{\noprint{#1}}
\newcommand{\figsetgrptitle}[1]{\noprint{#1}}
\newcommand{\figsetplot}[1]{\noprint{#1}}
\newcommand{\figsetgrpnote}[1]{\noprint{#1}}
\begin{document}
\title{Connecting X-ray and Infrared Variability among Young Stellar Objects: Ruling out potential sources of disk fluctuations}
\author{Flaherty, K.M. \altaffilmark{1,6}, Muzerolle, J.\altaffilmark{2}, Wolk, S.J.\altaffilmark{3}, Rieke, G. \altaffilmark{1}, Gutermuth, R. \altaffilmark{4}, Balog, Z.\altaffilmark{5}, Herbst, W. \altaffilmark{6}, Megeath, S.T. \altaffilmark{7}, Furlan, E.\altaffilmark{8,9}}
\email{kflaherty@wesleyan.edu}

\altaffiltext{1}{Steward Observatory, University of Arizona, Tucson, AZ 85721}
\altaffiltext{2}{Space Telescope Science Institute, 3700 San Martin Dr., Baltimore, MD, 21218}
\altaffiltext{3}{Harvard-Smithsonian Center for Astrophysics, 60 Garden Street, Cambridge, MA 02138, USA}
\altaffiltext{4}{Department of Astronomy, University of Massachusetts, Amherst, MA 01003}
\altaffiltext{5}{Max-Planck Institut f\"{u}r Astronomie, K\"{o}nigstuhl 17, D-69117 Heidelberg, Germany}
\altaffiltext{6}{Department of Astronomy, Wesleyan University, Middletown, CT 06459}
\altaffiltext{7}{Department of Physics and Astronomy, University of Toledo, Toledo, OH}
\altaffiltext{8}{National Optical Astronomy Observatory, 950 North Cherry Avenue, Tucson, AZ 85719, USA}
\altaffiltext{9}{Visitor at the Infrared Processing and Analysis Center, Caltech, 770 South Wilson Avenue, Pasadena, CA 91125, USA}

\begin{abstract}
Variability in the infrared emission from disks around pre-main sequence stars over the course of days to weeks appears to be common, but the physical cause of the changes in disk structure are not constrained. Here we present coordinated monitoring of one young cluster with the {\it Spitzer} and {\it Chandra} space telescopes aimed at studying the physical source of the variability. In fall 2011 we obtained ten epochs of Chandra ACIS photometry over a period of 30 days with a roughly 3 day cadence contemporaneous with 20 epochs of Spitzer [3.6],[4.5] photometry over 40 days with a roughly 2 day cadence of the IC 348 cluster. This cadence allows us to search for week to month long responses of the infrared emission to changes in the high-energy flux. We find no strong evidence for a direct link between the X-ray and infrared variability on these timescales among 39 cluster members with circumstellar disks. There is no significant correlation between the shape of the infrared and X-ray light curves, or between the size of the X-ray and infrared variability. Among the stars with an X-ray flare none showed evidence of a correlated change in the infrared photometry on timescales of days to weeks following the flare. This lack of connection implies that X-ray heating of the planet forming region of the disk is not significant, although we cannot rule out rapid or instantaneous changes in infrared emission. 
\end{abstract}

\section{Introduction}
The inner AU around a pre-main sequence star is a complex region. Dust and gas flow inward through the disk, possibly accompanied by massive planets built in the outer disk \citep[e.g.][]{lin96}. Close to the star the dust will sublimate, creating an inner wall at a few tenths of an AU \citep{kam09}. The gas will flow within this radius \citep{eis09} until it is loaded on the large dipolar stellar magnetic field and lifted out of the midplane. Once locked onto the stellar field this material free-falls onto the star, creating a shock as it strikes the stellar surface and heating a small spot to well above the photospheric temperature \citep{ing13}.  Add in the possibility of a significant wind being driven outward from close to the star \citep[e.g.][]{zan13}, and one can begin to understand the wide range of physical processes at work close to the star.

Synoptic observations point toward even more complexity. Models \citep[e.g.][]{kul08} and observations \citep[e.g.][]{fan13} of accretion variability indicate that the flow of material onto the star fluctuates on short timescales. Hot and cold spots dot the stellar surface leading to rapid optical variability \citep[e.g.][]{alen12}. Large scale disk instabilities lead to outbursts lasting years to decades \citep[e.g.][]{zhu10,hil13} and frequent fluctuations at a wide range of infrared wavelengths indicate structural changes at or near the sublimation edge of the circumstellar disk on weekly timescales \citep[e.g.][]{mor11,wol13,cod14}. The frequency of moderate X-ray flares, believed to arise from magnetic reconnection events in the coronae \citep{fei07}, indicate that the magnetic field is not only complex but constantly varying. Even single epoch measurements of the magnetic field find that not all young stars have strictly dipolar magnetic fields; some are dominated by the octopolar or quadropolar components close to the stellar surface \citep{gre12}. 

Here we attempt to take advantage of the infrared fluctuations to study the disk structure in more detail. By searching for correlations between the infrared variability and other forms of variability (X-ray, accretion, etc.) we can directly study the influence of various factors that may set the structure of the dusty disk. Recent observations have found that the occurrence rate of infrared variability increases with X-ray luminosity \citep{fla13}, suggestive of some sort of connection between the two. Enhanced X-ray emission was observed following the $\sim$few magnitude outburst of the FU Ori Object V1647 Ori \citep{ham10}, suggesting a direct connection between the change in disk structure and enhanced X-ray emission during these large accretion bursts. X-rays are also likely an important contributor to the clearing of the disk by photo-evaporation within the first few million years \citep{owe10}. High-energy X-ray emission may influence the structure of the disk directly through heating and ionization \citep{ski13} or may serve as an indirect tracer of rearrangements of the magnetic field \citep[e.g.][]{goo99}. Previous theoretical studies have focused mainly on the gas, where the X-ray illumination is most strongly felt \citep{are11} but it may penetrate into the dust layer and heat and/or ionize the dust. The frequent flaring of X-ray emission, with factors of 10 increase in X-ray emission a common occurrence \citep{fei07}, means that this illumination of the dust is constantly, and strongly, variable.  \citet{ste07} find that roughly half of the stars in Taurus exhibit X-ray variability, while \citet{wol05} find that solar-type stars in Orion spend $\sim$30\%\ of their time in an elevated state. This X-ray variability is a mix of rotational modulation \citep{fla05} and flares \citep{ste07} with larger fluctuations among stars with disks \citep{fla12}. Young stellar objects have been found to have very strong magnetic fields \citep{joh07,val04} that likely interact with the disk, although this interaction has been difficult to observe directly. The structure of the magnetic field can be very complex \citep[e.g.][]{don12,don11} and even among well studied stars it can be hard to characterize observationally \citep[e.g.][]{sch13}. 

Given the complex, but prominent, nature of the magnetic field, studying an indirect tracer like X-ray emission may shed light on its structure and how these structural changes affect the planet forming region in a large sample of young stellar objects. Contemporaneous changes in X-ray flux and infrared emission from the disk would point towards either a direct influence of the X-ray photons on the disk, or changes in magnetic field structure that affect both the high-energy flux and the emission from the disk. \citet{fla12} report a difference in the relative size of X-ray fluctuations between stars with disks and diskless systems. \citet{for07} did not find a direct correlation between X-ray and near-infrared variability in the Coronet cluster, but this survey was limited to observations over the course of only one week of a small handful of objects. The diversity in magnetic field and circumstellar disk properties among young stellar objects suggests that a small sample may not be able to capture the full range of interactions. By extending this type of analysis to a larger sample ($\sim$100 vs 10 objects) on longer timescales (40 days vs one week) at a wavelength that has less contamination from stellar photospheric emission (3-5\micron\ vs. 1-2\micron) we can broadly address the interactions of the dusty disk and the magnetic field/X-ray emission. To do this we obtained coordinated Chandra and Spitzer observations of the IC 348 cluster. This 2-3 Myr old cluster contains $\sim$300 known cluster members, 50\%\ of which contain an infrared excess indicative of circumstellar material \citep{lad06}. In section 2 we present our observations, in section 3 we search for correlated variability and in section 4 we discuss the implications of these results.

\section{Data}
\subsection{X-Ray Data Reduction}

The field was observed by $Chandra$ 10 times between  Oct 17, 2011 and Nov 17, 2011.  The nominal 4-chip ACIS-I array was used in "Very Faint" mode.  Each observation was about 10 ks and centered at 03:44:31.50 +32:08:33.70 (J2000.)  Roll was left nominal so not all sources were in the field of view for all exposures  (see Table~\ref{xray_log} for full details).

The data  were processed through the standard CIAO pipeline at the
Chandra X-ray Center, using their software version D8.4.  This version of the
pipeline automatically incorporated a noise correction for low energy
events.   Background was nominal and non-variable. 

Since the focus was on the bright sources which may vary significantly, sources were identified in the individual frames. 
To identify point sources, photons
with energies below 300 eV and above 8.0 keV were filtered out from
this merged event list.  
This excluded energies which generally lack a stellar contribution.  By filtering the data as
described, contributions from hard, non-stellar sources  such as X-ray binaries and AGN are
attenuated, as is noise.  A monochromatic exposure map was generated in the standard way using an
energy of 1.49 keV which is a reasonable match to the expected 
peak energy of the stellar sources and the $Chandra$ mirror transmission.  The CIAO tool 
WavDetect was then run on a series of flux
corrected images binned by 1, 2 and 4 pixels. The thresholds were set to limit false detections to about 1 per 100 detections. 
The output resulted in the detection of between 119 -139 sources (mean 135.5)
%134.133,139,119, 137, 135,138 139 124, 136
%\footnote{The
% src\_significance given by the CIAO tool WavDetect is not in
% units of $\sigma$ so no false alarm probability is associated. While the
% relative value of src\_significance is meaningful, 
%  it is not clearly defined in a
%  statistical sense (V.~Kashyap Private communication).}.

At each source position an extraction ellipse was calculated following
\citet{wol06} updated for the appropriate roll.  This provided an
extraction ellipse containing 95\% of the source flux within the 0.3-8 keV range. 
For each of the sources, a background ellipse was identified. 
The background was an annular ellipse with the same center, eccentricity, and 
rotation as the source.  The outer radius was 6 times the radius of the source 
and the inner radius was 3 times larger than the source.  From this region any 
nearby sources were subtracted with ellipses 3 times the size of the source 
ellipse.  The net counts were calculated by subtracting the background
counts  (corrected for area) and multiplying the result by 1.053 to
correct for the use of a 95\% encircled energy radius.

\subsection{Spitzer}
We also repeatedly observed IC 348 with the {\it Spitzer} Space Telescope using the IRAC instrument during the fall 2011 observing window (PID60160). The observing strategy was identical to that described in \citet{fla13}, and here we provide a brief summary. Both [3.6] and [4.5] photometry were obtained simultaneously over the course of 20 epochs from Oct 15, 2011 to Nov 23, 2011 (Table~\ref{obs_log}). The cadence was chosen to trace the daily to weekly fluctuations in infrared emission with one observation every two days . The cluster was mapped in a 5 by 5 grid, with each grid point separated by 360" and the total field centered at 03:44:20 +32:03:01. Since the fields of view of the [3.6] and [4.5] arrays do not completely overlap, our final map covered by both bands was not a square and instead had a total area of roughly 0.4$^{\circ}$ by 0.25$^{\circ}$. Images were taken in HDR mode with a frame time of 12 seconds and 3 cycles at each position. The IRAC data reduction pipeline is described in \citet{gut09} and \citet{mor12} with updates appropriate for the warm Spitzer mission described in Gutermuth et al. in prep.

\section{Results}
\subsection{Demographics and selection of variables}

As a nearby ($\sim$320 pc), young (2-3 Myr), relatively compact cluster IC 348 has proven to be a valuable laboratory for studying young stellar object variability \citep{coh04,fla13}. With over 300 cluster members, only one of which is earlier than A0, this region of low mass star formation contains a wide variety of circumstellar structures from pre-natal envelopes to disks with many-AU wide gaps \citep{lad06}.  In the infrared, almost 60\%\ of the sources with an infrared excess are variable at [3.6], [4.5] \citep{fla13}, while 89\%\ of the pre-main sequence stars brighter than I=14 are variable in the optical \citep{coh04} on timescales of weeks to months. The frequency of these fluctuations indicates that the majority of pre-main sequence stars experience variability on their stellar surface,  probed by the optical data, and within an AU of the star, probed by the infrared data. 

\begin{figure}
\includegraphics[scale=.31]{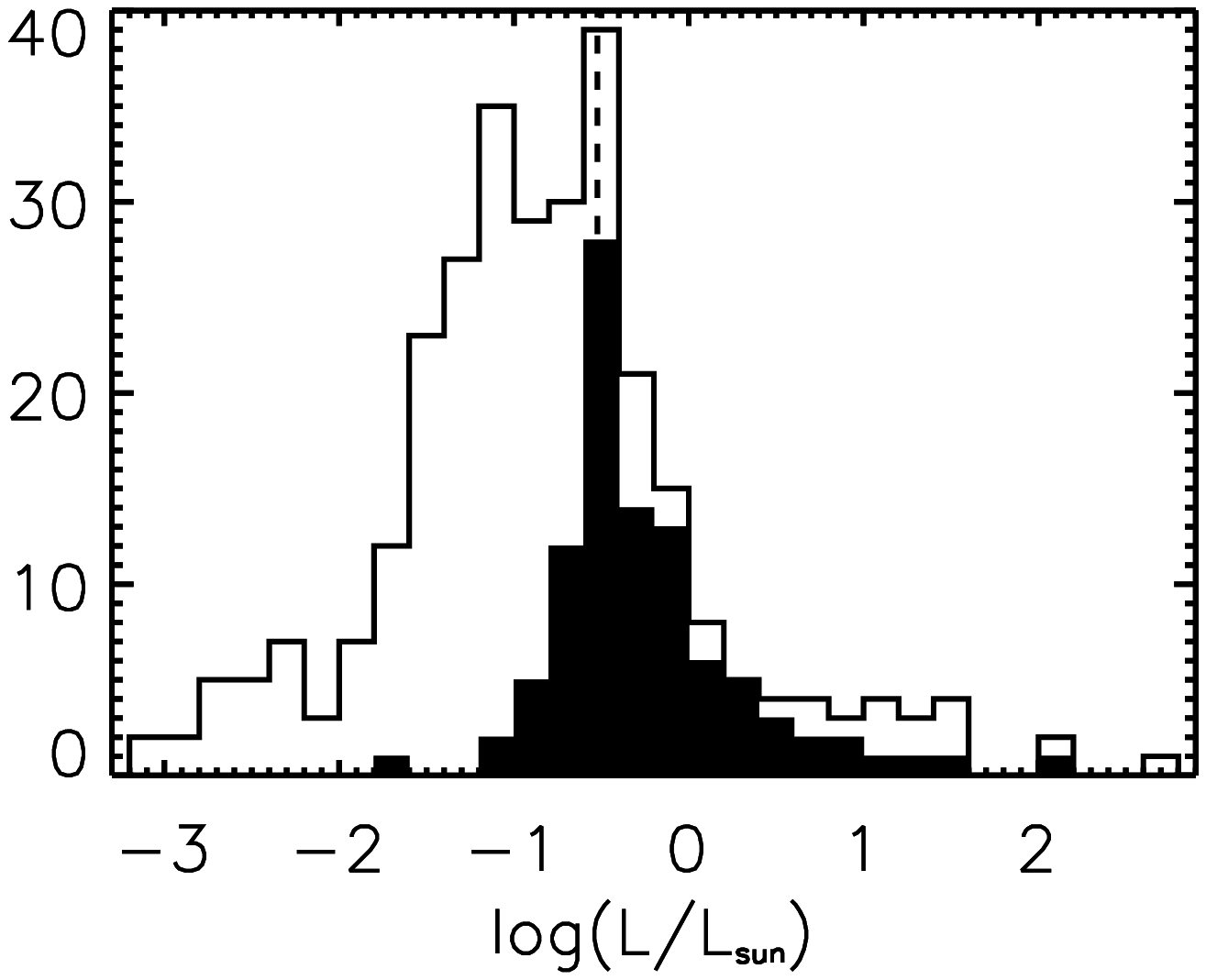}
\includegraphics[scale=.31]{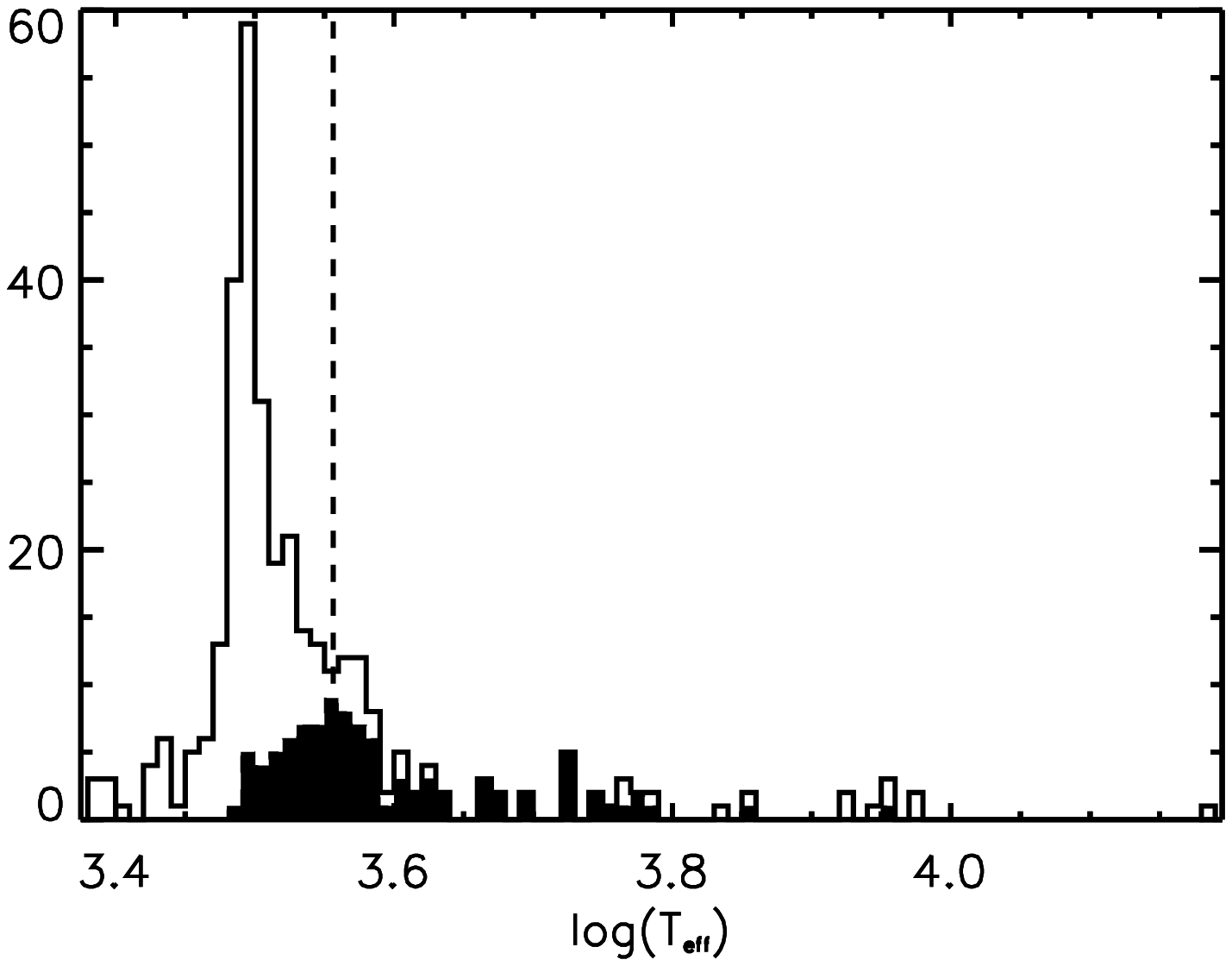}
\includegraphics[scale=.31]{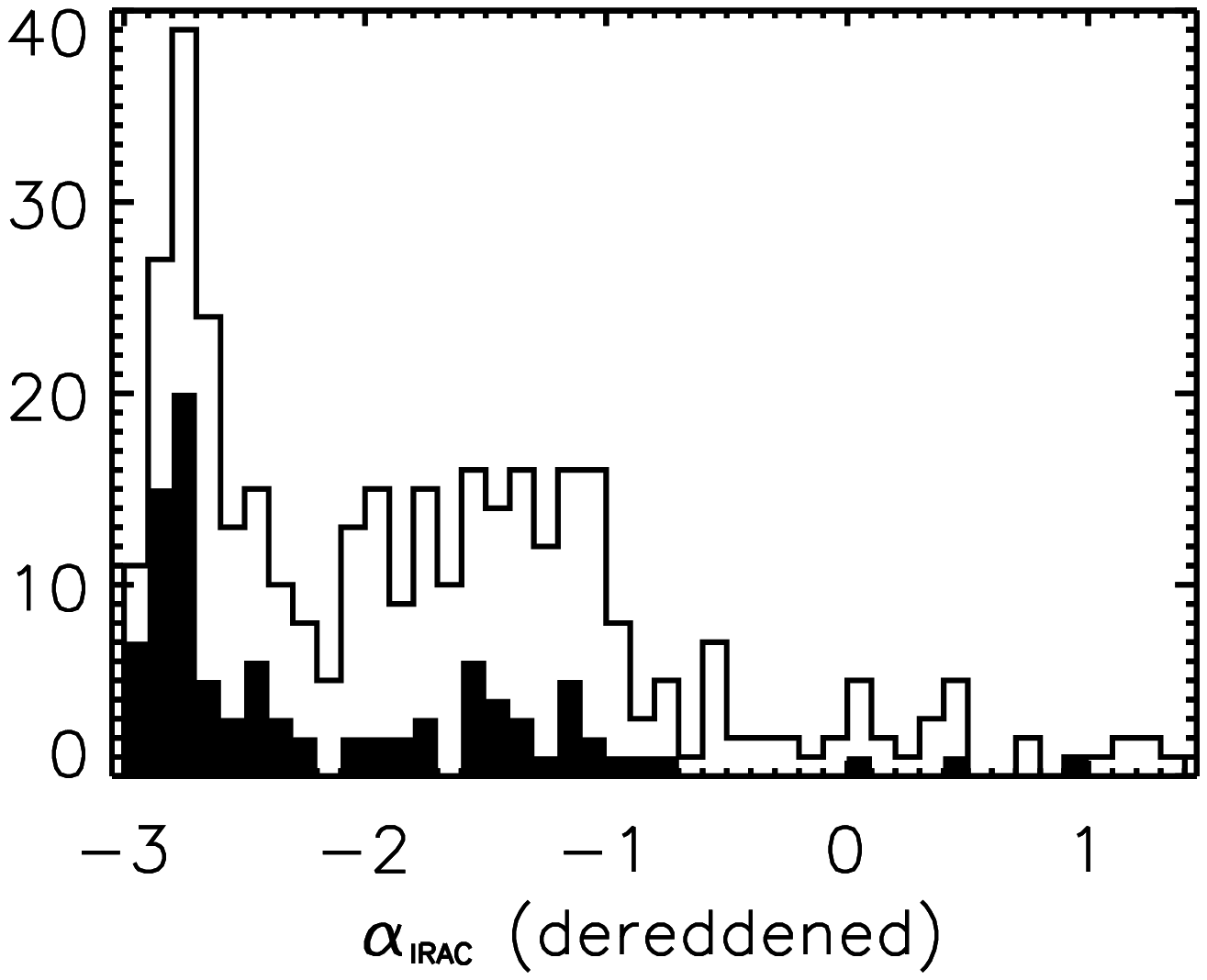}
\caption{Demographics of sources detected in more than 4 X-ray epochs (filled histogram) compared to the entire membership sample (open histogram) for stellar luminosity (left), effective temperature (middle) and dereddened infrared SED slope (right). We are incomplete to members below T$_{eff}$=3600K and L$_*$=0.3L$_{\odot}$, and are more sensitive to sources without disks ($\alpha_{IRAC}<-2.56$) than sources with disks.\label{demographics}}
\end{figure}

In our search for correlated X-ray and infrared variability, we focus on cluster members that have the most complete infrared and X-ray time-series data. Cluster members, and their stellar properties (L$_*$,T$_{\rm eff}$, etc.), are drawn from \citet{luh03,lad06,mue07,luh05} as compiled by \citet{fla13}. There are a total of 133 cluster members detected on more than one X-ray epoch, with 107 detected on more than four epochs. We restrict ourselves to these 107 stars because they have enough epochs for studying variability. While we include X-ray fluxes from previous studies of IC 348 \citep{pre02,ale12}, our analysis focuses on the data that are contemporaneous with the infrared data. As a sub-sample of the cluster, the X-ray detected sources are incomplete below L=0.3L$_{\odot}$ and T$_{eff}$=3600 K (Fig~\ref{demographics}). Within IC 348, the X-ray luminosity is typically 3 to 4 orders of magnitude lower than the bolometric luminosity \citep{pre02,ale12}. X-ray emission is known to be brighter among sources without active accretion \citep{ste01,fla03,pre05} and in our sample this will present itself as a detection efficiency that is a function of the infrared SED slope (defined as the logarithmic slope of $\lambda F_{\lambda}$ versus $\lambda$ over the 3-8\micron\ range) since the presence of a circumstellar disk as measured with the infrared excess is correlated with accretion. Using the dereddened flux, the SED slope varies from $\alpha_{IRAC}<-2.56$ for sources without a disk to $\alpha_{IRAC}>0$ for sources with an envelope and we use $\alpha_{IRAC}=-2.56$ as the boundary between a source with or without a circumstellar disk.  \citet{ste12} find that within IC 348 the median X-ray luminosity among stars with disks is $\sim$0.3 dex lower than comparable stars without disks. Figure~\ref{demographics} shows the distribution of SED slope for X-ray detections compared with the entire sample, demonstrating our inefficient sensitivity to sources with disks. For the majority of our analysis we restrict ourselves to the sources with a disk. 

The presence of infrared variability is assessed using the techniques described in \citet{fla13}. In short, we use the reduced chi-square statistic for each band, and well as the stetson index S, which is sensitive to correlated variability in the two IRAC bands. The use of the stetson index does require us to restrict the analysis to sources detected in both bands, but the added sensitivity of this index compared to the reduced chi-square offsets the reduction in sample size. A star is marked as an infrared variable if $\chi^2_{\nu}([3.6])>3$ or $\chi^2_{\nu}([4.5])>3$ or S$>$0.45. Based on prior analysis, we are sensitive to $>99\%$ of fluctuations down to 0.04 mag for sources brighter than 14th magnitude \citep{fla13}; this sensitivity limit is relatively independent of source brightness because most of the uncertainty is due to systematic effects for these bright targets \citep{cod14}. While Spitzer monitoring data is available from multiple observing campaigns, we focus on the contemporaneous fall 2011 data when assessing variability since we are mainly concerned with the behavior during the same time frame as any X-ray fluctuations. The vast majority of stars variable in one season are also variable in other seasons, and our results would not significantly change if we included the entire dataset. Since the 3-5$\micron$ emission probed by this photometry is dominated by the inner edge of the dusty disk \citep{mcc13}, we are not sensitive to the dynamics of the regions of the disk beyond this point.

\begin{figure}
\includegraphics[scale=.5]{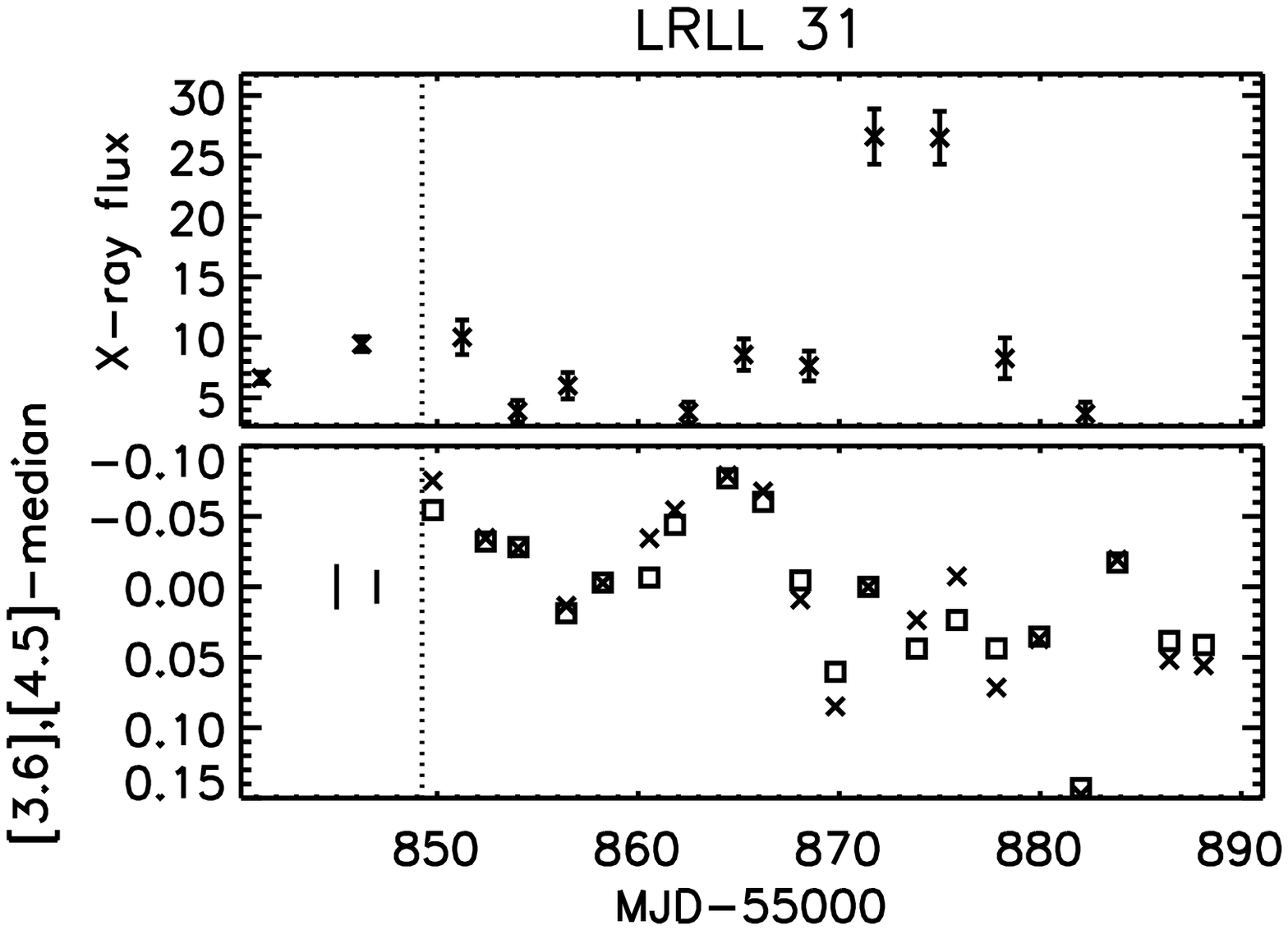}
\includegraphics[scale=.5]{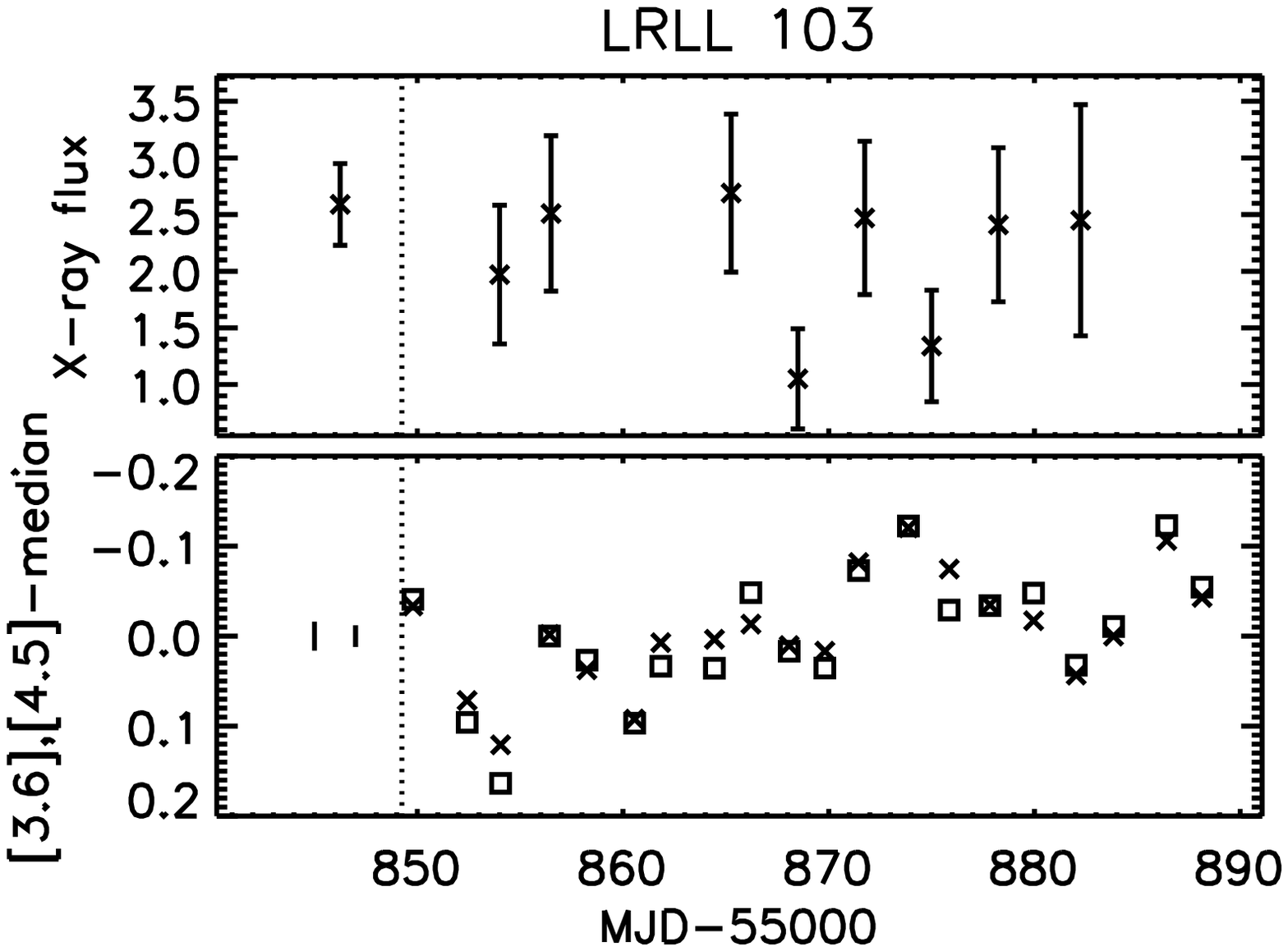}
\includegraphics[scale=.5]{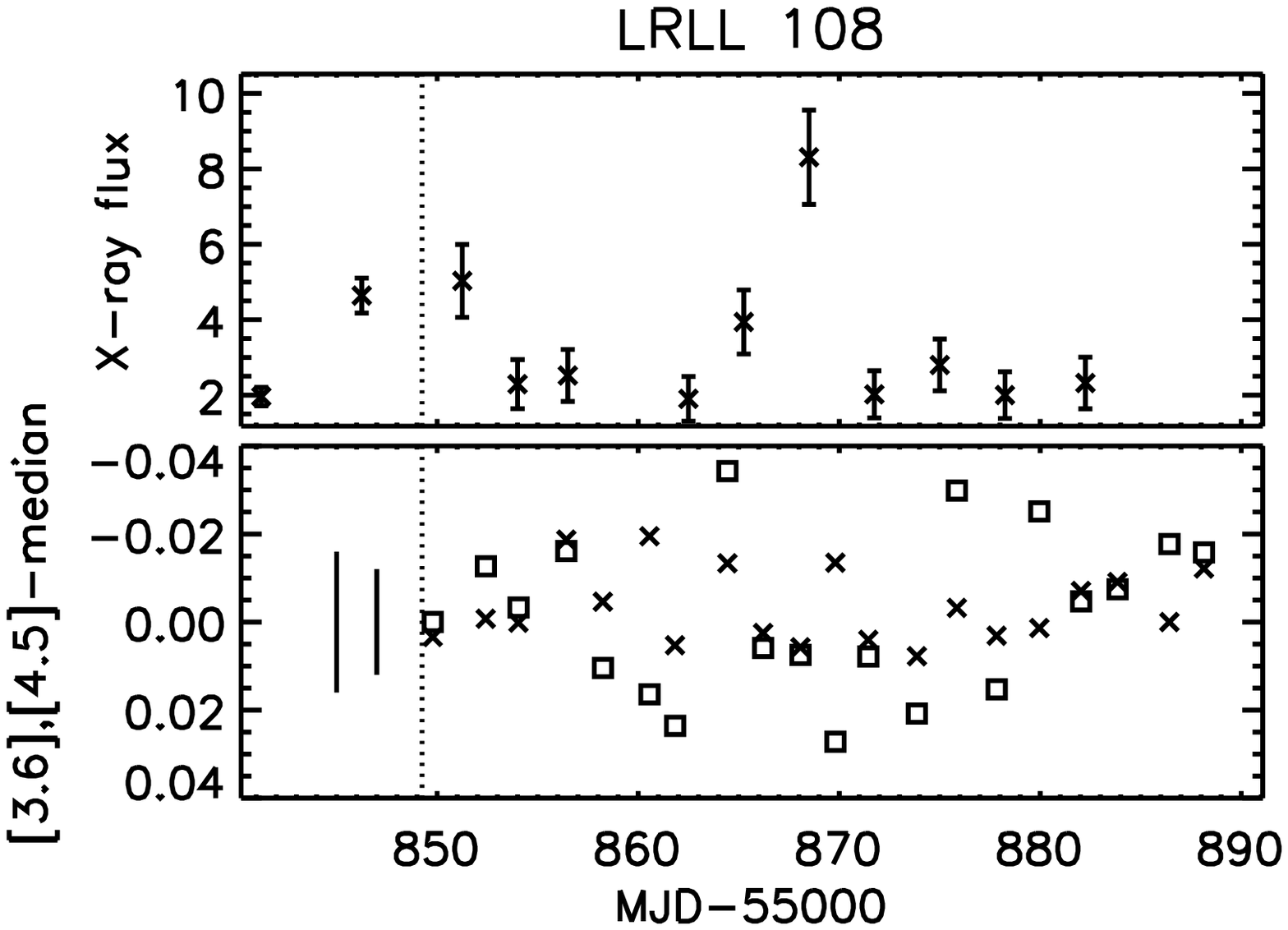}
\includegraphics[scale=.5]{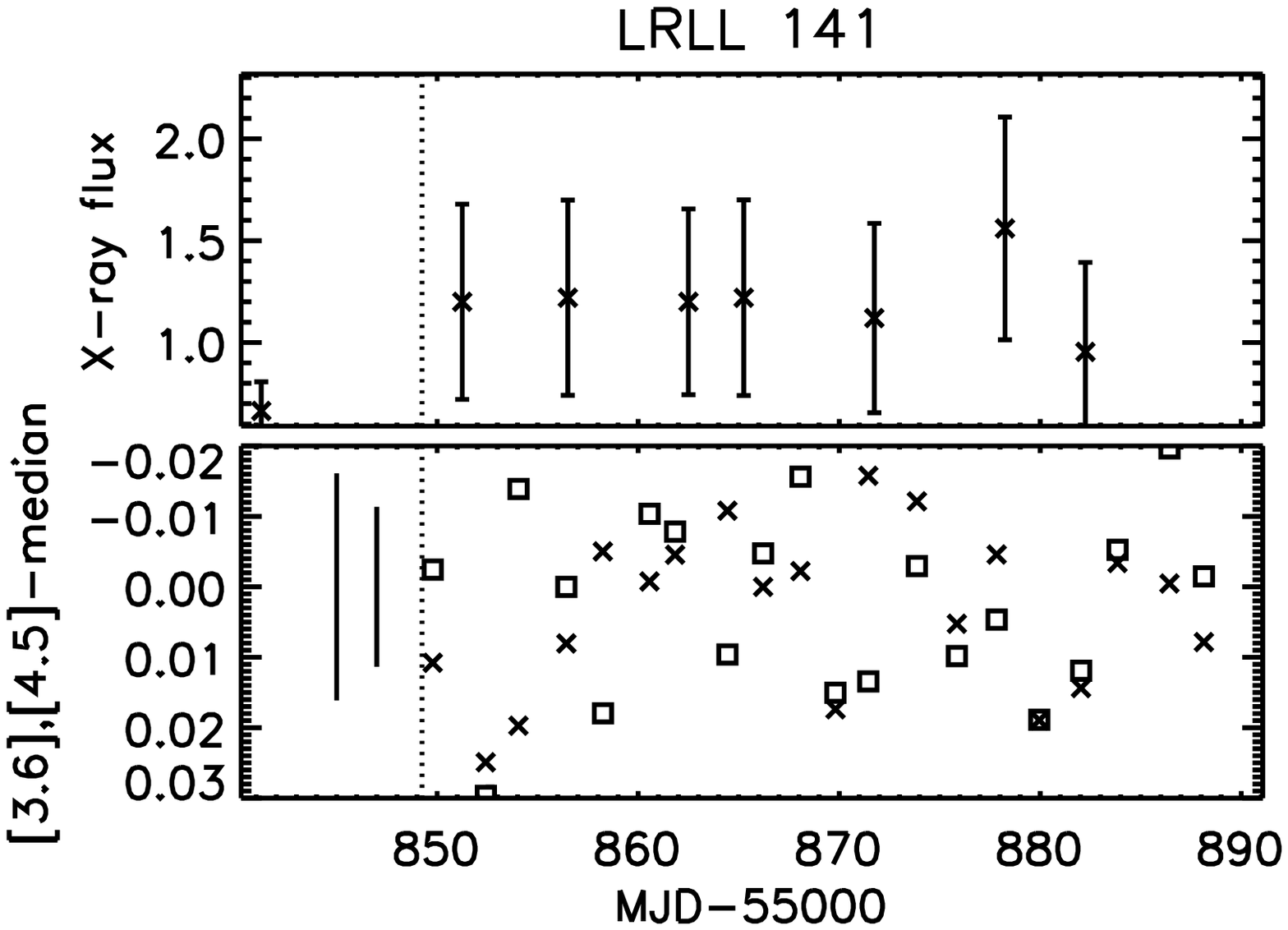}
\caption{Examples of X-ray and infrared light curves with each quadrant presenting a different star. Top panels show the X-ray light curve, in units of 1e-6 photons per cm$^2$ per sec. Points to the left of the vertical dashed line are derived from earlier Chandra observations of IC 348 \citep{pre02,ale12}. The infrared light curves are shown in the corresponding bottom panels. Error bars for [3.6] (squares) and [4.5] (crosses) photometry are shown as the left and right vertical lines. {\bf Upper Left:} LRLL 31, a source that is variable in both X-ray and infrared. {\bf Upper Right:} LRLL 103, a star that is variable in the infrared, but not in the X-ray. {\bf Lower left:} LRLL 108, a star that is variable in the X-ray but not the infrared. {\bf Lower right:} LRLL 141, a star that is not variable in either the X-ray or infrared. Light curves for all stars with disks are included in Figure~\ref{all_lc}.\label{lc_examples}}
\end{figure}

Given that we are mainly concerned with the change in X-ray flux, rather than its overall level, we do not convert photon fluxes to luminosities, which requires assumptions about the distance to the cluster and the underlying X-ray spectrum. While the arrival time of each photon is recorded, we instead use the photon flux averaged over each 10ks observing run, with each epoch comprising one point on our light curve. We chose to do this to increase sensitivity at the expense of knowledge about the X-ray variability on very short timescales.  We restrict ourselves to sources detected on at least four X-ray epochs. Photon fluxes range from 1e-6 to 1e-4 photons/s/cm$^2$, or 1 to 1200 net photons per 10ks block. We can roughly estimate the X-ray luminosities by assuming an average photon energy of 1.5keV, to convert from photon flux to energy flux, and a distance of 320pc; we find luminosities range from $\sim$10$^{28}$ to  $\sim$10$^{31}$ erg/sec with a typical luminosity of  $\sim$10$^{29}$erg/sec similar to previous studies \citep{ale12,ste12}. Our final sample consists of 39 stars (Table~\ref{var_table}) with circumstellar disks with sufficient X-ray photometry to examine variability, all of which were detected in the infrared. Figure~\ref{lc_examples} show examples of light curves for stars with and without X-ray and infrared variability to demonstrate the fidelity of our data.

Our sensitivity to infrared variability is fairly uniform over our entire sample because the uncertainties are dominated by systematics for the vast majority of sources. This is not the case for the X-ray emission whose uncertainties are dominated by poisson statistics and scale with the square root of the flux. This leads to substantially nonuniform sensitivities throughout our sample. As with the infrared flux we use the reduced chi-square to select variable X-ray emission, using $\chi^2_{\nu}=3$ as the boundary between variables and non-variables. This picks out variations that are significant relative to the noise, which varies substantially between the brightest and the faintest sources. As a result, the size of the fluctuations selected as variable will strongly depend on the mean X-ray flux. On average this $\chi^2_{\nu}$ boundary corresponds to a factor of 2 increase in the flux relative to the mean with fluctuations reaching close to a factor of 10. Despite the non-continuous nature of our observations and the use of different selection criteria for selecting variables as compared to these previous studies, we pick out stars with similar fluctuations. \citet{fla12} find that among stars with disks, on timescales of a week, fluctuations are typically a factor of 3, similar to our results. Overall our sample consists of 39 sources with (1) detections on four or more X-ray epochs, (2) detections in both IRAC bands on all epochs, (3) an infrared excess indicative of a disk. Among these 39 stars, 31 have significant fluctuations in the infrared, while 28 have detectable X-ray variability. 

Our observing cadence was designed to probe long term effects of X-ray fluctuations on disk emission, at the expense of knowledge about any short, instantaneous changes in disk emission. While very rapid changes in infrared emission have been observed \citep{sta14}, week to month long variations make up a substantial fraction of the infrared variability \citep{fla13,cod14}. The time between an X-ray epoch and the nearest infrared observation ranges from 0.1 to 2.1 days (12-180 ks) and we are completely insensitive to changes in infrared emission on timescale shorter than this. We instead focus on responses that cover more than 2 infrared epochs ($>$300ks = 4 days). This caveat should be kept in mind when interpreting our results, especially as they apply to physical connections between X-ray and infrared variability.

\subsection{The lack of correlation between X-ray and Infrared variability}

We first look at the 28 stars with disks that are variable in both the X-ray and infrared to see if the light curves themselves are correlated. This is assessed using the Kendall's tau and Spearman's rho statistics. Both of these coefficients serve as non-parametric tests of whether increases or decreases in the X-ray are mirrored in infrared.  We find no significant correlation between the X-ray and infrared light curves suggesting that there is no strong correlation between the infrared and X-ray variability. We also find that the size of the infrared fluctuations is not correlated with the size of the X-ray fluctuations among these variable stars. 

\citet{fla10} find correlated variations in optical flux and soft, but not hard, X-ray flux of CTTS in NGC 2264 between observations separated by two weeks, suggesting that there is a difference between the behavior of the soft and hard X-ray emission. We examine how the median X-ray photon energy changes over the course of the light curve to see if it changes in concert with the infrared emission. If the soft X-ray emission tracks the infrared emission while the hard X-ray flux is constant, then the median photon energy will decrease as the infrared emission increases. Given that we typically only detect 24 photons per source per 10ks block we use the median flux instead of just the soft X-ray photons to reduce the errors and maximize our sensitivity to fluctuations. Using both the Kendall's tau and Spearman's rho metrics, we do not find a correlation between the median photon energy and the infrared flux in any of our disk-bearing targets. We also find no significant correlation in the size of the infrared variability and the size of the variability in median photon energy. Among the two brightest X-ray sources, LRLL 2 and 6, for which we measure $\sim$400 photons per 10 ks, while there is some evidence for a change in median photon energy this change does not track with the infrared emission.  %12/36 disk-bearing sources show variability in median photon energy. 3/30 diskless sources show variability in median photon energy.

\begin{figure}
\includegraphics[scale=.5]{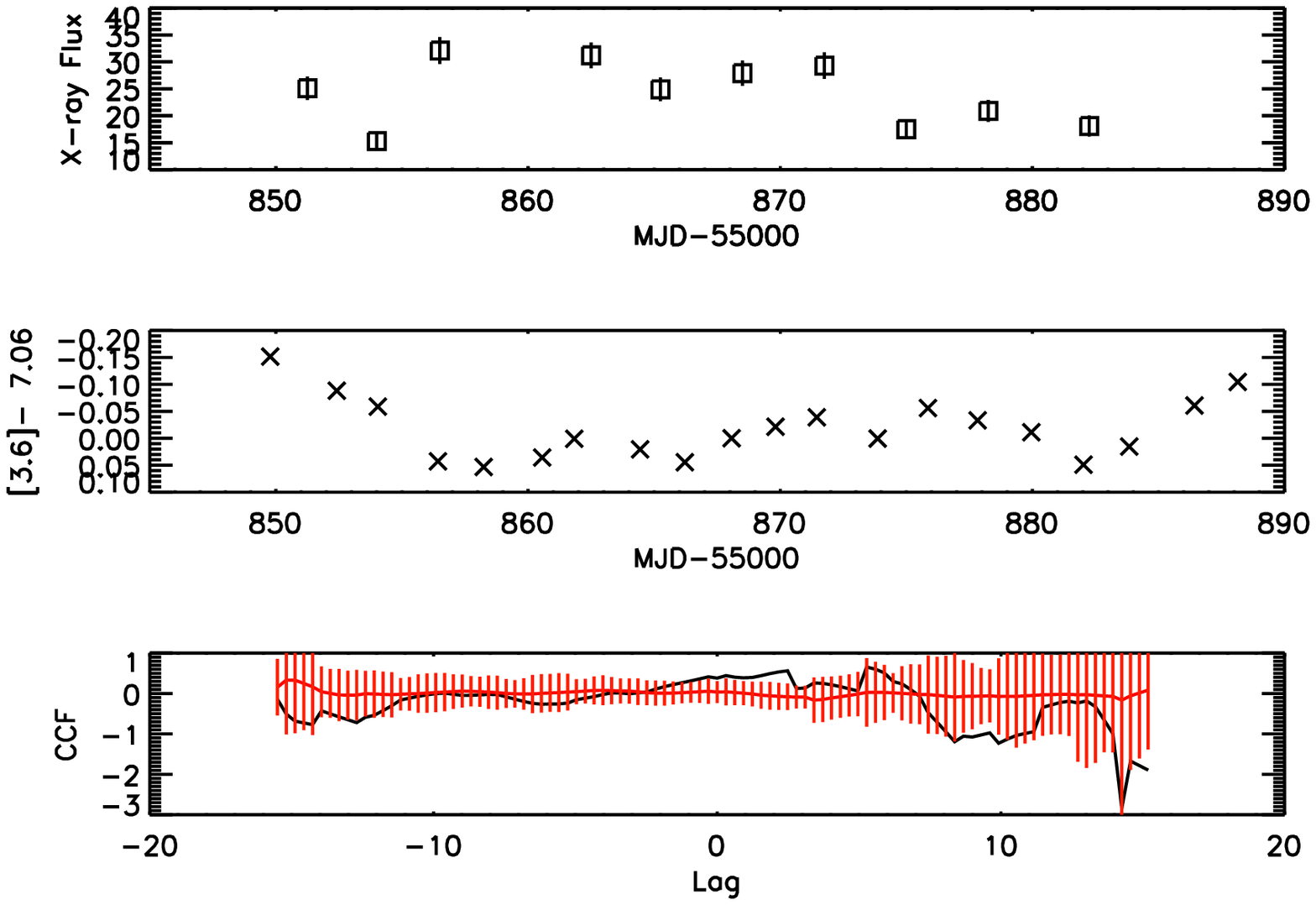}
\includegraphics[scale=.5]{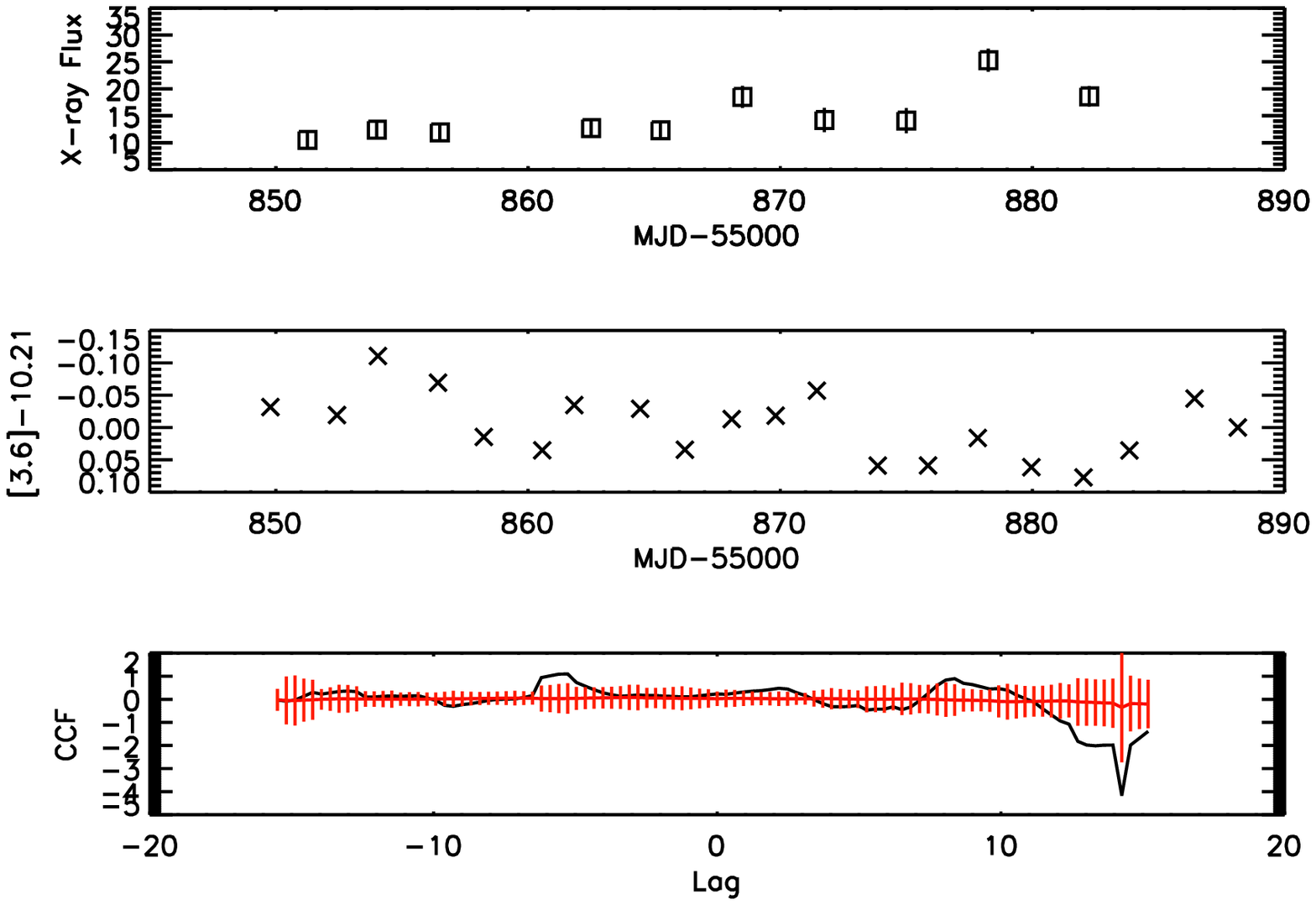}
\caption{Two examples (LRLL 5 on the left and LRLL 58 on the right) of our cross-correlation analysis. The top row is the X-ray light curve while the middle row is the infrared light curve. The bottom row shows the cross-correlation (black line) as a function of the lag in days between the X-ray and infrared light curves. The red lines are our one-sigma uncertainties on the cross-correlation. In these two examples, as well as the rest of the light curves, there is no significant cross-correlation signal at any lag, indicating that the X-ray and infrared light curves do not have similar shapes with a simple time delay between their fluctuations.\label{ccf}}
\end{figure}

The Kendall's tau and Spearman's rho statistics are effective at selecting correlated variations when there is no time-delay between the infrared and X-ray fluctuations and if the light curve is monotonic.  \citet{fla05} find that roughly a tenth of the stars in Orion with known optical periods have detectable X-ray periods and any corresponding, but phase shifted, variability in the infrared would be missed by the previous metrics. To account for a time delay between the two light curves we employ the cross-correlation function, defined as:
\begin{equation}
CCF(\tau)=\frac{1}{N}\Sigma^N_{i=1}\frac{[x(t_i)-\bar{x}][y(t_i-\tau)-\bar{y}]}{\sigma_x\sigma_y}
\end{equation}
where $x(t_i)$ and $y(t_i)$ represent the X-ray and infrared light curves, $\bar{x}$, $\sigma_x$, $\bar{y}$, $\sigma_y$ are the respective means and standard deviations of the two light curves and $\tau$ is the time lag between the two light curves \citep{gas87}. This effectively shifts the infrared light curves by $\tau$ days and calculates how well the two light curves match after this shift. A positive value of $\tau$ implies that the IR light curve leads the X-ray light curve, while a negative lag corresponds to the X-ray light curve leading the IR light curve. Such an effect might be expected if there is an appreciable time delay between the occurrence of an X-ray fluctuation and the response of the disk or if the X-ray and infrared flux show a rotational modulation that is out of phase. Given that the data are not continuously sampled, we interpolate the more heavily sampled infrared light curve onto the X-ray light curve in our analysis. We assess the uncertainty in the CCF at different lags by computing the CCF of a particular X-ray light curve with infrared light curves from 100 random cluster members, and use the dispersion in the value of the CCF at each lag as its uncertainty. Figure~\ref{ccf} shows our results for LRL 5 and 58 as examples. When examining all 28 cluster members with variability in both the X-ray and infrared, we find no evidence of a strong cross-correlation signal between a lag of -20 and 20 days, with our strongest sensitivity to lags between -10 and 10 days. Again there is no strong evidence for a direct connection between the infrared and X-ray variability.

\begin{figure}
\includegraphics[scale=.5]{lrll90_xray.eps}
\includegraphics[scale=.5]{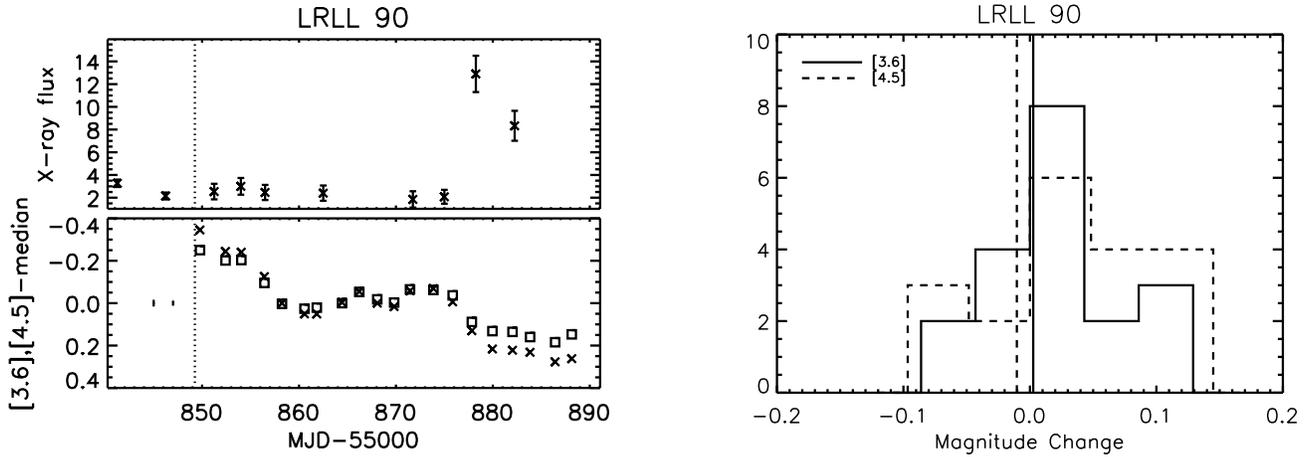}
\caption{(Left) X-ray (top) and infrared (bottom) light curves for LRLL 90, a star which has a bright X-ray flare. X-ray points to the left of the dotted line are from previous surveys. Infrared error bars for [3.6] and [4.5] are shown as vertical tick marks (left and right respectively) to the left of dotted line. While the star is variable in the infrared, its variability does not seem to change in response to the X-ray flare.  (Right) Distribution of epoch-to-epoch changes in [3.6] (squares) and [4.5] (crosses) magnitude for LRLL 90. The change experienced between the two epochs surrounding the X-ray flare are shown as vertical lines. The IR flux variation experienced due to the X-ray flare is not significantly different than that seen during other epochs. A similar behavior is seen in the other stars that have an X-ray flare, indicating that the observed X-ray flares do not drive large changes in infrared flux.\label{flare}}
\end{figure}

The CCF, while effective at picking out light curves with a time lag, breaks down in the presence of flares.  \citet{ste07} find that roughly a quarter of stars within Taurus exhibit flares, with these flaring sources making up half of the variable sample. While these outlier points within a light curve can easily disrupt any underlying correlation, they could also be a source of changes in infrared emission if the flare is powerful enough. We define a source as having a flare if it is variable in the X-ray ($\chi^2_{\nu}>3$) and it has one epoch whose flux is greater than three times larger than the median flux.  We can also look for X-ray flares within a single 10 ks  epoch among the brightest sources since each photon is tracked when taking X-ray observation. Only one source (LRLL 36) shows evidence for an X-ray flare during one of these epochs, and this star has been included in our analysis. There are a total of 13 cluster members with an X-ray flare, all but one of which is variable in the infrared. In none of these stars do we see strong evidence for an extended change in the IR flux following the X-ray flare. Figure~\ref{flare} shows the X-ray and infrared light curves of one star with an X-ray flare demonstrating the lack of response in the infrared light curve. If the infrared flux is changing before the flare, it continues along its course after the flare. 

We examine in detail the change in IR flux associated with an X-ray flare by calculating the change in [3.6] and [4.5] magnitude between the two epochs (i.e. about 2 days) surrounding the X-ray flare and comparing this to all consecutive epoch magnitude changes in the infrared light curve. Figure~\ref{flare} shows an example of this analysis. This allows us to quantify the infrared fluctuation closest to the X-ray flare and how it compares to the fluctuations seen at other times. In none of the light curves do we see evidence that the IR fluctuation near the X-ray flare is substantially different than what is seen in any other part of the light curve. This is consistent with our visual inspection which found no evidence for a change in infrared flux in response to an X-ray flare. 

Again we note that given our observing cadence we are severely limited in assessing the immediate response of the infrared flux to an X-ray flare. The time between an X-ray epoch and an infrared epoch ranges from 0.1 to 2.1 days (12-180ks). The majority of X-ray flares have decay times less than 60ks with an average decay time of 10ks \citep{ste07} and almost all X-ray flares would have decayed away by the time the system was observed in the infrared. YSOs typically exhibit one flare per star per 500ks with the time between bright flares (L$_X>10^{32}$erg/sec) a factor of a few longer than this \citep{ste07}. \citet{wol05} find similar results for solar-type stars, with roughly one flare per star per 650ks. A theory for the week to month long infrared variability that relies solely on X-ray flares would require the infrared response to last at least $\sim$500ks(=5.8days) to extend the infrared variability throughout the quiescent period between flares. Even with our limited cadence we can exclude these types of large, long infrared responses.

\section{Physical Mechanisms}
We find no direct connection between X-ray fluctuations and infrared variability. There is no clear correlation between the shapes of the infrared and X-ray light curves and there are no large, long-term changes in infrared emission following X-ray flares. This is consistent with the lack of correlated X-ray/infrared variability seen in the Coronet cluster \citep{for07}. 
Given the cadence of our observations we cannot comment on short (less than a few days), immediate responses of the disk to the changing X-ray flux; we can only constrain the type of interactions that lead to sustained week-long fluctuations. This caveat should be kept in mind in the context of the different physical situations that we discuss below. 

Interpreting this result depends on how X-ray variability could be physically connected to changes in disk structure. Our [3.6],[4.5] infrared observations are dominated by emission from the inner edge of the disk, which is made up of a slightly rounded wall of dust a few tenths of an AU from the star at the boundary where the dust becomes hot enough to sublimate \citep{kam09}. There are three main possibilities connecting X-ray variability to changes in the structure of this wall, which we will consider in turn: (1) heating of dust by X-ray photons, (2) ionization of dust by X-ray photons, (3) interactions between the circumstellar disk and the hot plasma created during the flare

{\it X-ray Heating: }
A rapid increase in the stellar flux illuminating the disk would lead to an almost instantaneous change in the temperature of the disk surface layers \citep{chi97}, causing the infrared emission to increase. Our observations show no evidence for a long-term change in infrared emission in response to changes in X-ray flux, suggesting that heating by X-rays is not an important factor in setting the dust temperature. This is not surprising since models predict that much of the X-ray flux will be absorbed by gas in the tenuous upper layers of the disk \citep{gla04,are11} and dust emission models are generally able to reproduce circumstellar disk emission in a wide range of stars without including illumination from stellar X-rays \citep{dal06}. In IC 348 the ratio of X-ray luminosity to bolometric luminosity ranges from 10$^{-3}-10^{-4.5}$ \citep{ste12} again suggesting that X-ray heating does not contribute significantly to the dust temperature.

{\it X-ray Ionization}
As with heating, an increase in X-ray flux may increase the ionization fraction of the dust through the increased collisions with electrons and ions \citep{tie05}, leading to observable dynamical effects due to interactions between this ionized material and the magnetic field. Even if X-ray heating is minimal, the increased depth at which X-ray ionization can penetrate the disk \citep{ige99} leaves open another avenue for X-ray photons to change the dust structure. \citet{ke12} provide detailed models of such a scenario and find that a large flare (L$_X\sim10^{32}$erg/s) that lasts for a day (=86ks) can substantially change the height of the inner disk. This effect was predicted to decay over the course of a week, leading to extended perturbations to the infrared emission. \citet{fav05} find that eight out of 32 flares have L$_X>10^{32}$erg/s while \citet{wol05} find similar numbers for the occurrence of large flares. In our sample we find no strong evidence for an extended response to an X-ray although we do not observe any flares as large as employed in these models and cannot comment on any immediate response to such outbursts.

{\it High temperature plasma}
X-ray flares are associated with magnetic reconnection events injecting magnetic energy into the upper reaches of the stellar atmosphere \citep{ben10}. This energy creates a $\sim$100 MK plasma that expands into the surrounding coronal loop, injecting thermal energy into the area surrounding the star, decaying on a timescale that is proportional to the size of the magnetic loop \citep{fav05}. Based on detailed modeling of the decay time and emission properties of X-ray flares in Orion Favata et al. find that 9 out of 22 flares reach larger than 5R$_*$ and could potentially interact with the disk in systems with very close in dust. \citet{orl11} model a different type of flare that originates at the interface between the disk and the stellar magnetic field that provides a direct interaction between the flare and the disk structure. In this model the disk height can change by a factor of two in response to a L$_X\approx10^{32}$erg/s sized flare. We see no evidence for long-term reactions to X-ray flares suggesting that these types of interactions do not drive the majority of the observed infrared variability.

\section{A brief discussion of variability among diskless sources}
As with the disked sources, in the diskless stars we see no evidence for a correlation between the X-ray and infrared variability. Without excess emission, the infrared flux in the diskless stars is tracing the stellar flux rather than the dust in the inner disk. Stellar variability in diskless stars is dominated by the rotation of star spots, which may be located at sites of amplified magnetic field strength. We see no response in the infrared light curve following X-ray flares, similar to the systems with protoplanetary disks. Analysis with the cross-correlation function as well as Kendall's tau and Spearman's rho statistics, confirms the lack of similarity in light curve shapes. Compared to stars with disks, diskless stars show weaker fluctuations in the infrared \citep{fla13} and in the X-ray \citep{fla12}. Previous studies have found that while X-ray and optical periods are sometimes similar \citep{fla05} there is often not a direct correlation between the X-ray and optical fluctuations \citep{sta07,fla10}. \citet{sta07}  conclude that even though X-ray production is connected to the presence of a magnetically active, spotted surface, as evidenced by the connection between optical variability and X-ray luminosity, the regions of X-ray production and the star spots are not spatially coincident.

\section{Conclusion}
We find no evidence for an increased frequency of infrared variability among stars with X-ray variability. We see no correlation in the shape of the light curves in stars that exhibit variability in both bands, which is especially true in the stars with moderate X-ray flares for which we see no evidence for a change in infrared flux in response to the flare. We put limits on the relation between X-ray variability and the week to month long infrared variability of dust at the inner edge of the disk that is common among young stellar objects. 

Based on the observational evidence, the influence of X-ray heating on the temperature of the dust close to the star is relatively minor, which in turn has implications for the overall structure of the inner edge of the disk.  Below z/r$\sim$0.1 the densities are high enough that the gas and dust temperature are highly coupled through collisions \citep{gla04}, allowing the dust temperature to set the gas temperature. The gas temperature in turn sets the pressure scale height of the disk, which is the basis of the vertical structure of the disk in a region where planet formation is expected to occur. If the X-rays cannot influence the dust thermal structure, then they will not influence the overall structure in this area of the disk. Compared to FUV and EUV photons, X-rays have a higher penetration depth because the opacity of the gas decreases towards higher energies and if X-ray photons cannot reach the dusty layers, then FUV and EUV photons contribute even less to the dust temperature. The dust temperature is instead set by a combination of viscous dissipation and irradiation by stellar optical emission \citep{dal99}, at least until the disk is significantly depleted \citep{ski13}. Beyond $\sim$1 AU, which is not probed by our infrared photometry, the X-rays may still have a strong influence on disk structure through e.g. MRI turbulence \citep{ige99}.

In terms of constraining the origin of the long-term infrared fluctuations, we can rule out the influence of X-ray ionization as well as interactions between the X-ray emitting plasma and the dust at the sublimation front close to the star from explaining the majority of the observed infrared variability. These factors may come into play on short timescales for a small handful of stars, or at levels below our detection limits, but generally they are not important in setting disk structure. More likely possibilities include perturbations related to accretion heating \citep{fla13}, MRI driven turbulence lifting dust out of the midplane \citep{tur10} or perturbations from a companion \citep{fra10,nag12,ata13} with observations pointing toward accretion bursts, variable obscuration of the central star and changes in the inner disk structure \citep{cod14,sta14,wol13}.

\acknowledgements
We would like to thank the anonymous referee for their substantive comments that greatly improved this paper. The scientific results reported in this article are based in part on observations made by the Chandra X-ray Observatory. Support for this work was provided by the National Aeronautics and Space Administration through Chandra Award Number G02-13028A issued by the Chandra X-ray Observatory Center, which is operated by the Smithsonian Astrophysical Observatory for and on behalf of the National Aeronautics Space Administration under contract NAS8-03060. This work is based in part on observations made with the Spitzer Space Telescope, which is operated by the Jet Propulsion Laboratory, California Institute of Technology under a contract with NASA. Support for the work was provided by NASA through an award issued by JPL/Caltech. RG gratefully acknowledges funding support from NASA ADAP grants NNX11AD14G and NNX13AF08G and Caltech/JPL awards 1373081, 142329 and 1440160 in support of Spitzer Space Telescope observing programs.

\begin{deluxetable}{ccccc}
\tablewidth{0pt}
\tablecaption{X-ray Observing Log\label{xray_log}}
\tablehead{\colhead{Obsid} & \colhead{Exposure start time} & \colhead{MJD} & \colhead{Time (ksec)} & \colhead{Roll Angle}}
\startdata
13425&  	2011-10-17 06:11:07  & 55851.3 &  9.91&    	118.1\\
13426&	2011-10-19 22:05:11  & 55853.9 & 9.91&	119.5\\
13427& 	2011-10-22 13:31:32   & 55856.5 & 9.91&	121.1\\
13429&	2011-10-28 09:52:52&	55862.4 & 10.41&	125.6\\
13430&	2011-10-31 06:54:51  & 55865.3 &  9.91&	128.3\\
13431&	2011-11-03 09:45:09   & 55868.4 & 9.92&	131.9\\
13432&	2011-11-06 18:22:08   & 55871.8 & 9.91&	136.7\\
13433&	2011-11-10 01:35:24  & 55875.0 & 10.79&	142.9\\
13434&  	2011-11-13 06:48:37 &  55878.3 &  9.91&    	150.6\\
13428&	2011-11-17 05:34:13  & 55882.2 &  9.22&	163.2\\
\enddata
\end{deluxetable}

\begin{deluxetable}{ccc}
\tablewidth{0pt}
\tablecaption{Infrared Observing Log\label{obs_log}}
\tablehead{\colhead{AORID}&\colhead{Exposure start time}&\colhead{MJD}}
\startdata
42463744 & 2011-10-15 18:22:53 & 55849.8\\
42464000 & 2011-10-18 09:44:53 & 55852.4\\
42464256 & 2011-10-20 00:45:16 & 55854.0\\
42464512 & 2011-10-22 09:56:54 & 55856.4\\
42464768 & 2011-10-24 05:31:16 & 55858.2\\
42465024 & 2011-10-26 13:08:49 & 55860.6\\
42465280 & 2011-10-27 19:42:07 & 55861.8\\
42465536 & 2011-10-30 10:18:35 & 55864.4\\
42465792 & 2011-11-1 04:52:37 & 55866.2 \\
42466048 & 2011-11-3 01:27:30 & 55868.1 \\
42466304 & 2011-11-4 19:14:03 & 55869.8 \\
42466560 & 2011-11-6 10:30:23 & 55871.5 \\
42466816 & 2011-11-8 20:29:04 & 55873.9 \\
42467072 & 2011-11-10 20:15:18 & 55875.9\\
42467328 & 2011-11-12 19:38:16 & 55877.8\\
42467584 & 2011-11-14 22:39:14 & 55880.0\\
42467840 & 2011-11-17 00:21:17 & 55882.0\\
42468096 & 2011-11-18 19:53:33 & 55883.8\\
42468352 & 2011-11-21 09:56:45 & 55886.4\\
42468608 & 2011-11-23 03:08:30 & 55888.1\\
\enddata
\end{deluxetable}

\begin{deluxetable}{ccccccccc}
\tablewidth{0pt}
\tabletypesize{\scriptsize}
\tablecaption{Variability among protoplanetary disk systems\label{var_table}}
\tablehead{\colhead{LRLL ID\#}&\colhead{RA (J2000)}&\colhead{Dec(J2000)}&\colhead{IR variable?}&\colhead{X-ray variable?}&\colhead{X-ray flare?}&\colhead{T$_{eff}$}&\colhead{L$_*$}&\colhead{$\alpha_{IRAC}$}}
\startdata
2 & 03:44:35.36 & 32:10:04.6 & y & y & n & 8970 & 137 & -1.47\\
5 & 03:44:26.03 & 32:04:30.4 & y & y & n & 5520 & 9.90 & -1.24\\
6 & 03:44:36.94 & 32:06:45.4 & y & y & n & 5830 & 17.0 & -2.08\\
15 & 03:44:44.73 & 32:04:02.6 & y & y & n & 3778 & 1.90 & -1.40\\
21 & 03:44:56.15 & 32:09:15.5 & y & y & y & 5250 & 3.90 & -1.98\\
31 & 03:44:18.16 & 32:04:57.0 & y & y & y & 5945 & 9.60 & -1.85\\
32 & 03:44:37.89 & 32:08:04.2 & y & y & n & 4060 & 1.40 & -1.56\\
36 & 03:44:38.47 & 32:07:35.7 & y & y & y & 4205 & 1.50 & -0.73\\
37 & 03:44:37.99 & 32:03:29.8 & y & y & y & 4205 & 0.99 & -1.03\\
41 & 03:44:21.61 & 32:10:37.6 & y & y & y & 4060 & 0.790 & -1.72\\
55 & 03:44:31.36 & 32:00:14.7 & y & y & y & 3778 & 0.680 & -1.19\\
58 & 03:44:38.55 & 32:08:00.7 & y & y & n & 3669 & 0.720 & -1.73\\
61 & 03:44:22.29 & 32:05:42.7 & y & y & y & 3955 & 0.540 & -1.45\\
65 & 03:44:33.98 & 32:08:54.1 & y & y & n & 3850 & 0.560 & -2.41\\
67 & 03:43:44.62 & 32:08:17.9 & y & y & n & 3741 & 0.480 & -2.04\\
72 & 03:44:22.57 & 32:01:53.7 & y & y & y & 3488 & 0.510 & -2.23\\
82 & 03:44:37.41 & 32:06:11.7 & y & y & n & 4060 & 0.510 & -2.43\\
83 & 03:44:37.41 & 32:09:00.9 & y & y & y & 3705 & 0.510 & -0.84\\
90 & 03:44:33.31 & 32:09:39.6 & y & y & y & 3560 & 0.410 & -2.39\\
91 & 03:44:39.21 & 32:09:44.7 & y & y & y & 3560 & 0.390 & -1.53\\
97 & 03:44:25.56 & 32:06:17.0 & y & y & y & 3524 & 0.540 & -1.75\\
110 & 03:44:37.40 & 32:12:24.3 & y & y & y & 3560 & 0.340 & -1.59\\
116 & 03:44:21.56 & 32:10:17.4 & y & y & y & 3632 & 0.290 & -2.44\\
9024 & 03:44:35.37 & 32:07:36.2 & y & y & n & ... & ... & -1.12\\
\hline
71 & 03:44:32.58 & 32:08:55.8 & y & n & n & 3415 & 0.470 & -2.01\\
75 & 03:44:43.78 & 32:10:30.6 & y & n & n & 3669 & 0.620 & -0.53\\
103 & 03:44:44.59 & 32:08:12.7 & y & n & n & 3560 & 0.380 & -1.16\\
113 & 03:44:37.19 & 32:09:16.1 & y & n & n & 4205 & 0.320 & -1.99\\
186 & 03:44:46.33 & 32:11:16.8 & y & n & n & 3560 & 0.310 & -2.07\\
203 & 03:44:18.10 & 32:10:53.5 & y & n & n & 3741 & 0.021 & -1.42\\
1872 & 03:44:43.31 & 32:01:31.6 & y & n & n & ... & ... & 0.43\\
\hline
88 & 03:44:32.77 & 32:09:15.8 & n & y & n & 3379 & 0.360 & -2.39\\
108 & 03:44:38.70 & 32:08:56.7 & n & y & n & 3379 & 0.240 & -2.44\\
146 & 03:44:42.63 & 32:06:19.5 & n & y & n & 3705 & 0.190 & -2.48\\
226 & 03:44:31.42 & 32:11:29.4 & n & y & y & 3091 & 0.078 & -1.33\\
\hline
52 & 03:44:43.53 & 32:07:43.0 & n & n & n & 3705 & 0.920 & -2.42\\
139 & 03:44:25.31 & 32:10:12.7 & n & n & n & 3161 & 0.200 & -1.22\\
141 & 03:44:30.54 & 32:06:29.7 & n & n & n & 3778 & 0.350 & -2.32\\
1401 & 03:44:54.69 & 32:04:40.3 & n & n & n & ... & ... & 0.10\\
\enddata
\end{deluxetable}

\begin{deluxetable}{ccccccccc}
\tablewidth{0pt}
\tabletypesize{\scriptsize}
\tablecaption{Variability among diskless stars\label{var_table2}}
\tablehead{\colhead{LRLL ID\#}&\colhead{RA (J2000)}&\colhead{Dec (J2000)}&\colhead{IR variable?}&\colhead{X-ray variable?}&\colhead{X-ray flare?}&\colhead{T$_{eff}$}&\colhead{L$_*$}&\colhead{$\alpha_{IRAC}$}}
\startdata
29 & 03:44:31.53 & 32:08:45.0 & y & y & y & 4900 & 2.10 & -2.88\\
45 & 03:44:24.29 & 32:10:19.4 & y & y & y & 4350 & 0.920 & -2.72\\
48 & 03:44:34.88 & 32:06:33.6 & y & y & y & 4278 & 0.990 & -2.70\\
66 & 03:44:28.47 & 32:07:22.4 & y & y & y & 4132 & 0.710 & -2.98\\
69 & 03:44:27.02 & 32:04:43.6 & y & y & n & 3705 & 0.460 & -2.95\\
86 & 03:44:27.88 & 32:07:31.6 & y & y & n & 3560 & 0.460 & -2.73\\
92 & 03:44:32.67 & 32:06:46.5 & y & y & y & 3488 & 0.390 & -2.87\\
125 & 03:44:21.66 & 32:06:24.8 & y & y & n & 3451 & 0.290 & -2.64\\
\hline
62 & 03:44:26.63 & 32:03:58.3 & y & n & n & 3161 & 0.390 & -2.88\\
105 & 03:44:11.26 & 32:06:12.1 & y & n & n & 3850 & 0.390 & -2.88\\
170 & 03:44:28.42 & 32:11:22.5 & y & n & n & 3415 & 0.190 & -2.77\\
\hline
9 & 03:44:29.17 & 32:09:18.3 & n & y & n & 5520 & 11.00 & -2.86\\
16 & 03:44:32.74 & 32:08:37.5 & n & y & n & 5700 & 0.750 & -2.72\\
23 & 03:44:38.72 & 32:08:42.0 & n & y & n & 4730 & 4.10 & -2.96\\
47 & 03:44:55.51 & 32:09:32.5 & n & y & n & 5250 & 1.90 & -2.89\\
50 & 03:44:55.63 & 32:09:20.2 & n & y & y & 4590 & 1.40 & -2.87\\
53 & 03:44:16.43 & 32:09:55.2 & n & y & y & 5250 & 1.40 & -2.80\\
74 & 03:44:34.27 & 32:10:49.7 & n & y & n & 3560 & 0.620 & -2.92\\
79 & 03:45:01.52 & 32:10:51.5 & n & y & n & 5250 & 0.940 & -2.81\\
142 & 03:43:56.20 & 32:08:36.3 & n & y & n & 3850 & 0.330 & -2.91\\
154 & 03:44:37.79 & 32:12:18.2 & n & y & n & 3198 & 0.190 & -2.79\\
171 & 03:44:44.85 & 32:11:05.8 & n & y & n & 3451 & 0.170 & -2.75\\
174 & 03:44:04.11 & 32:07:17.1 & n & y & n & 3632 & 0.180 & -2.86\\
178 & 03:44:48.83 & 32:13:22.1 & n & y & y & 3451 & 0.130 & -2.60\\
184 & 03:44:53.76 & 32:06:52.2 & n & y & y & 3270 & 0.180 & -2.96\\
188 & 03:44:56.12 & 32:05:56.7 & n & y & y & 3451 & 0.130 & -2.73\\
\hline
4 & 03:44:31.19 & 32:06:22.1 & n & n & n & 7200 & 34.00 & -2.72\\
10 & 03:44:24.66 & 32:10:15.0 & n & n & n & 6890 & 13.0 & -2.89\\
30 & 03:44:19.13 & 32:09:31.4 & n & n & n & 7200 & 6.50 & -2.84\\
33 & 03:44:32.59 & 32:08:42.5 & n & n & n & 3488 & 0.870 & -2.58\\
35 & 03:44:39.26 & 32:07:35.5 & n & n & n & 4730 & 3.60 & -2.92\\
38 & 03:44:23.99 & 32:11:00.0 & n & n & n & 6030 & 3.10 & -2.77\\
56 & 03:44:05.00 & 32:09:53.8 & n & n & n & 4660 & 0.930 & -2.87\\
59 & 03:44:40.13 & 32:11:34.3 & n & n & n & 4900 & 1.400 & -2.76\\
98 & 03:44:38.62 & 32:05:06.5 & n & n & n & 3270 & 0.330 & -2.59\\
151 & 03:44:34.83 & 32:11:18.0 & n & n & n & 3560 & 0.260 & -2.83\\
163 & 03:44:35.45 & 32:08:56.3 & n & n & n & 3091 & 0.140 & -2.67\\
\enddata
\end{deluxetable}

\figsetstart
\figsetnum{3}
\figsettitle{X-ray and Infrared light curves}

\figsetgrpstart
\figsetgrpnum{3.1}
\figsetgrptitle{LRLL 2 light curves}
\figsetplot{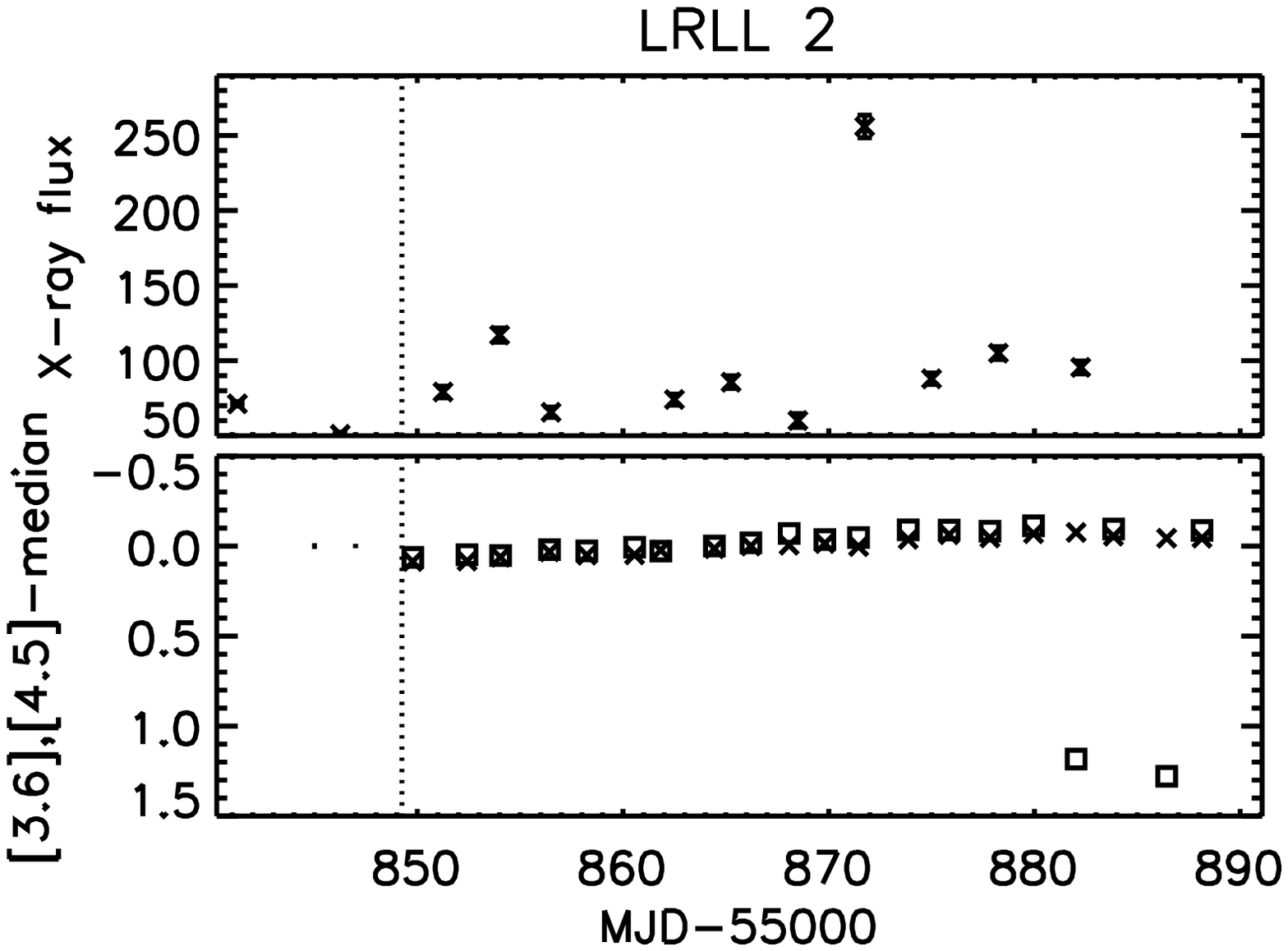}
\figsetgrpnote{X-ray and infrared light curves. The top panel show the X-ray light curve, in units of 1e-6 photons per cm$^2$ per sec. Points to the left of the vertical dashed line are derived from earlier Chandra observations of IC 348 \citep{pre02,ale12}. The infrared light curves are shown in the bottom panel. Error bars for [3.6] (squares) and [4.5] (crosses) photometry are shown as the left and right vertical lines.}
\figsetgrpend

\figsetgrpstart
\figsetgrpnum{3.2}
\figsetgrptitle{LRLL 5 light curves}
\figsetplot{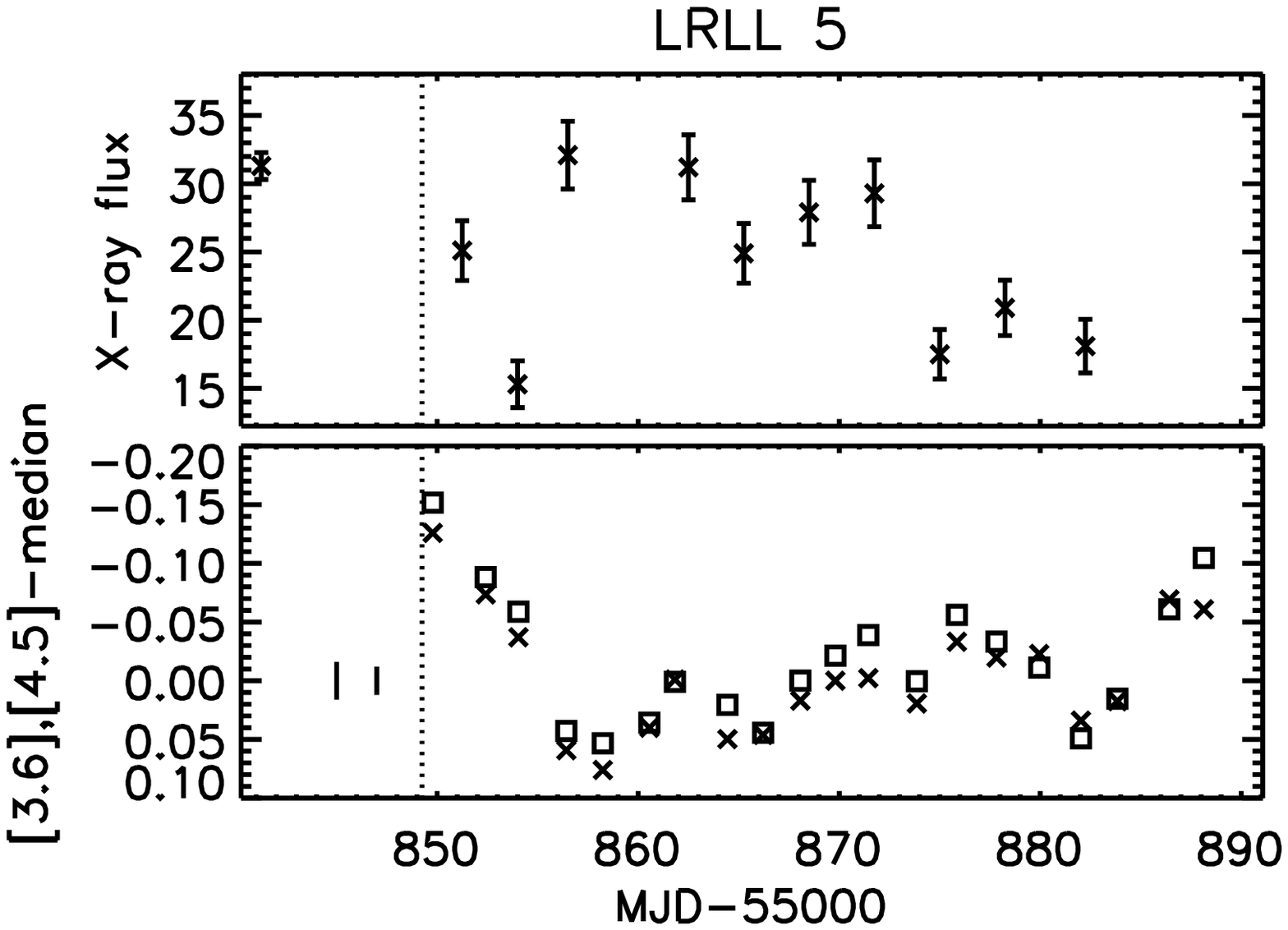}
\figsetgrpnote{X-ray and infrared light curves. The top panel show the X-ray light curve, in units of 1e-6 photons per cm$^2$ per sec. Points to the left of the vertical dashed line are derived from earlier Chandra observations of IC 348 \citep{pre02,ale12}. The infrared light curves are shown in the bottom panel. Error bars for [3.6] (squares) and [4.5] (crosses) photometry are shown as the left and right vertical lines.}
\figsetgrpend

\figsetgrpstart
\figsetgrpnum{3.3}
\figsetgrptitle{LRLL 6 light curves}
\figsetplot{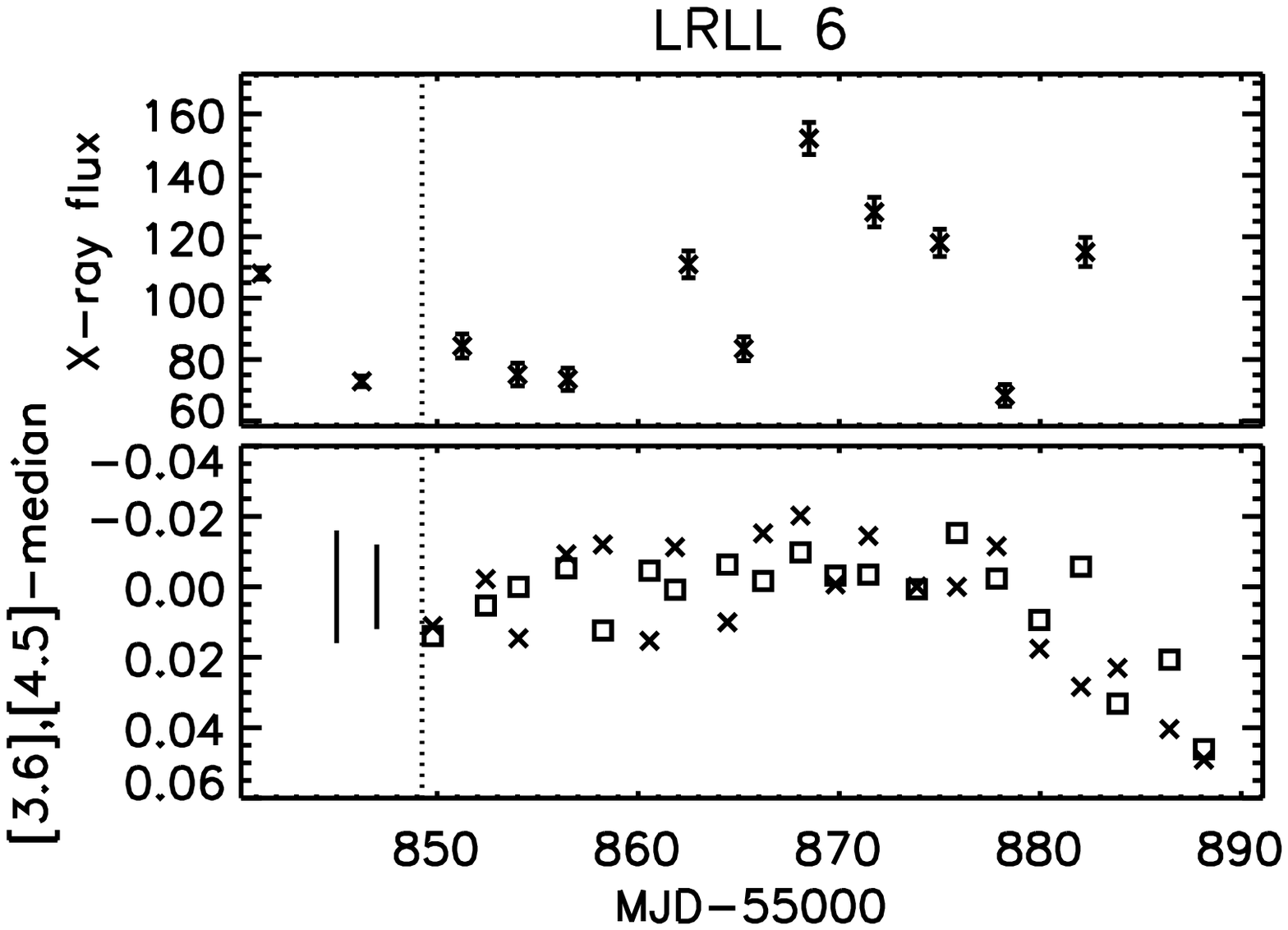}
\figsetgrpnote{X-ray and infrared light curves. The top panel show the X-ray light curve, in units of 1e-6 photons per cm$^2$ per sec. Points to the left of the vertical dashed line are derived from earlier Chandra observations of IC 348 \citep{pre02,ale12}. The infrared light curves are shown in the bottom panel. Error bars for [3.6] (squares) and [4.5] (crosses) photometry are shown as the left and right vertical lines.}
\figsetgrpend

\figsetgrpstart
\figsetgrpnum{3.4}
\figsetgrptitle{LRLL 15 light curves}
\figsetplot{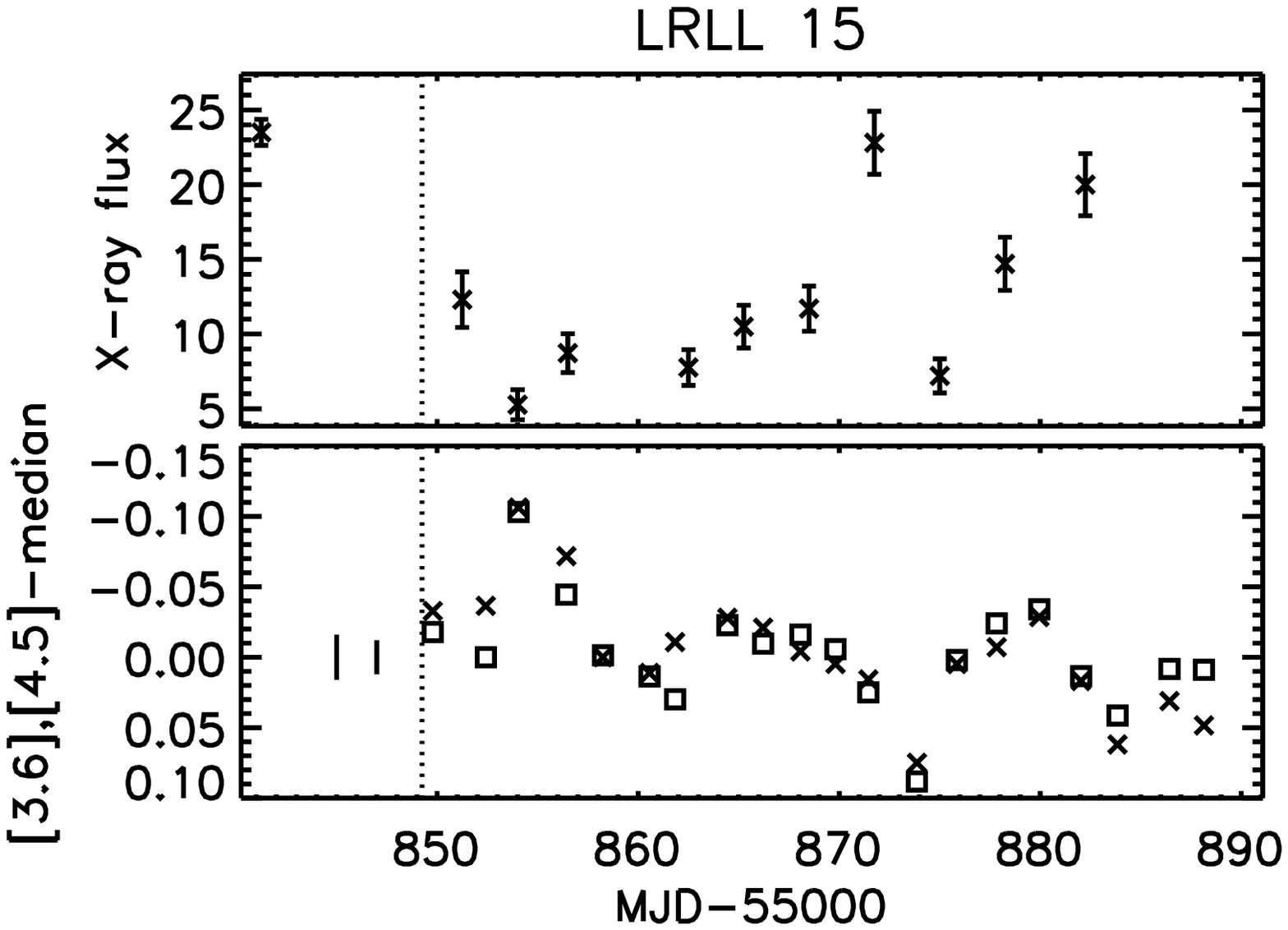}
\figsetgrpnote{X-ray and infrared light curves. The top panel show the X-ray light curve, in units of 1e-6 photons per cm$^2$ per sec. Points to the left of the vertical dashed line are derived from earlier Chandra observations of IC 348 \citep{pre02,ale12}. The infrared light curves are shown in the bottom panel. Error bars for [3.6] (squares) and [4.5] (crosses) photometry are shown as the left and right vertical lines.}
\figsetgrpend

\figsetgrpstart
\figsetgrpnum{3.5}
\figsetgrptitle{LRLL 21 light curves}
\figsetplot{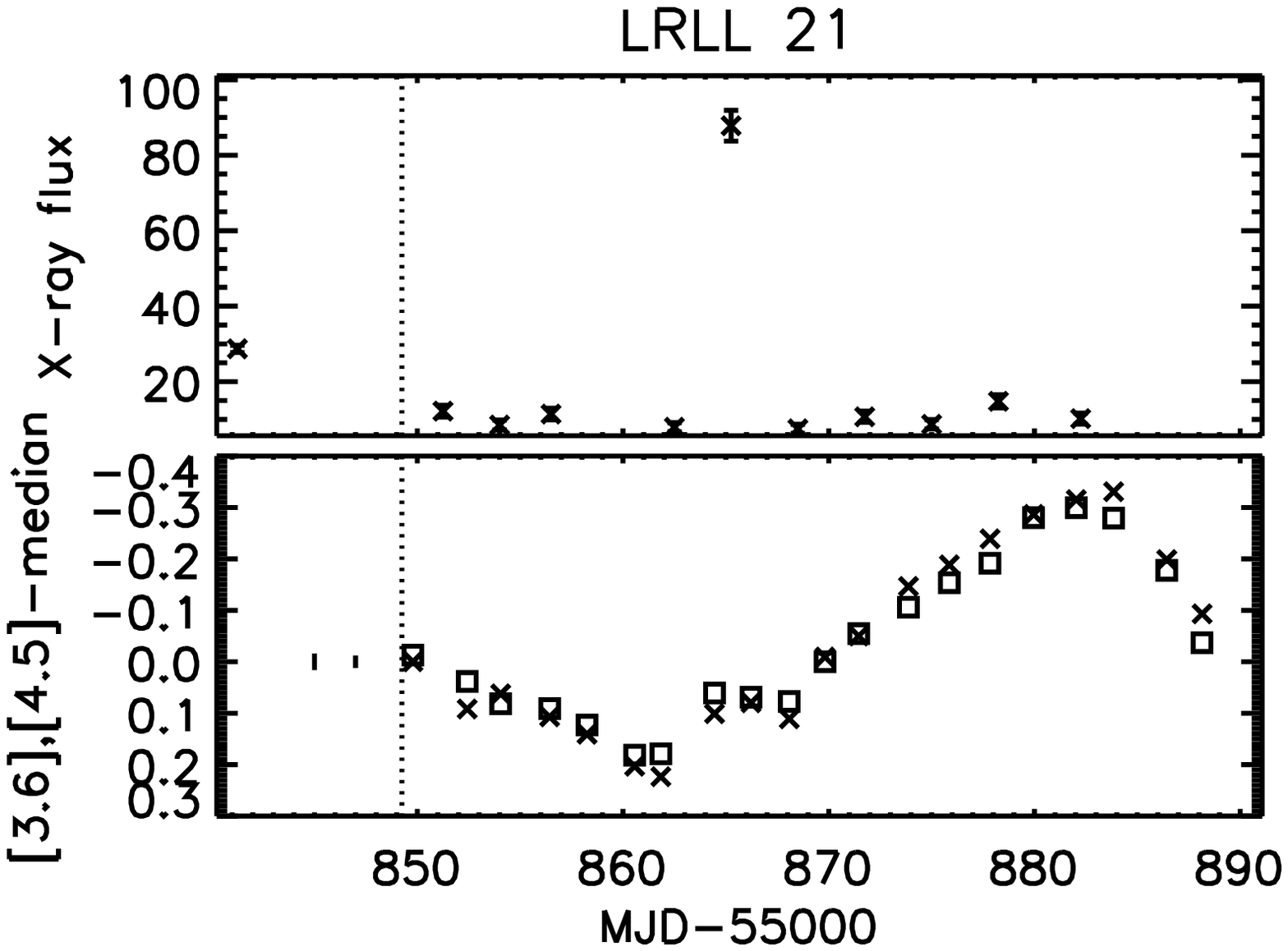}
\figsetgrpnote{X-ray and infrared light curves. The top panel show the X-ray light curve, in units of 1e-6 photons per cm$^2$ per sec. Points to the left of the vertical dashed line are derived from earlier Chandra observations of IC 348 \citep{pre02,ale12}. The infrared light curves are shown in the bottom panel. Error bars for [3.6] (squares) and [4.5] (crosses) photometry are shown as the left and right vertical lines.}
\figsetgrpend

\figsetgrpstart
\figsetgrpnum{3.6}
\figsetgrptitle{LRLL 31 light curves}
\figsetplot{lrll31_xray.eps}
\figsetgrpnote{X-ray and infrared light curves. The top panel show the X-ray light curve, in units of 1e-6 photons per cm$^2$ per sec. Points to the left of the vertical dashed line are derived from earlier Chandra observations of IC 348 \citep{pre02,ale12}. The infrared light curves are shown in the bottom panel. Error bars for [3.6] (squares) and [4.5] (crosses) photometry are shown as the left and right vertical lines.}
\figsetgrpend

\figsetgrpstart
\figsetgrpnum{3.7}
\figsetgrptitle{LRLL 32 light curves}
\figsetplot{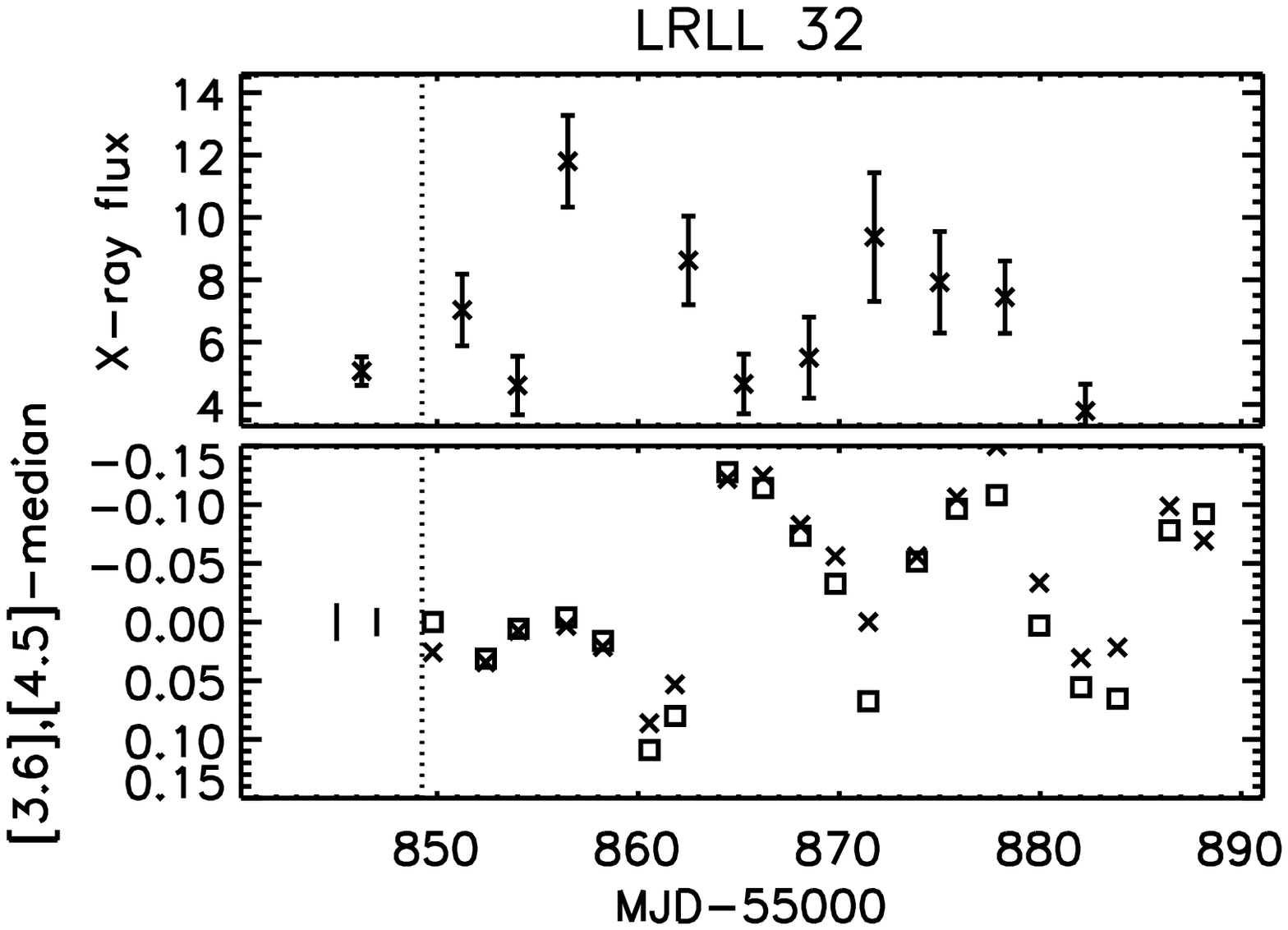}
\figsetgrpnote{X-ray and infrared light curves. The top panel show the X-ray light curve, in units of 1e-6 photons per cm$^2$ per sec. Points to the left of the vertical dashed line are derived from earlier Chandra observations of IC 348 \citep{pre02,ale12}. The infrared light curves are shown in the bottom panel. Error bars for [3.6] (squares) and [4.5] (crosses) photometry are shown as the left and right vertical lines.}
\figsetgrpend

\figsetgrpstart
\figsetgrpnum{3.8}
\figsetgrptitle{LRLL 36 light curves}
\figsetplot{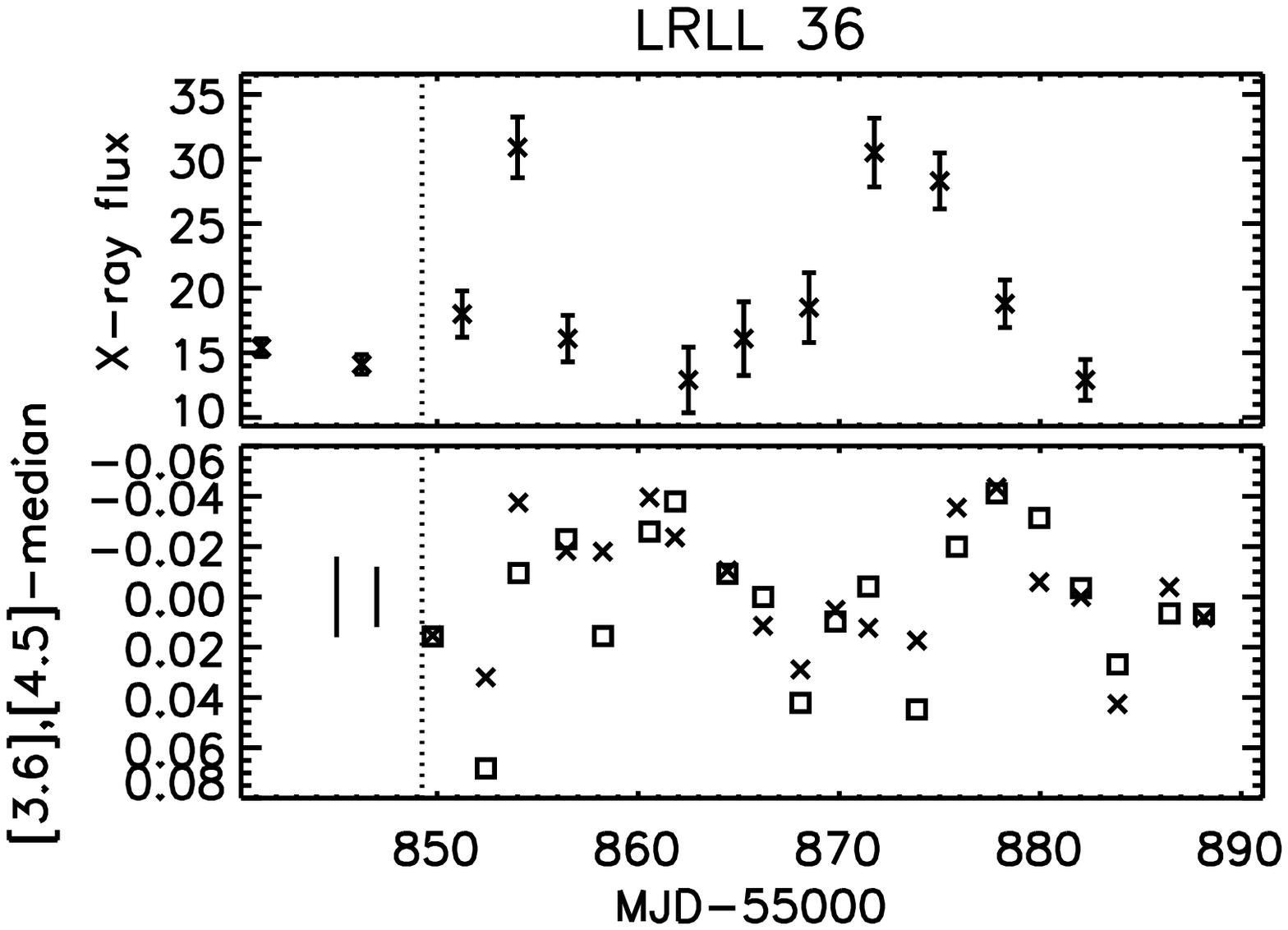}
\figsetgrpnote{X-ray and infrared light curves. The top panel show the X-ray light curve, in units of 1e-6 photons per cm$^2$ per sec. Points to the left of the vertical dashed line are derived from earlier Chandra observations of IC 348 \citep{pre02,ale12}. The infrared light curves are shown in the bottom panel. Error bars for [3.6] (squares) and [4.5] (crosses) photometry are shown as the left and right vertical lines.}
\figsetgrpend

\figsetgrpstart
\figsetgrpnum{3.9}
\figsetgrptitle{LRLL 37 light curves}
\figsetplot{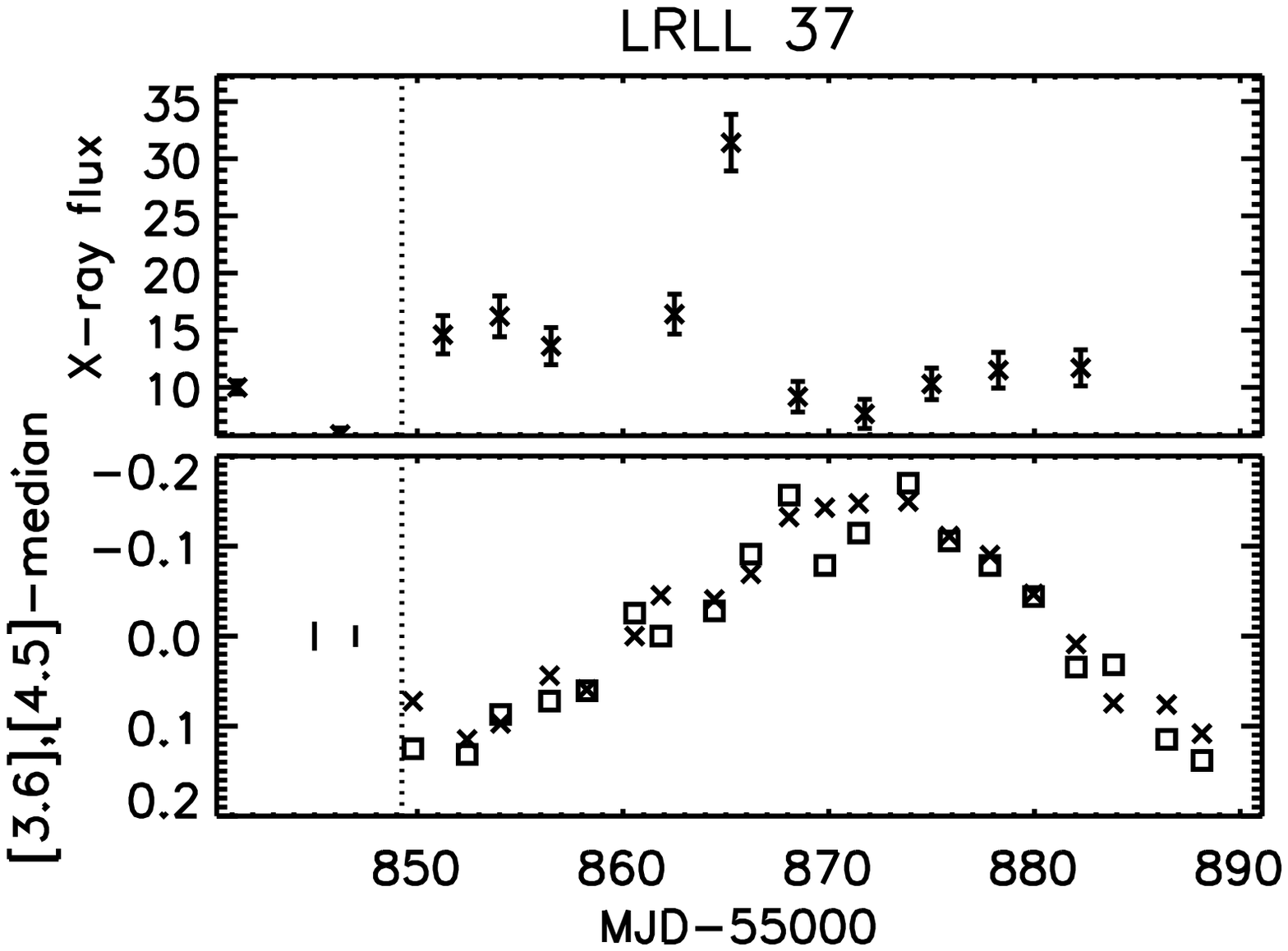}
\figsetgrpnote{X-ray and infrared light curves. The top panel show the X-ray light curve, in units of 1e-6 photons per cm$^2$ per sec. Points to the left of the vertical dashed line are derived from earlier Chandra observations of IC 348 \citep{pre02,ale12}. The infrared light curves are shown in the bottom panel. Error bars for [3.6] (squares) and [4.5] (crosses) photometry are shown as the left and right vertical lines.}
\figsetgrpend

\figsetgrpstart
\figsetgrpnum{3.10}
\figsetgrptitle{LRLL 41 light curves}
\figsetplot{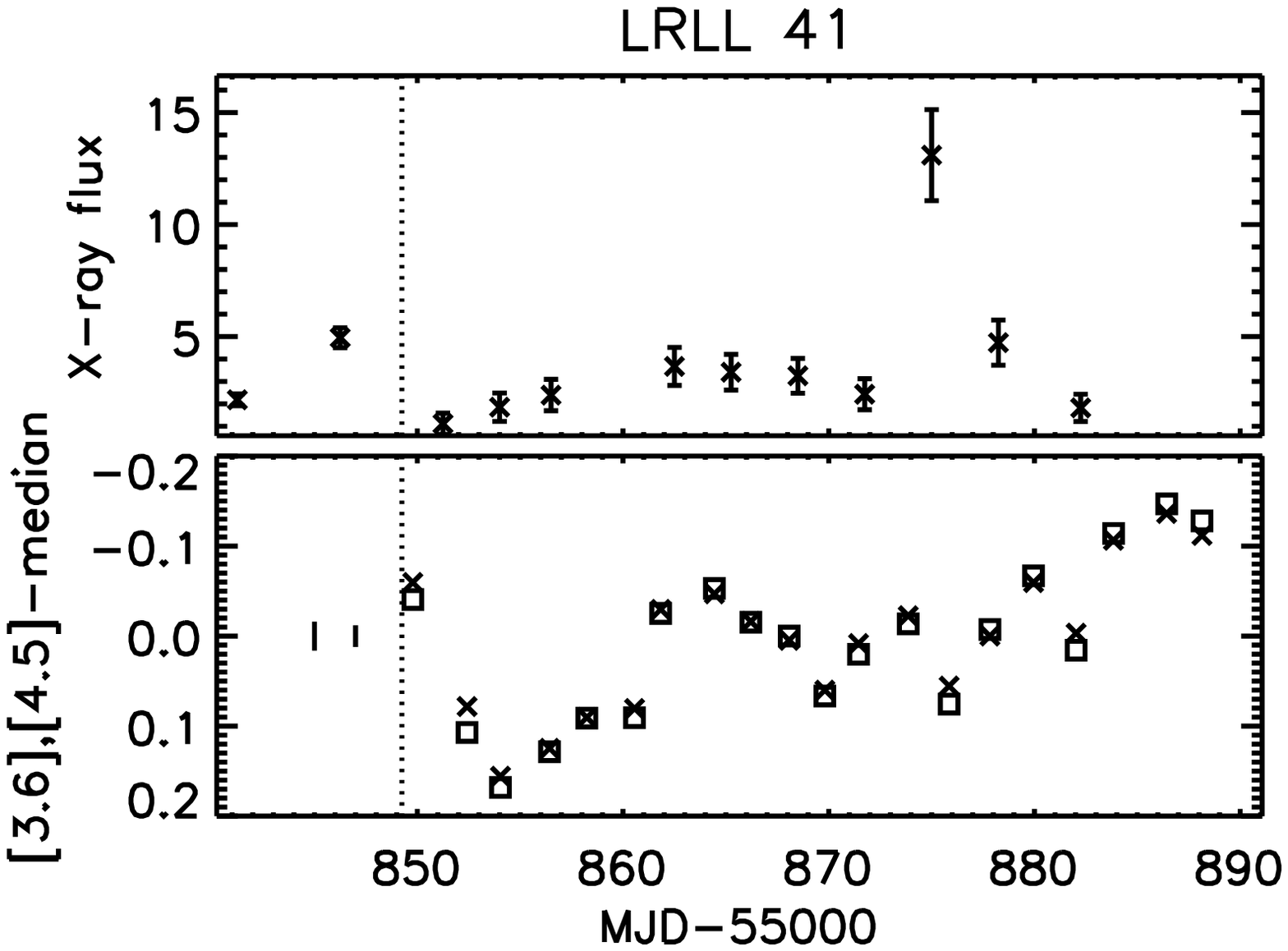}
\figsetgrpnote{X-ray and infrared light curves. The top panel show the X-ray light curve, in units of 1e-6 photons per cm$^2$ per sec. Points to the left of the vertical dashed line are derived from earlier Chandra observations of IC 348 \citep{pre02,ale12}. The infrared light curves are shown in the bottom panel. Error bars for [3.6] (squares) and [4.5] (crosses) photometry are shown as the left and right vertical lines.}
\figsetgrpend

\figsetgrpstart
\figsetgrpnum{3.11}
\figsetgrptitle{LRLL 52 light curves}
\figsetplot{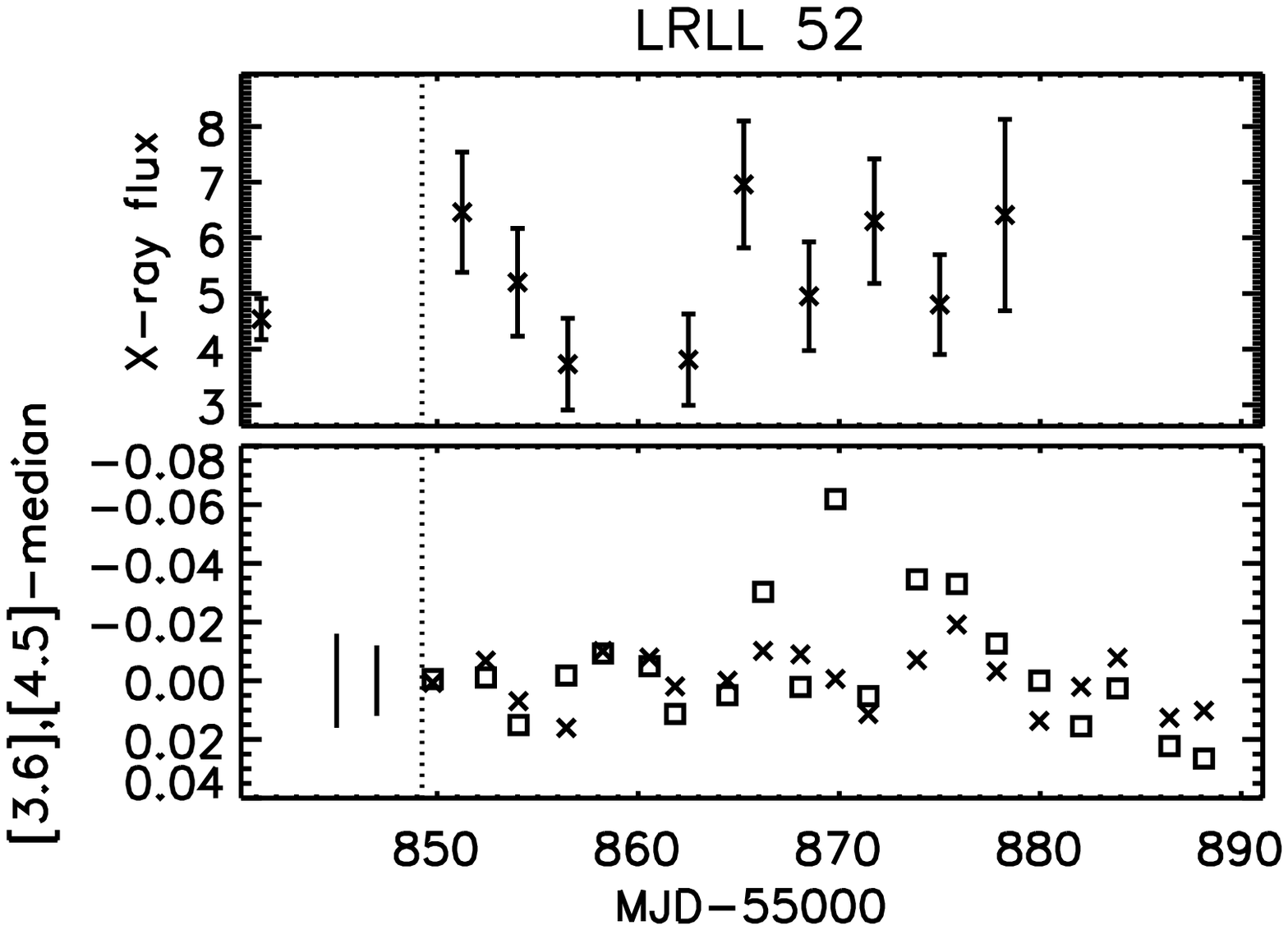}
\figsetgrpnote{X-ray and infrared light curves. The top panel show the X-ray light curve, in units of 1e-6 photons per cm$^2$ per sec. Points to the left of the vertical dashed line are derived from earlier Chandra observations of IC 348 \citep{pre02,ale12}. The infrared light curves are shown in the bottom panel. Error bars for [3.6] (squares) and [4.5] (crosses) photometry are shown as the left and right vertical lines.}
\figsetgrpend

\figsetgrpstart
\figsetgrpnum{3.12}
\figsetgrptitle{LRLL 55 light curves}
\figsetplot{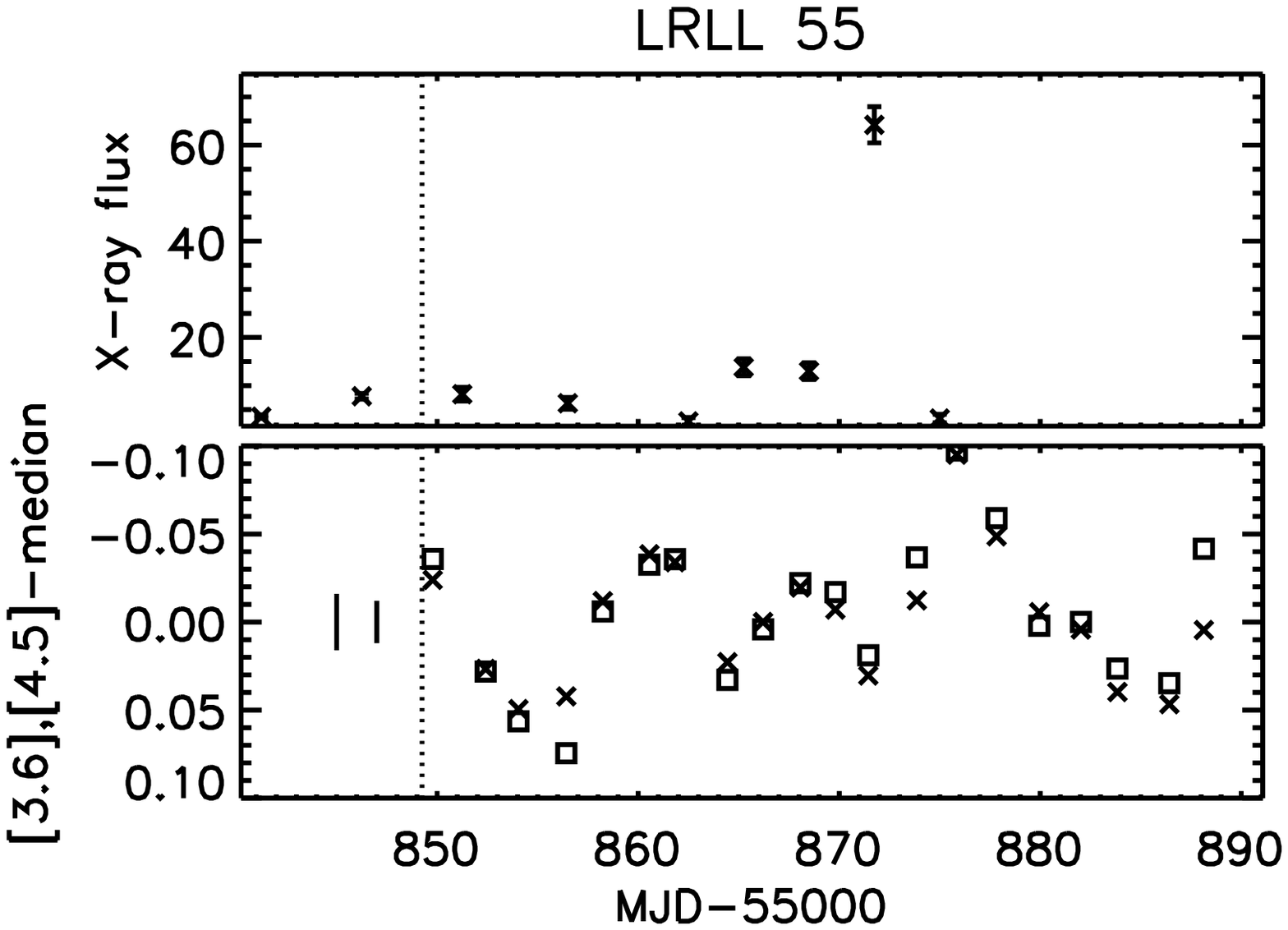}
\figsetgrpnote{X-ray and infrared light curves. The top panel show the X-ray light curve, in units of 1e-6 photons per cm$^2$ per sec. Points to the left of the vertical dashed line are derived from earlier Chandra observations of IC 348 \citep{pre02,ale12}. The infrared light curves are shown in the bottom panel. Error bars for [3.6] (squares) and [4.5] (crosses) photometry are shown as the left and right vertical lines.}
\figsetgrpend

\figsetgrpstart
\figsetgrpnum{3.13}
\figsetgrptitle{LRLL 58 light curves}
\figsetplot{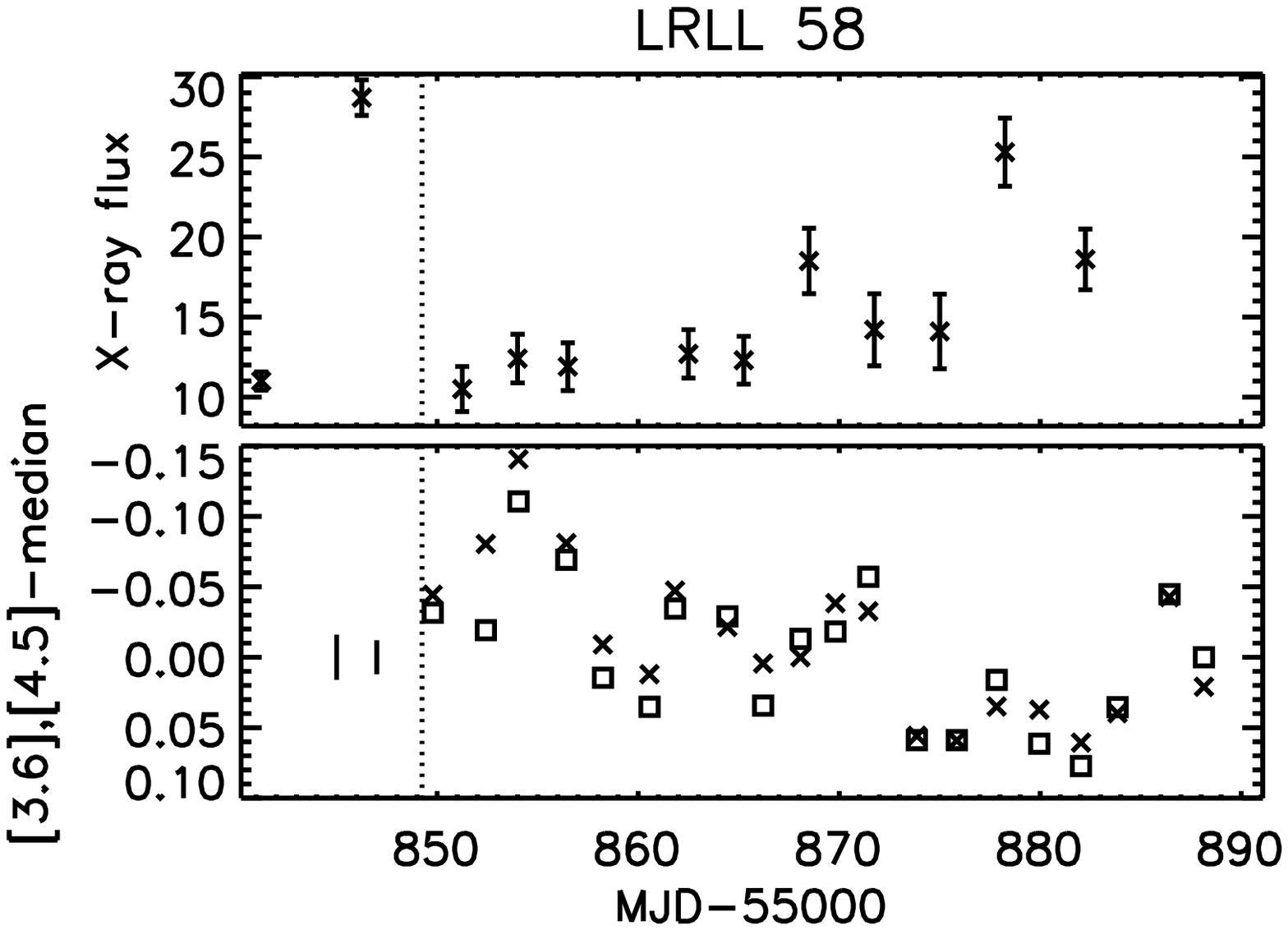}
\figsetgrpnote{X-ray and infrared light curves. The top panel show the X-ray light curve, in units of 1e-6 photons per cm$^2$ per sec. Points to the left of the vertical dashed line are derived from earlier Chandra observations of IC 348 \citep{pre02,ale12}. The infrared light curves are shown in the bottom panel. Error bars for [3.6] (squares) and [4.5] (crosses) photometry are shown as the left and right vertical lines.}
\figsetgrpend

\figsetgrpstart
\figsetgrpnum{3.14}
\figsetgrptitle{LRLL 61 light curves}
\figsetplot{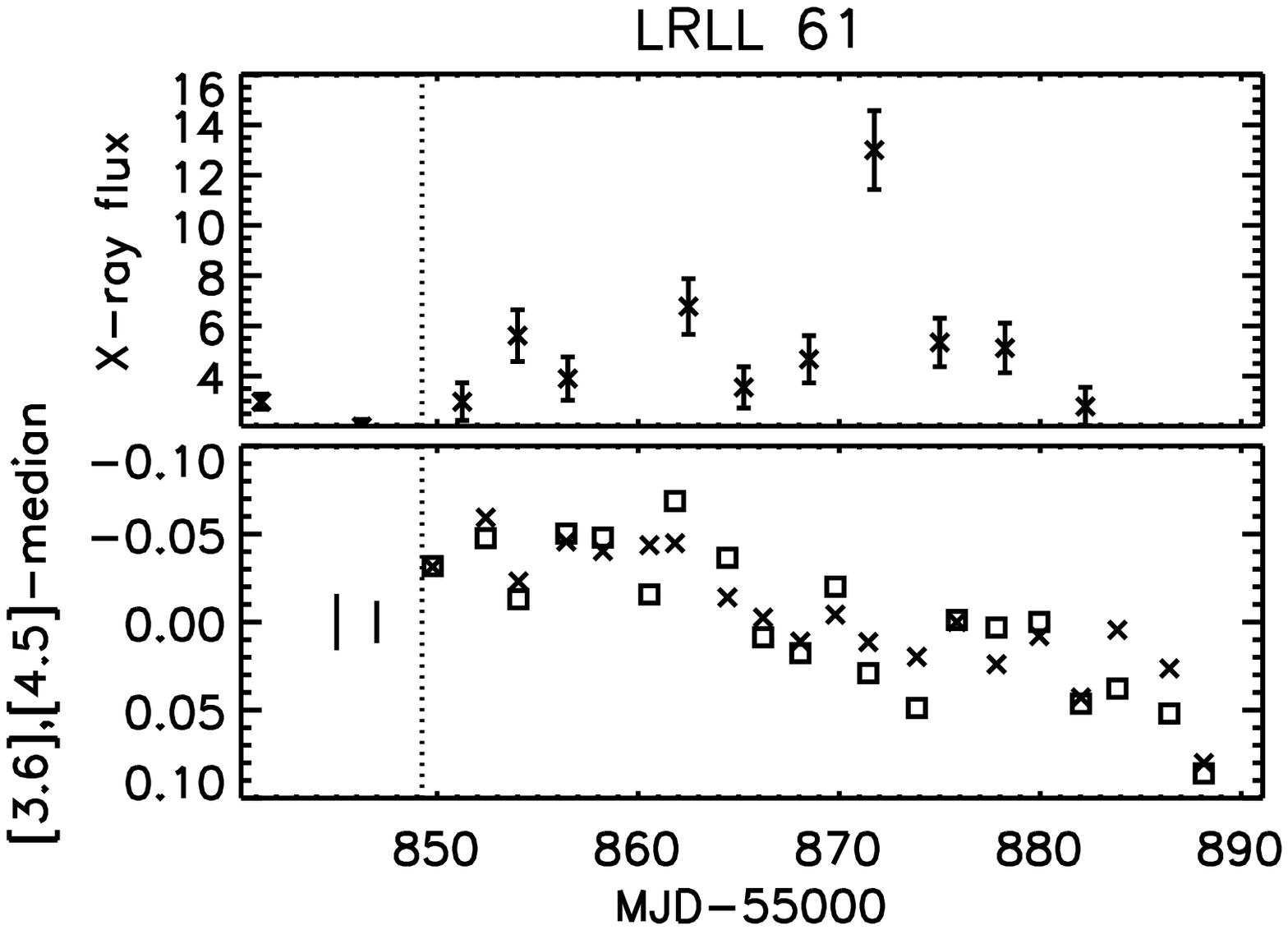}
\figsetgrpnote{X-ray and infrared light curves. The top panel show the X-ray light curve, in units of 1e-6 photons per cm$^2$ per sec. Points to the left of the vertical dashed line are derived from earlier Chandra observations of IC 348 \citep{pre02,ale12}. The infrared light curves are shown in the bottom panel. Error bars for [3.6] (squares) and [4.5] (crosses) photometry are shown as the left and right vertical lines.}
\figsetgrpend

\figsetgrpstart
\figsetgrpnum{3.15}
\figsetgrptitle{LRLL 65 light curves}
\figsetplot{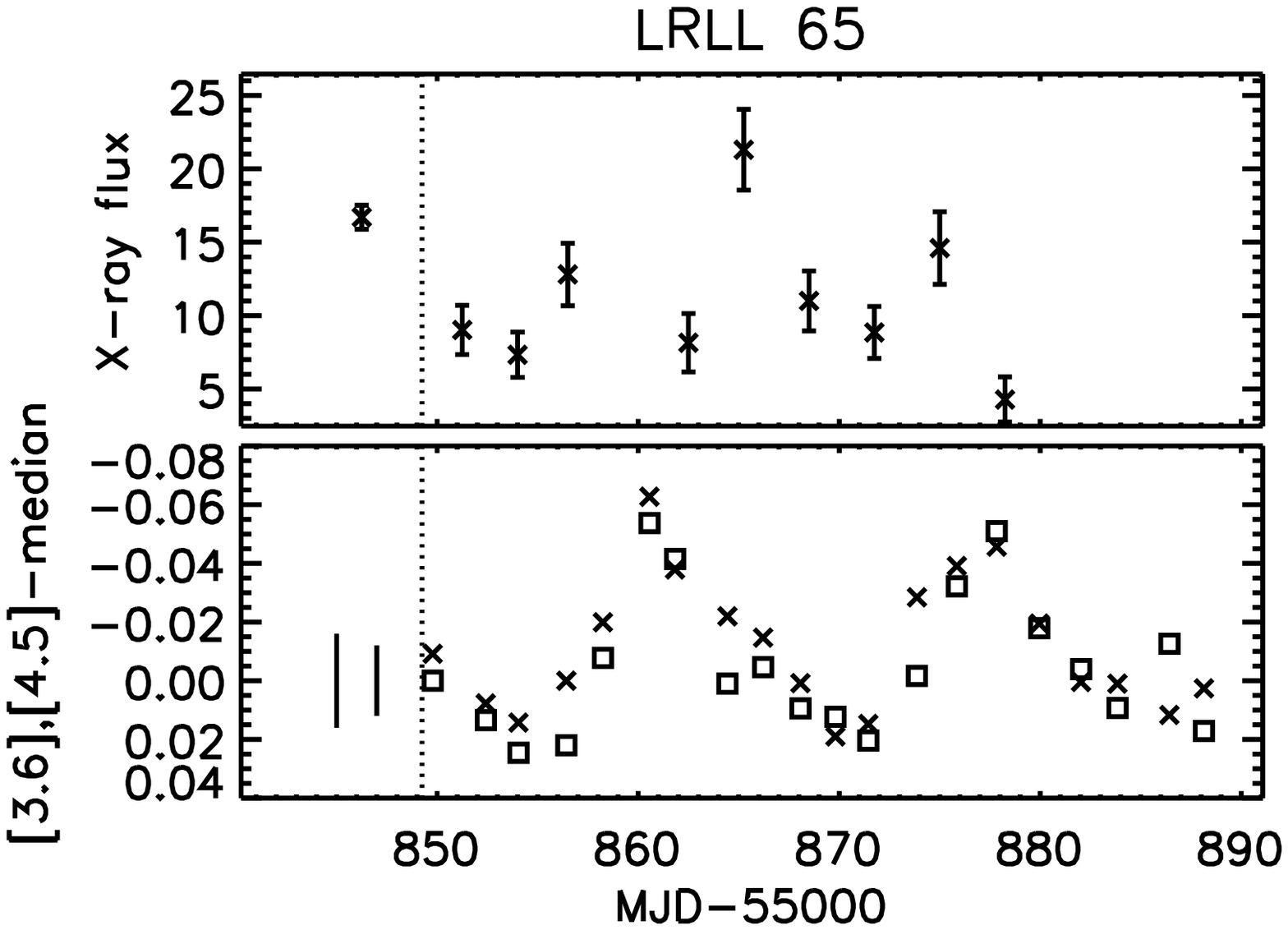}
\figsetgrpnote{X-ray and infrared light curves. The top panel show the X-ray light curve, in units of 1e-6 photons per cm$^2$ per sec. Points to the left of the vertical dashed line are derived from earlier Chandra observations of IC 348 \citep{pre02,ale12}. The infrared light curves are shown in the bottom panel. Error bars for [3.6] (squares) and [4.5] (crosses) photometry are shown as the left and right vertical lines.}
\figsetgrpend

\figsetgrpstart
\figsetgrpnum{3.16}
\figsetgrptitle{LRLL 67 light curves}
\figsetplot{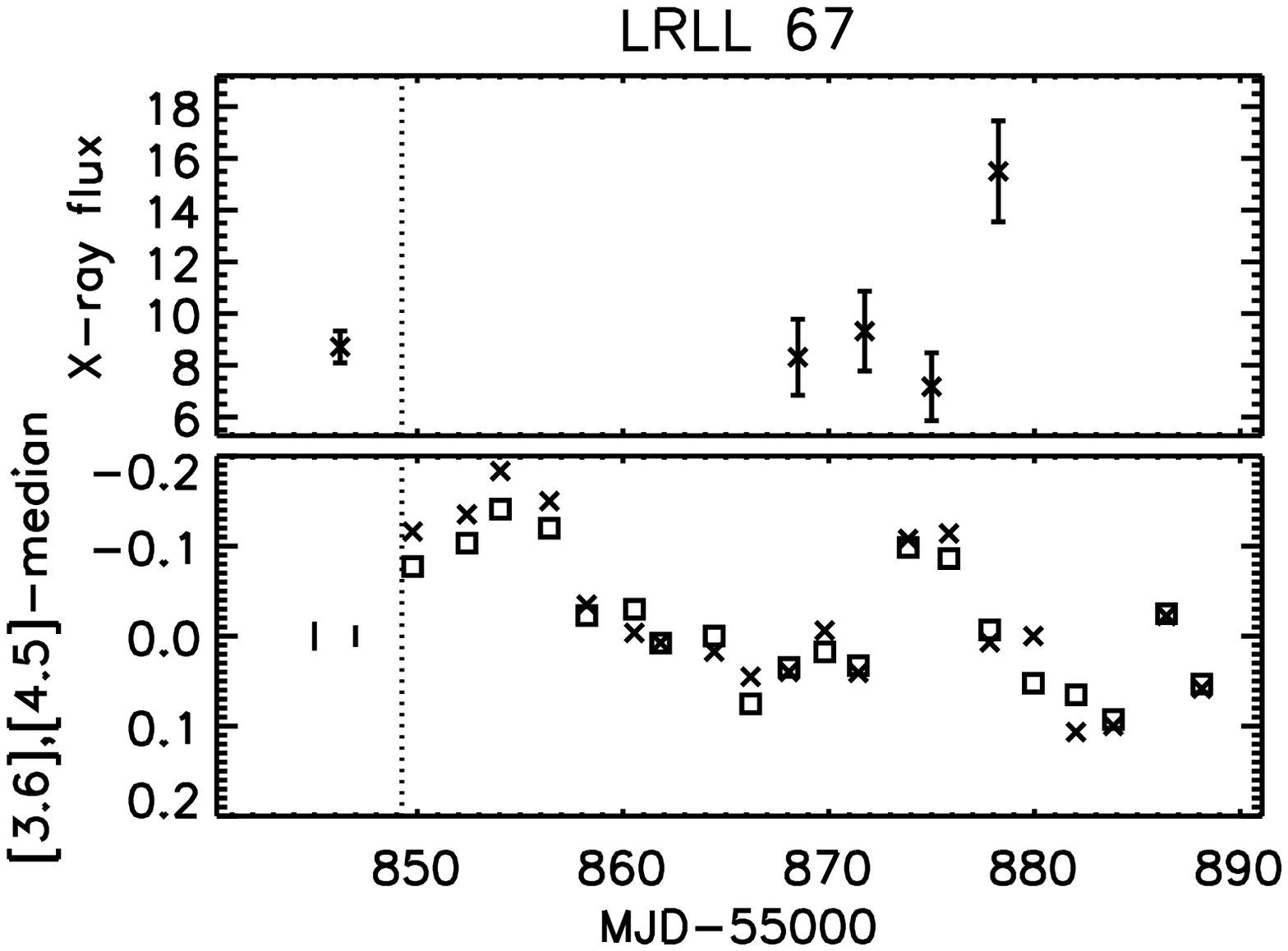}
\figsetgrpnote{X-ray and infrared light curves. The top panel show the X-ray light curve, in units of 1e-6 photons per cm$^2$ per sec. Points to the left of the vertical dashed line are derived from earlier Chandra observations of IC 348 \citep{pre02,ale12}. The infrared light curves are shown in the bottom panel. Error bars for [3.6] (squares) and [4.5] (crosses) photometry are shown as the left and right vertical lines.}
\figsetgrpend

\figsetgrpstart
\figsetgrpnum{3.17}
\figsetgrptitle{LRLL 71 light curves}
\figsetplot{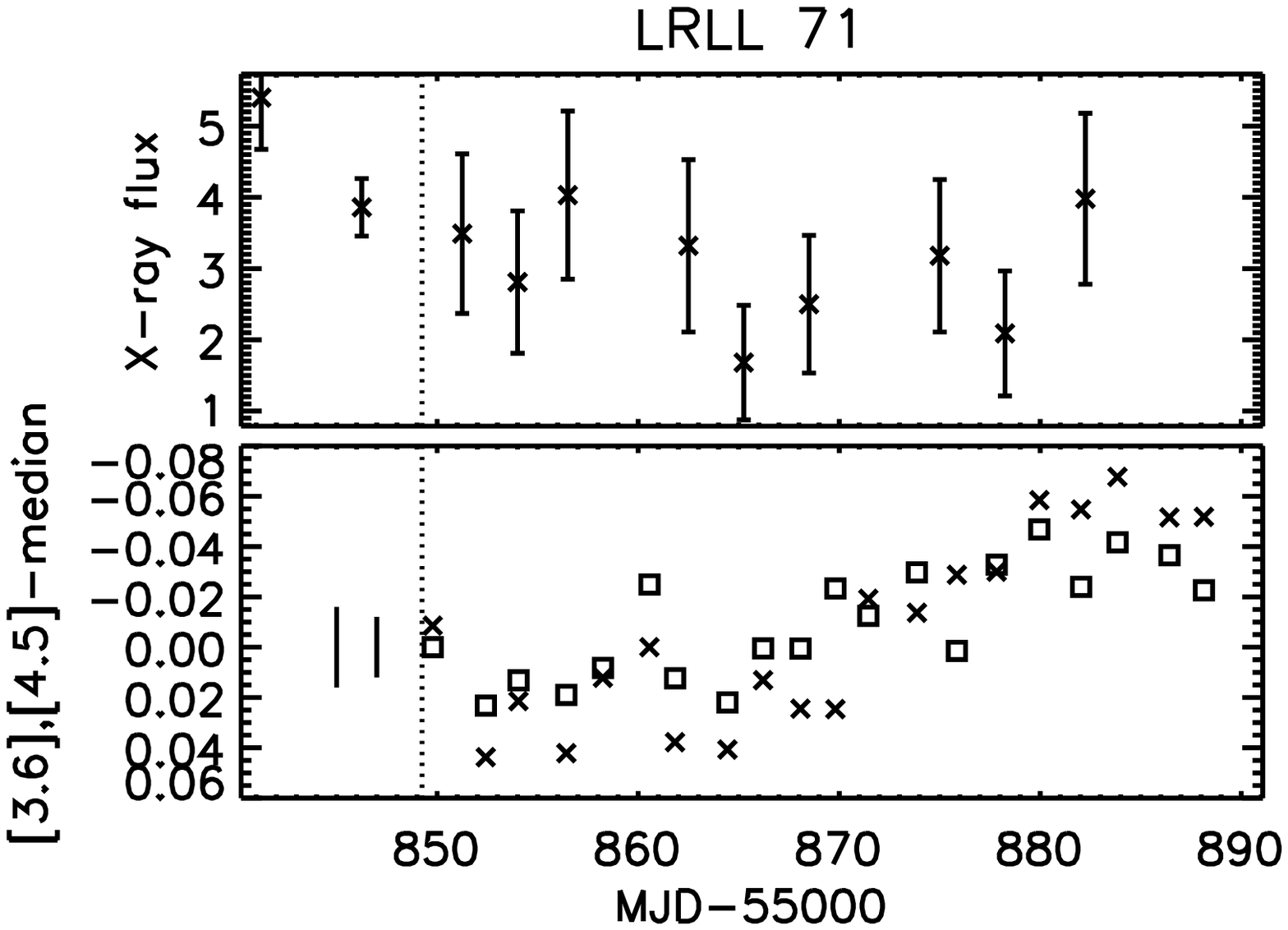}
\figsetgrpnote{X-ray and infrared light curves. The top panel show the X-ray light curve, in units of 1e-6 photons per cm$^2$ per sec. Points to the left of the vertical dashed line are derived from earlier Chandra observations of IC 348 \citep{pre02,ale12}. The infrared light curves are shown in the bottom panel. Error bars for [3.6] (squares) and [4.5] (crosses) photometry are shown as the left and right vertical lines.}
\figsetgrpend

\figsetgrpstart
\figsetgrpnum{3.18}
\figsetgrptitle{LRLL 72 light curves}
\figsetplot{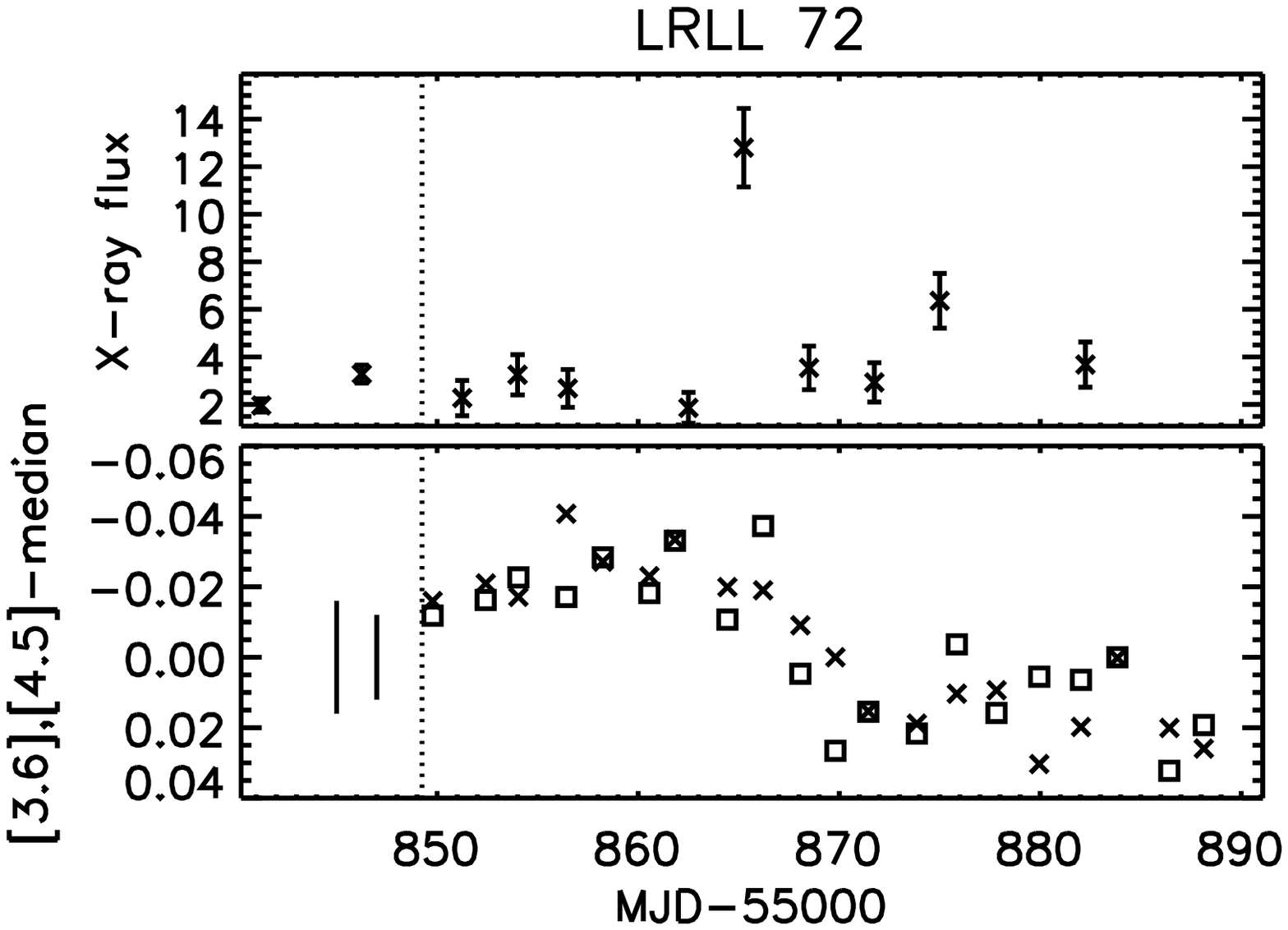}
\figsetgrpnote{X-ray and infrared light curves. The top panel show the X-ray light curve, in units of 1e-6 photons per cm$^2$ per sec. Points to the left of the vertical dashed line are derived from earlier Chandra observations of IC 348 \citep{pre02,ale12}. The infrared light curves are shown in the bottom panel. Error bars for [3.6] (squares) and [4.5] (crosses) photometry are shown as the left and right vertical lines.}
\figsetgrpend

\figsetgrpstart
\figsetgrpnum{3.19}
\figsetgrptitle{LRLL 75 light curves}
\figsetplot{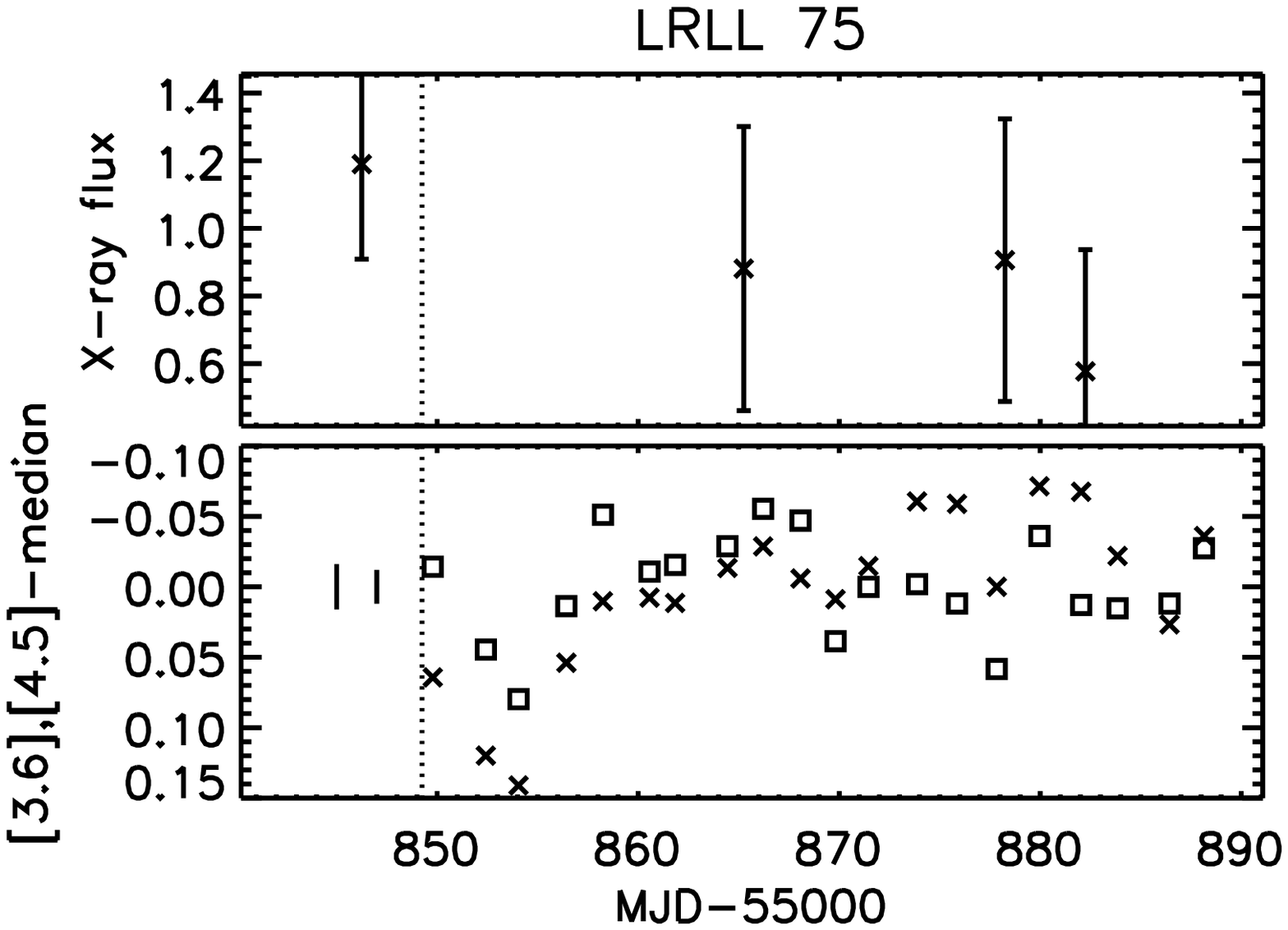}
\figsetgrpnote{X-ray and infrared light curves. The top panel show the X-ray light curve, in units of 1e-6 photons per cm$^2$ per sec. Points to the left of the vertical dashed line are derived from earlier Chandra observations of IC 348 \citep{pre02,ale12}. The infrared light curves are shown in the bottom panel. Error bars for [3.6] (squares) and [4.5] (crosses) photometry are shown as the left and right vertical lines.}
\figsetgrpend

\figsetgrpstart
\figsetgrpnum{3.20}
\figsetgrptitle{LRLL 82 light curves}
\figsetplot{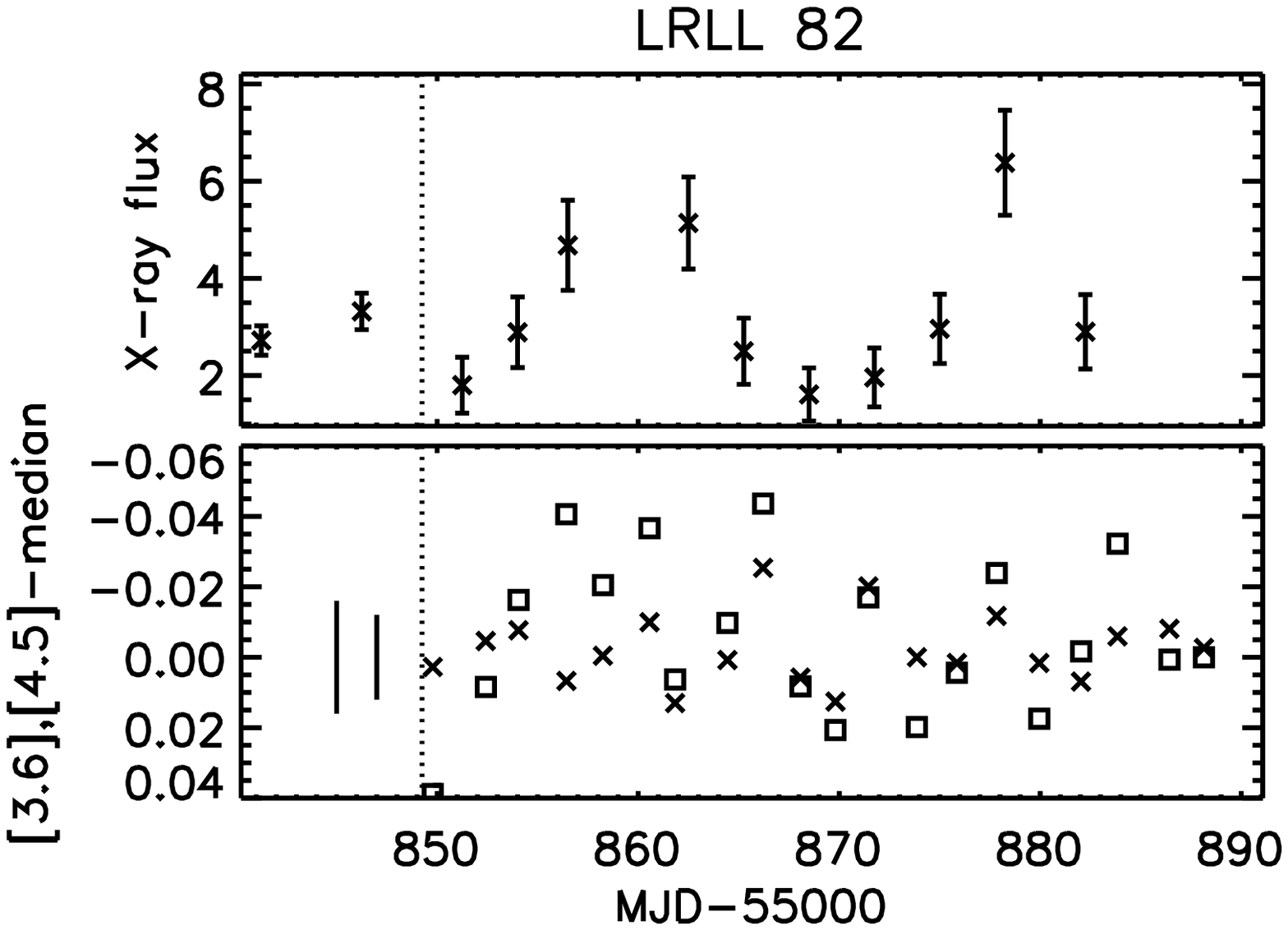}
\figsetgrpnote{X-ray and infrared light curves. The top panel show the X-ray light curve, in units of 1e-6 photons per cm$^2$ per sec. Points to the left of the vertical dashed line are derived from earlier Chandra observations of IC 348 \citep{pre02,ale12}. The infrared light curves are shown in the bottom panel. Error bars for [3.6] (squares) and [4.5] (crosses) photometry are shown as the left and right vertical lines.}
\figsetgrpend

\figsetgrpstart
\figsetgrpnum{3.21}
\figsetgrptitle{LRLL 83 light curves}
\figsetplot{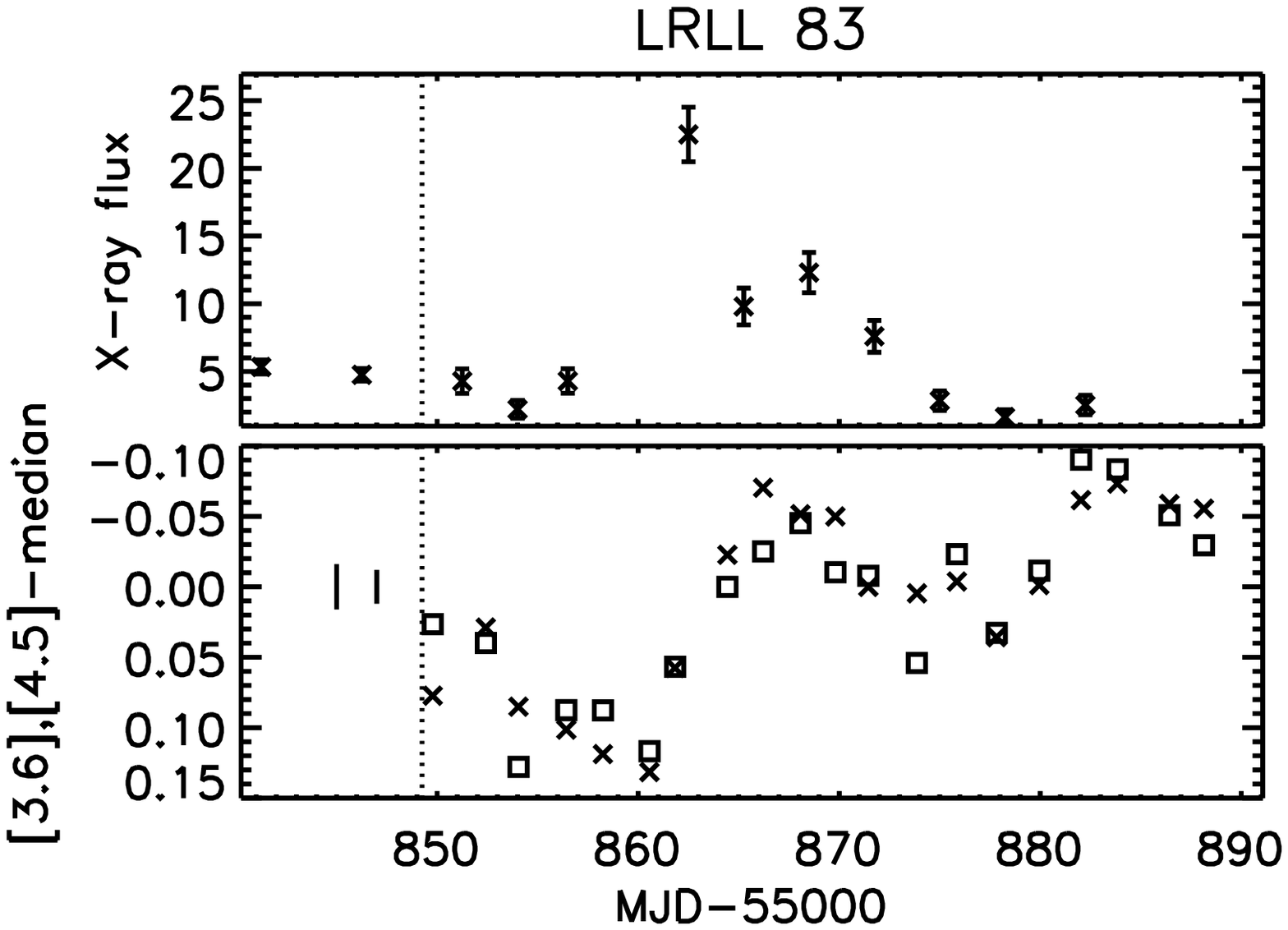}
\figsetgrpnote{X-ray and infrared light curves. The top panel show the X-ray light curve, in units of 1e-6 photons per cm$^2$ per sec. Points to the left of the vertical dashed line are derived from earlier Chandra observations of IC 348 \citep{pre02,ale12}. The infrared light curves are shown in the bottom panel. Error bars for [3.6] (squares) and [4.5] (crosses) photometry are shown as the left and right vertical lines.}
\figsetgrpend

\figsetgrpstart
\figsetgrpnum{3.22}
\figsetgrptitle{LRLL 88 light curves}
\figsetplot{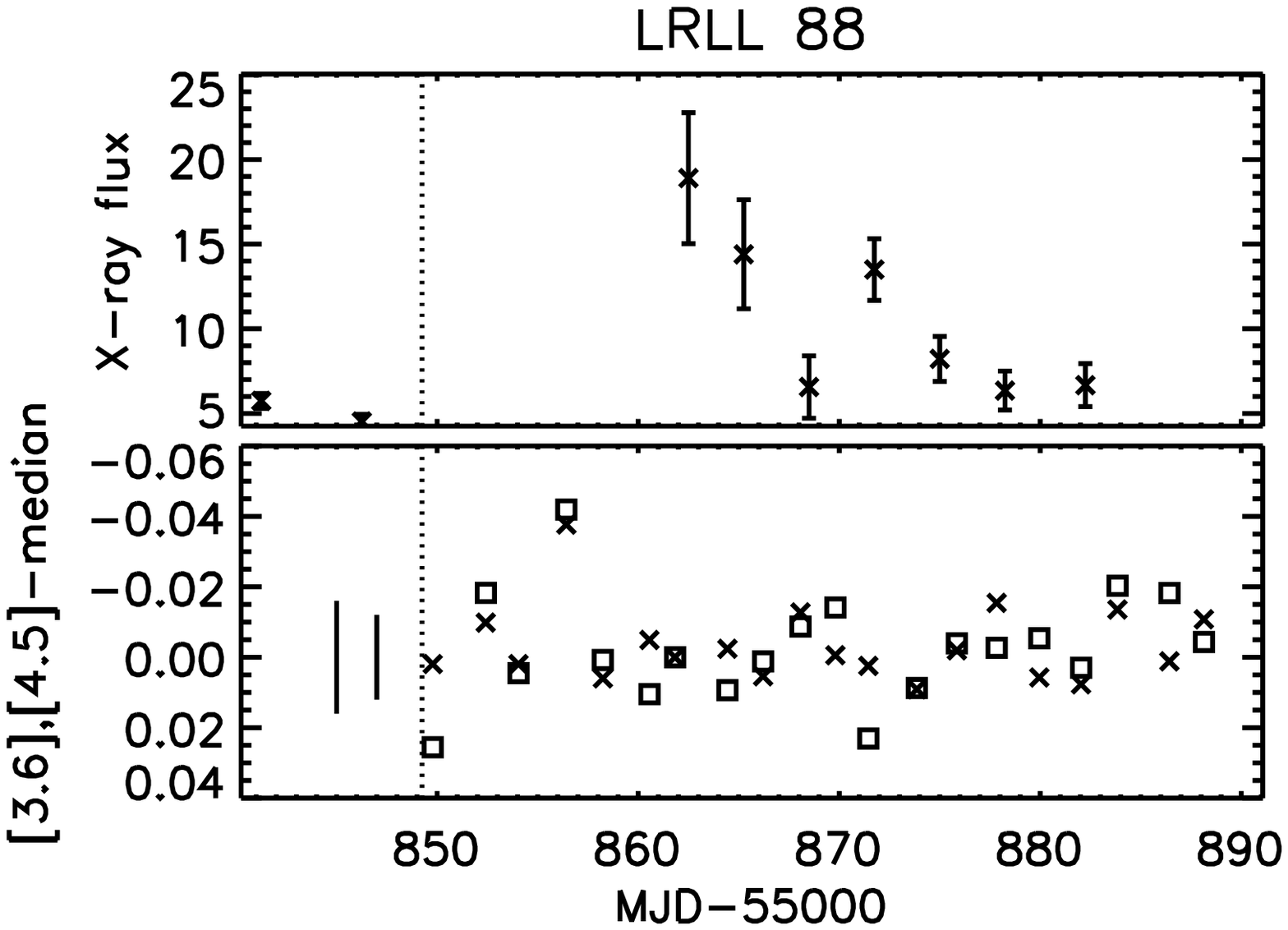}
\figsetgrpnote{X-ray and infrared light curves. The top panel show the X-ray light curve, in units of 1e-6 photons per cm$^2$ per sec. Points to the left of the vertical dashed line are derived from earlier Chandra observations of IC 348 \citep{pre02,ale12}. The infrared light curves are shown in the bottom panel. Error bars for [3.6] (squares) and [4.5] (crosses) photometry are shown as the left and right vertical lines.}
\figsetgrpend

\figsetgrpstart
\figsetgrpnum{3.23}
\figsetgrptitle{LRLL 90 light curves}
\figsetplot{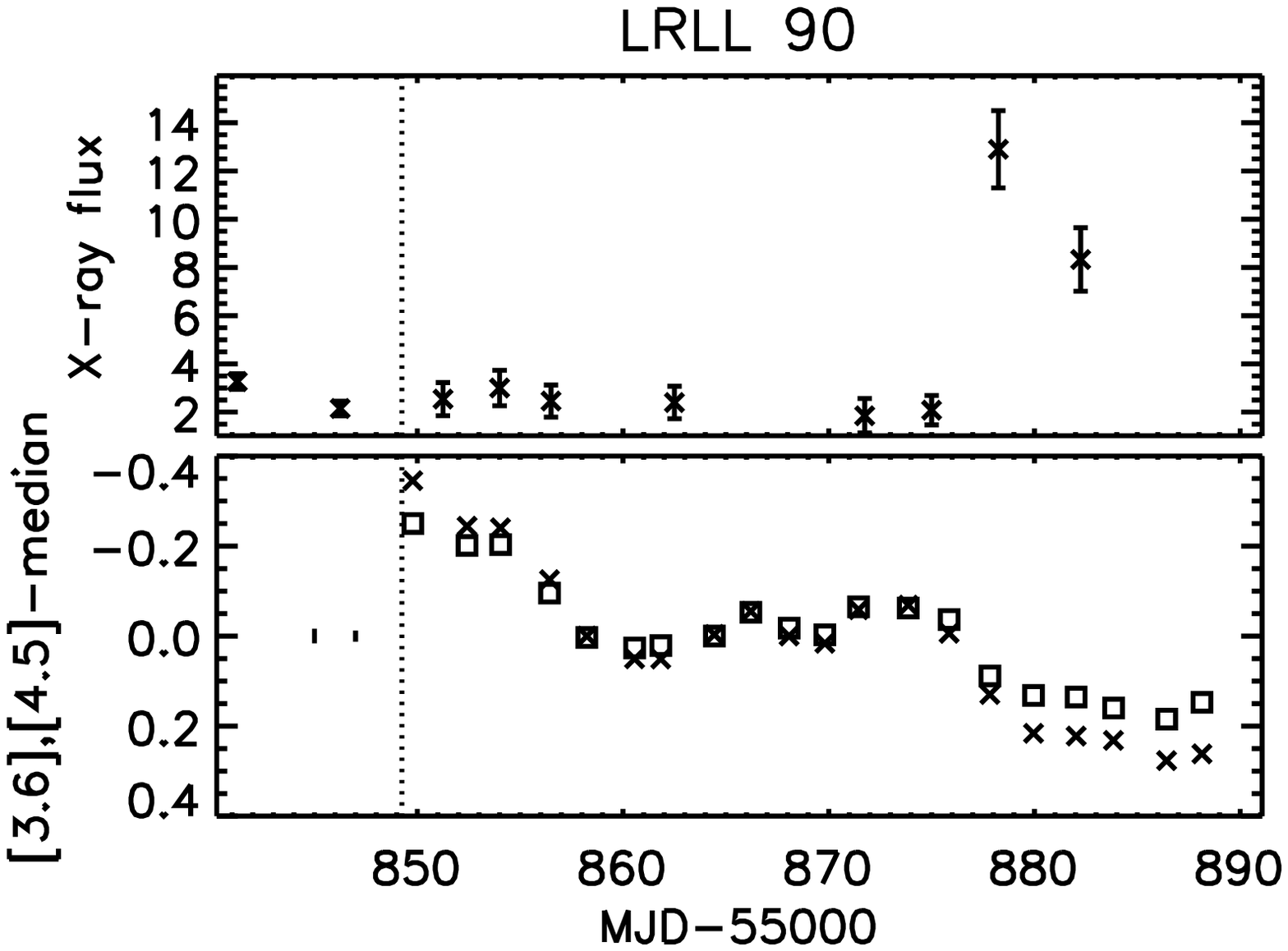}
\figsetgrpnote{X-ray and infrared light curves. The top panel show the X-ray light curve, in units of 1e-6 photons per cm$^2$ per sec. Points to the left of the vertical dashed line are derived from earlier Chandra observations of IC 348 \citep{pre02,ale12}. The infrared light curves are shown in the bottom panel. Error bars for [3.6] (squares) and [4.5] (crosses) photometry are shown as the left and right vertical lines.}
\figsetgrpend

\figsetgrpstart
\figsetgrpnum{3.24}
\figsetgrptitle{LRLL 91 light curves}
\figsetplot{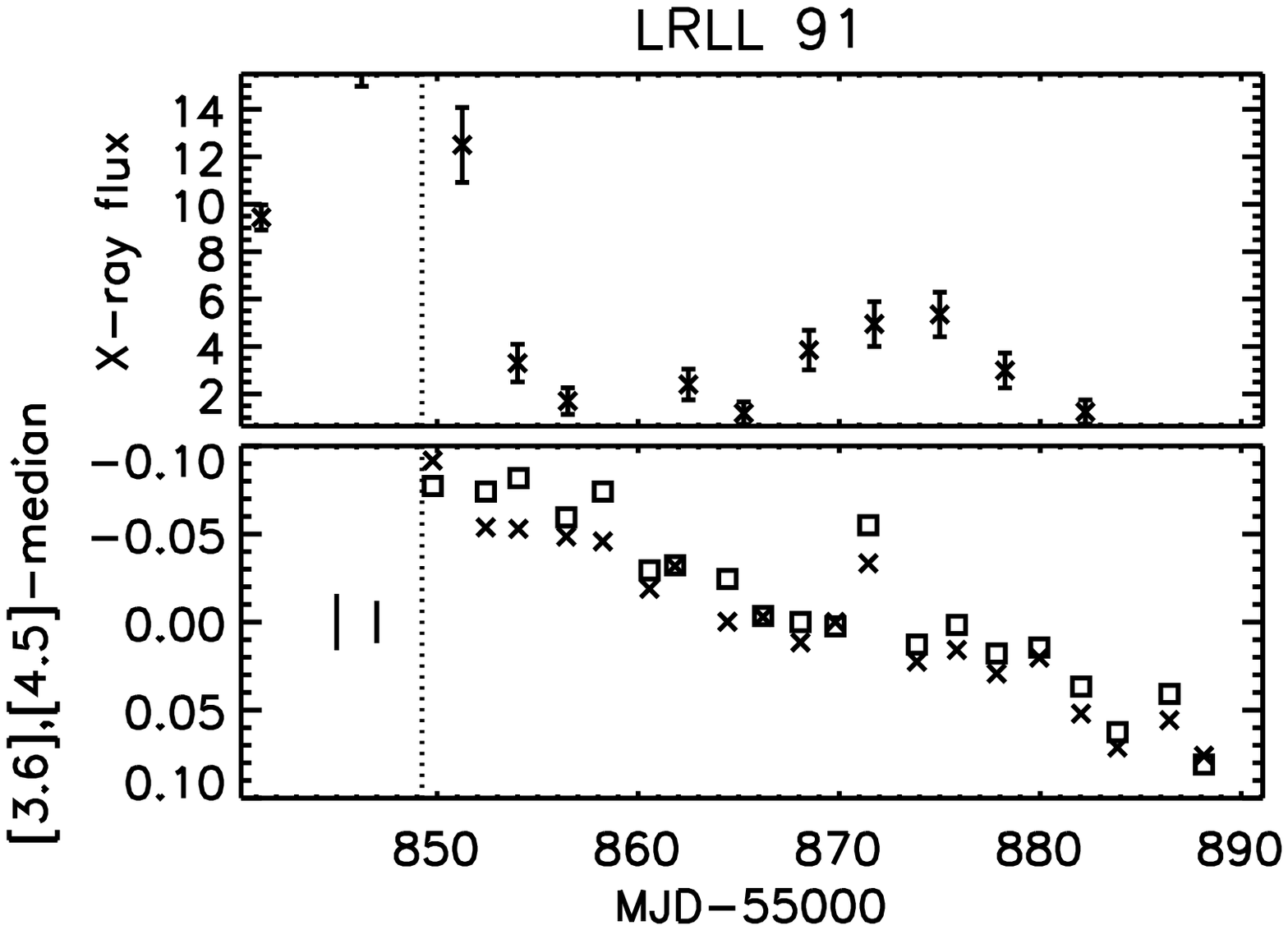}
\figsetgrpnote{X-ray and infrared light curves. The top panel show the X-ray light curve, in units of 1e-6 photons per cm$^2$ per sec. Points to the left of the vertical dashed line are derived from earlier Chandra observations of IC 348 \citep{pre02,ale12}. The infrared light curves are shown in the bottom panel. Error bars for [3.6] (squares) and [4.5] (crosses) photometry are shown as the left and right vertical lines.}
\figsetgrpend

\figsetgrpstart
\figsetgrpnum{3.25}
\figsetgrptitle{LRLL 97 light curves}
\figsetplot{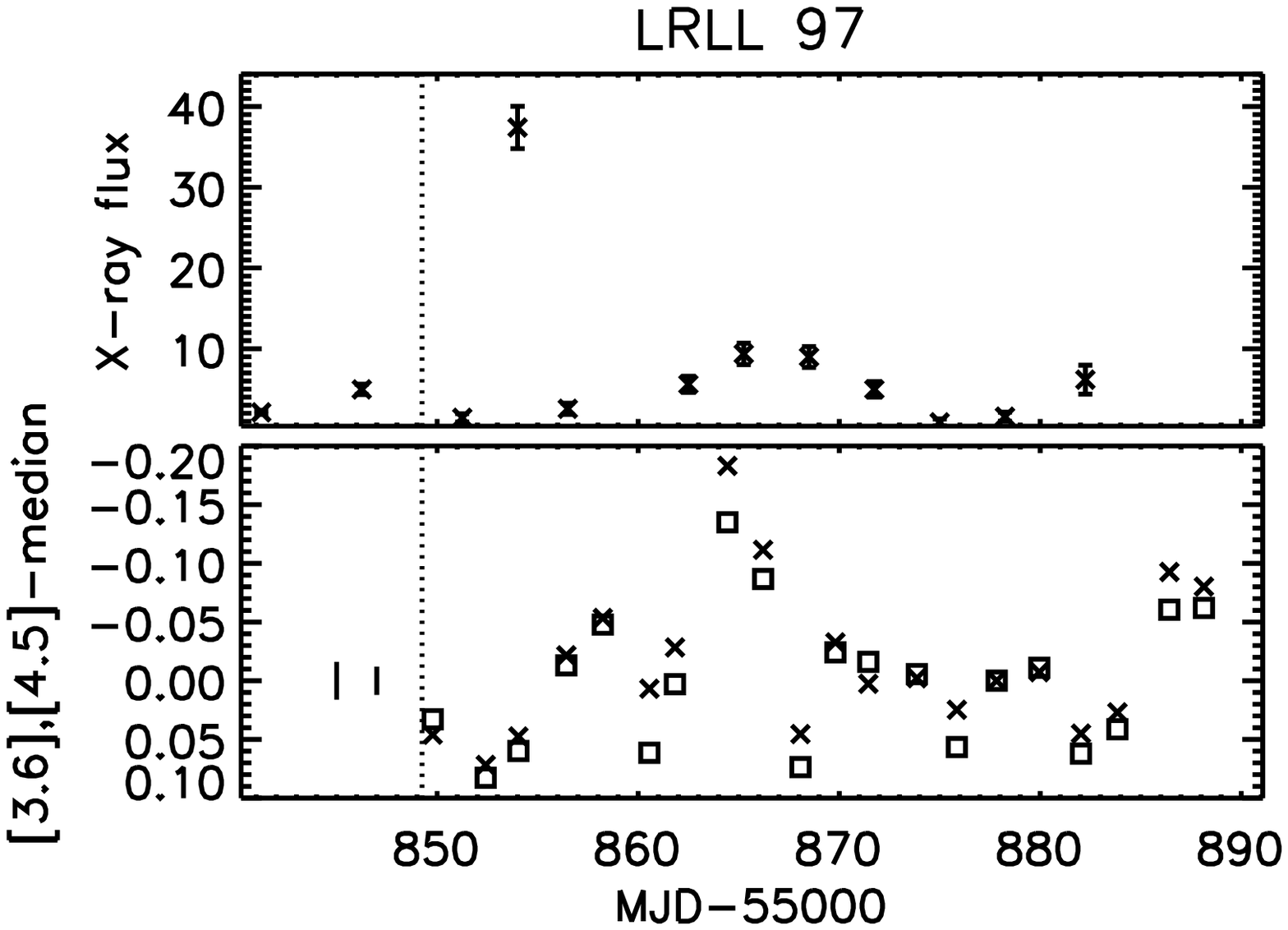}
\figsetgrpnote{X-ray and infrared light curves. The top panel show the X-ray light curve, in units of 1e-6 photons per cm$^2$ per sec. Points to the left of the vertical dashed line are derived from earlier Chandra observations of IC 348 \citep{pre02,ale12}. The infrared light curves are shown in the bottom panel. Error bars for [3.6] (squares) and [4.5] (crosses) photometry are shown as the left and right vertical lines.}
\figsetgrpend

\figsetgrpstart
\figsetgrpnum{3.26}
\figsetgrptitle{LRLL 103 light curves}
\figsetplot{lrll103_xray.eps}
\figsetgrpnote{X-ray and infrared light curves. The top panel show the X-ray light curve, in units of 1e-6 photons per cm$^2$ per sec. Points to the left of the vertical dashed line are derived from earlier Chandra observations of IC 348 \citep{pre02,ale12}. The infrared light curves are shown in the bottom panel. Error bars for [3.6] (squares) and [4.5] (crosses) photometry are shown as the left and right vertical lines.}
\figsetgrpend

\figsetgrpstart
\figsetgrpnum{3.27}
\figsetgrptitle{LRLL 108 light curves}
\figsetplot{lrll108_xray.eps}
\figsetgrpnote{X-ray and infrared light curves. The top panel show the X-ray light curve, in units of 1e-6 photons per cm$^2$ per sec. Points to the left of the vertical dashed line are derived from earlier Chandra observations of IC 348 \citep{pre02,ale12}. The infrared light curves are shown in the bottom panel. Error bars for [3.6] (squares) and [4.5] (crosses) photometry are shown as the left and right vertical lines.}
\figsetgrpend

\figsetgrpstart
\figsetgrpnum{3.28}
\figsetgrptitle{LRLL 110 light curves}
\figsetplot{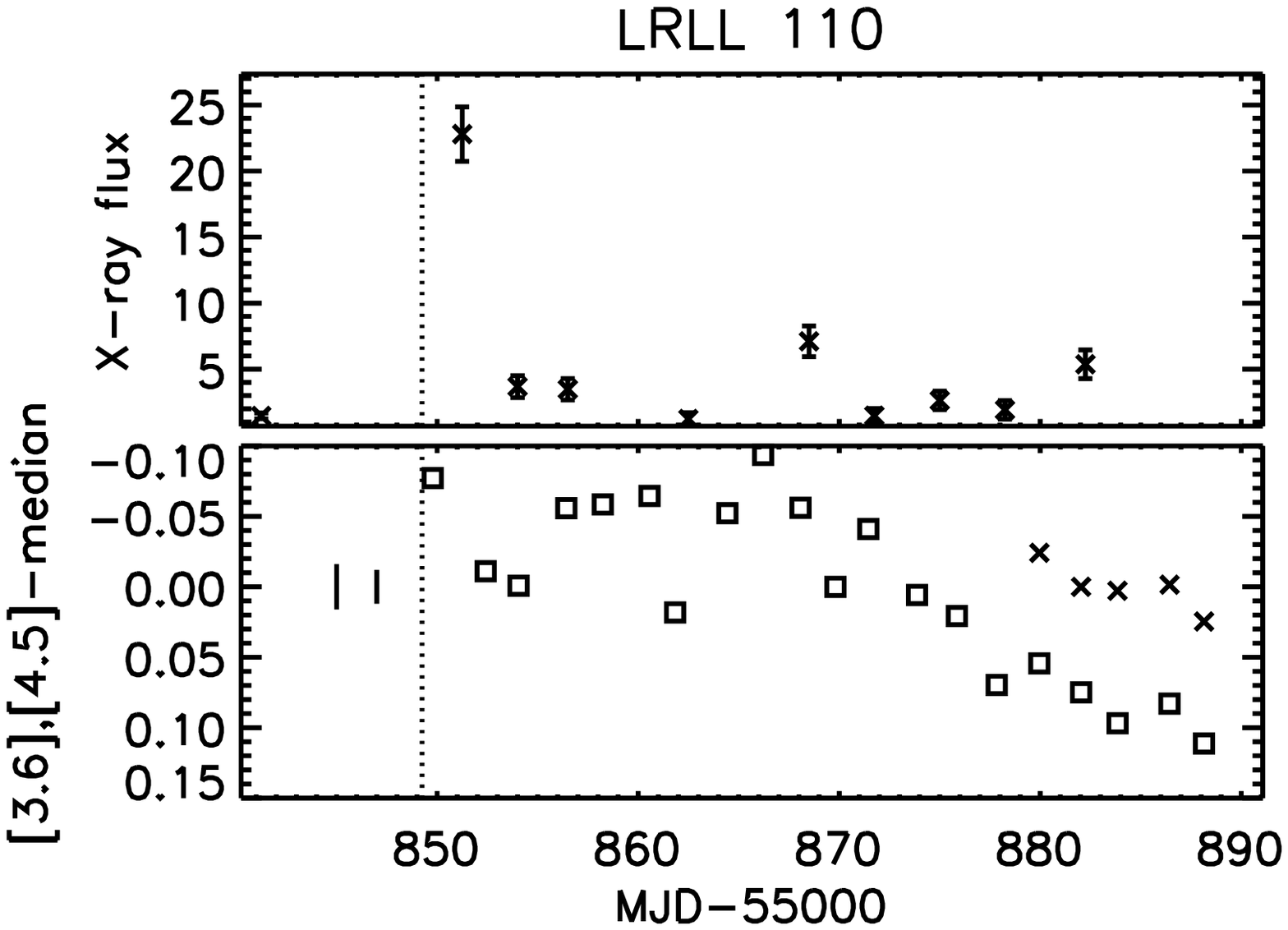}
\figsetgrpnote{X-ray and infrared light curves. The top panel show the X-ray light curve, in units of 1e-6 photons per cm$^2$ per sec. Points to the left of the vertical dashed line are derived from earlier Chandra observations of IC 348 \citep{pre02,ale12}. The infrared light curves are shown in the bottom panel. Error bars for [3.6] (squares) and [4.5] (crosses) photometry are shown as the left and right vertical lines.}
\figsetgrpend

\figsetgrpstart
\figsetgrpnum{3.29}
\figsetgrptitle{LRLL 113 light curves}
\figsetplot{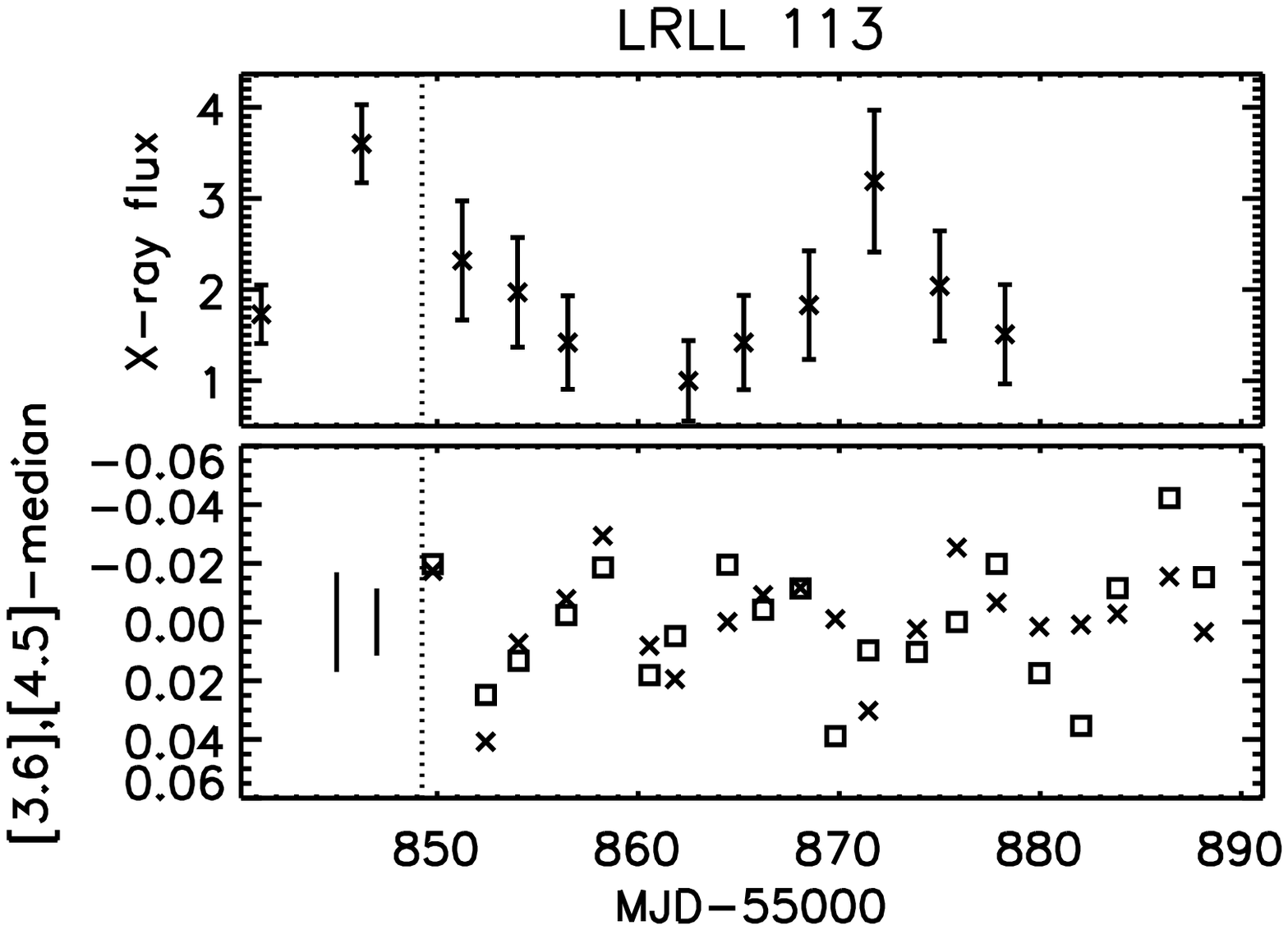}
\figsetgrpnote{X-ray and infrared light curves. The top panel show the X-ray light curve, in units of 1e-6 photons per cm$^2$ per sec. Points to the left of the vertical dashed line are derived from earlier Chandra observations of IC 348 \citep{pre02,ale12}. The infrared light curves are shown in the bottom panel. Error bars for [3.6] (squares) and [4.5] (crosses) photometry are shown as the left and right vertical lines.}
\figsetgrpend

\figsetgrpstart
\figsetgrpnum{3.30}
\figsetgrptitle{LRLL 116 light curves}
\figsetplot{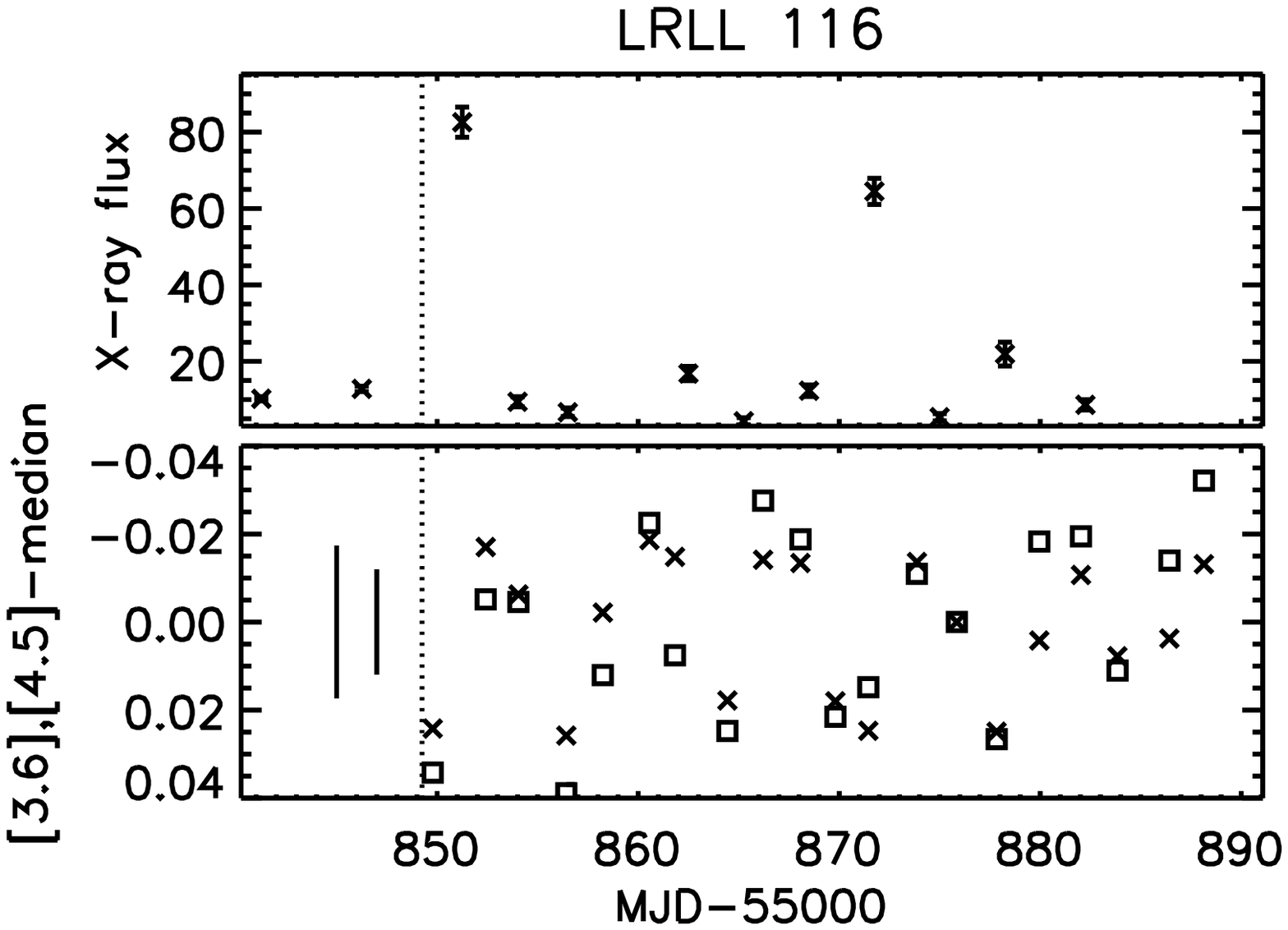}
\figsetgrpnote{X-ray and infrared light curves. The top panel show the X-ray light curve, in units of 1e-6 photons per cm$^2$ per sec. Points to the left of the vertical dashed line are derived from earlier Chandra observations of IC 348 \citep{pre02,ale12}. The infrared light curves are shown in the bottom panel. Error bars for [3.6] (squares) and [4.5] (crosses) photometry are shown as the left and right vertical lines.}
\figsetgrpend

\figsetgrpstart
\figsetgrpnum{3.31}
\figsetgrptitle{LRLL 139 light curves}
\figsetplot{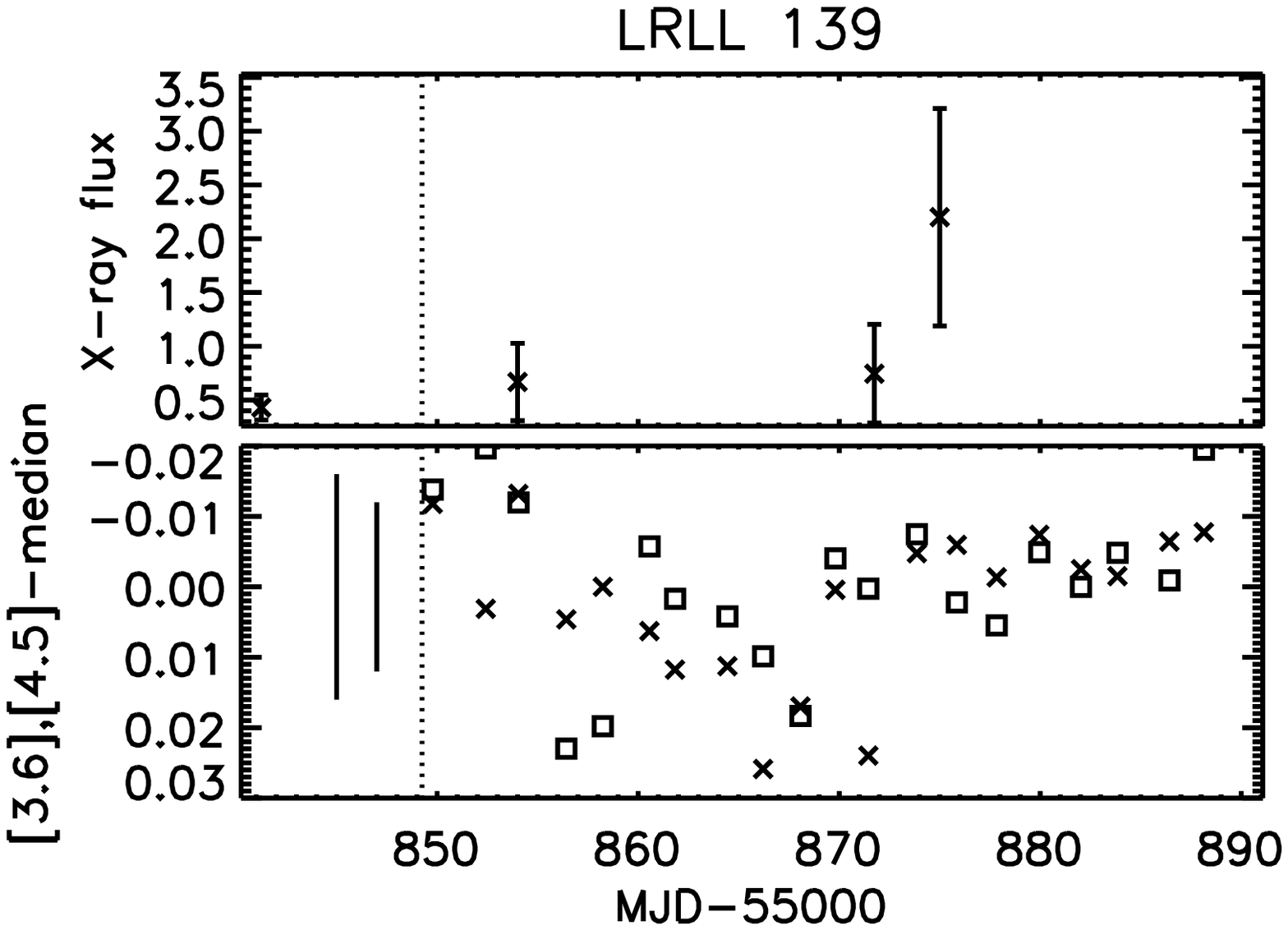}
\figsetgrpnote{X-ray and infrared light curves. The top panel show the X-ray light curve, in units of 1e-6 photons per cm$^2$ per sec. Points to the left of the vertical dashed line are derived from earlier Chandra observations of IC 348 \citep{pre02,ale12}. The infrared light curves are shown in the bottom panel. Error bars for [3.6] (squares) and [4.5] (crosses) photometry are shown as the left and right vertical lines.}
\figsetgrpend

\figsetgrpstart
\figsetgrpnum{3.32}
\figsetgrptitle{LRLL 141 light curves}
\figsetplot{lrll141_xray.eps}
\figsetgrpnote{X-ray and infrared light curves. The top panel show the X-ray light curve, in units of 1e-6 photons per cm$^2$ per sec. Points to the left of the vertical dashed line are derived from earlier Chandra observations of IC 348 \citep{pre02,ale12}. The infrared light curves are shown in the bottom panel. Error bars for [3.6] (squares) and [4.5] (crosses) photometry are shown as the left and right vertical lines.}
\figsetgrpend

\figsetgrpstart
\figsetgrpnum{3.33}
\figsetgrptitle{LRLL 146 light curves}
\figsetplot{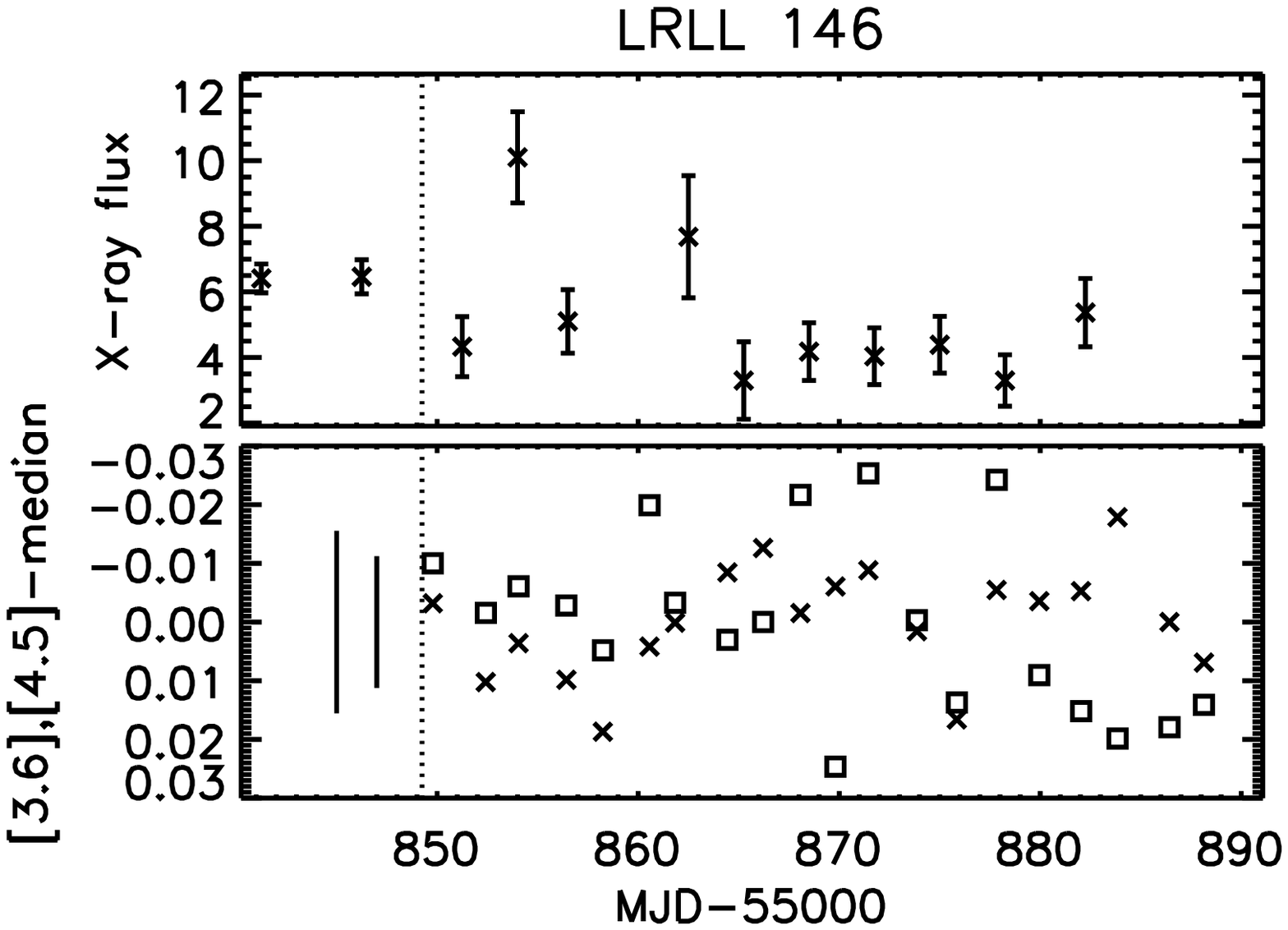}
\figsetgrpnote{X-ray and infrared light curves. The top panel show the X-ray light curve, in units of 1e-6 photons per cm$^2$ per sec. Points to the left of the vertical dashed line are derived from earlier Chandra observations of IC 348 \citep{pre02,ale12}. The infrared light curves are shown in the bottom panel. Error bars for [3.6] (squares) and [4.5] (crosses) photometry are shown as the left and right vertical lines.}
\figsetgrpend

\figsetgrpstart
\figsetgrpnum{3.34}
\figsetgrptitle{LRLL 186 light curves}
\figsetplot{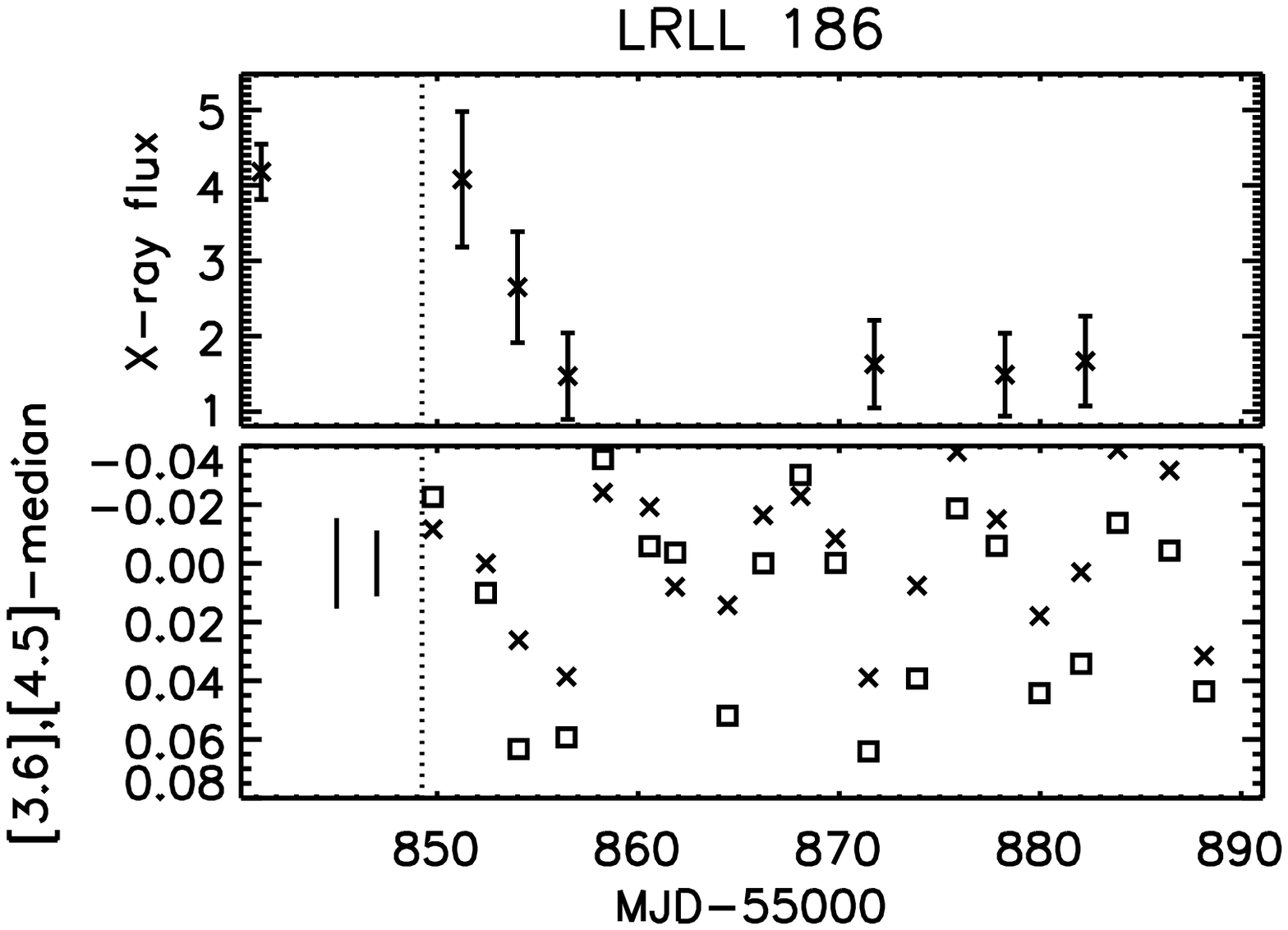}
\figsetgrpnote{X-ray and infrared light curves. The top panel show the X-ray light curve, in units of 1e-6 photons per cm$^2$ per sec. Points to the left of the vertical dashed line are derived from earlier Chandra observations of IC 348 \citep{pre02,ale12}. The infrared light curves are shown in the bottom panel. Error bars for [3.6] (squares) and [4.5] (crosses) photometry are shown as the left and right vertical lines.}
\figsetgrpend

\figsetgrpstart
\figsetgrpnum{3.35}
\figsetgrptitle{LRLL 203 light curves}
\figsetplot{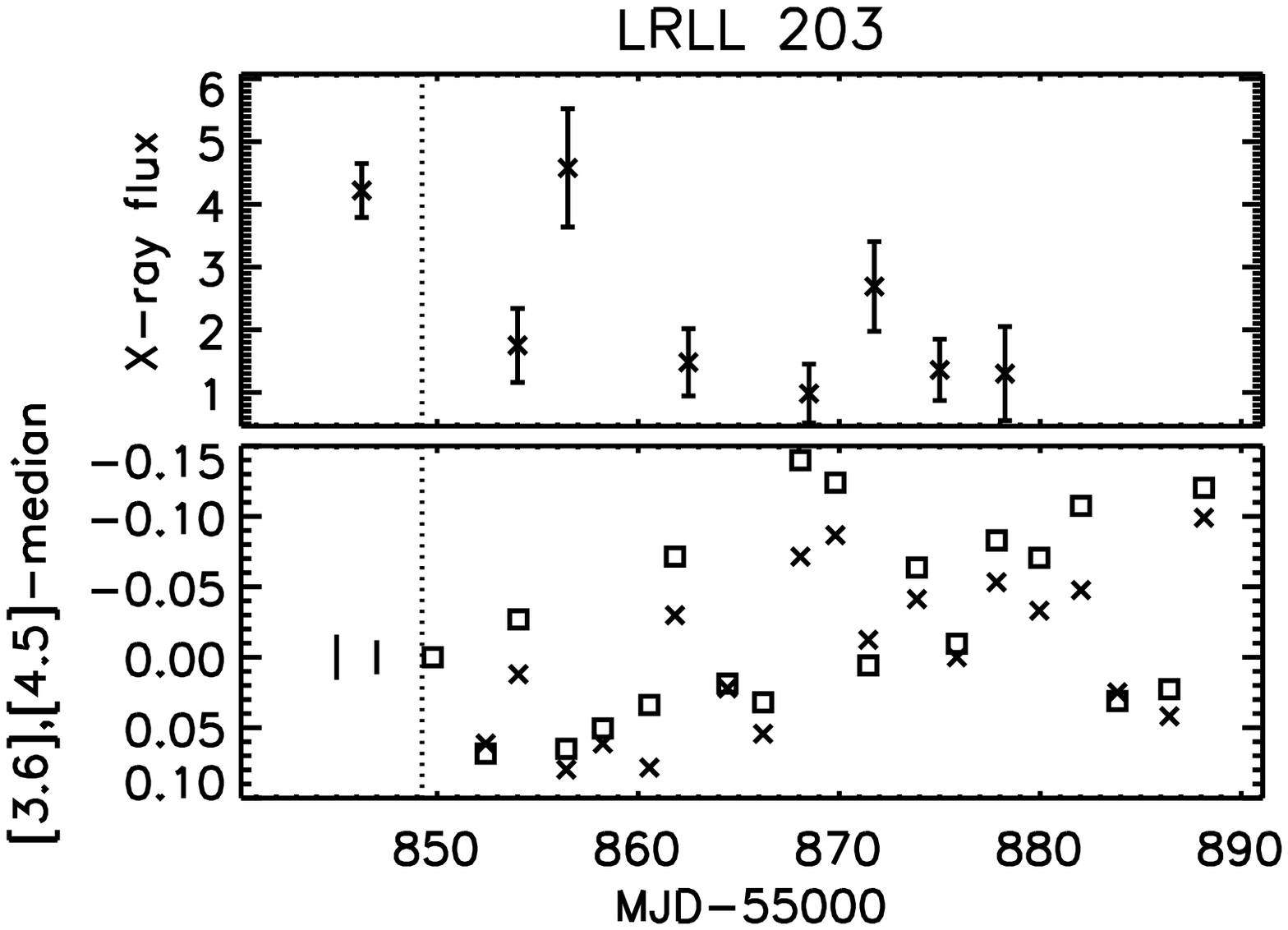}
\figsetgrpnote{X-ray and infrared light curves. The top panel show the X-ray light curve, in units of 1e-6 photons per cm$^2$ per sec. Points to the left of the vertical dashed line are derived from earlier Chandra observations of IC 348 \citep{pre02,ale12}. The infrared light curves are shown in the bottom panel. Error bars for [3.6] (squares) and [4.5] (crosses) photometry are shown as the left and right vertical lines.}
\figsetgrpend

\figsetgrpstart
\figsetgrpnum{3.36}
\figsetgrptitle{LRLL 226 light curves}
\figsetplot{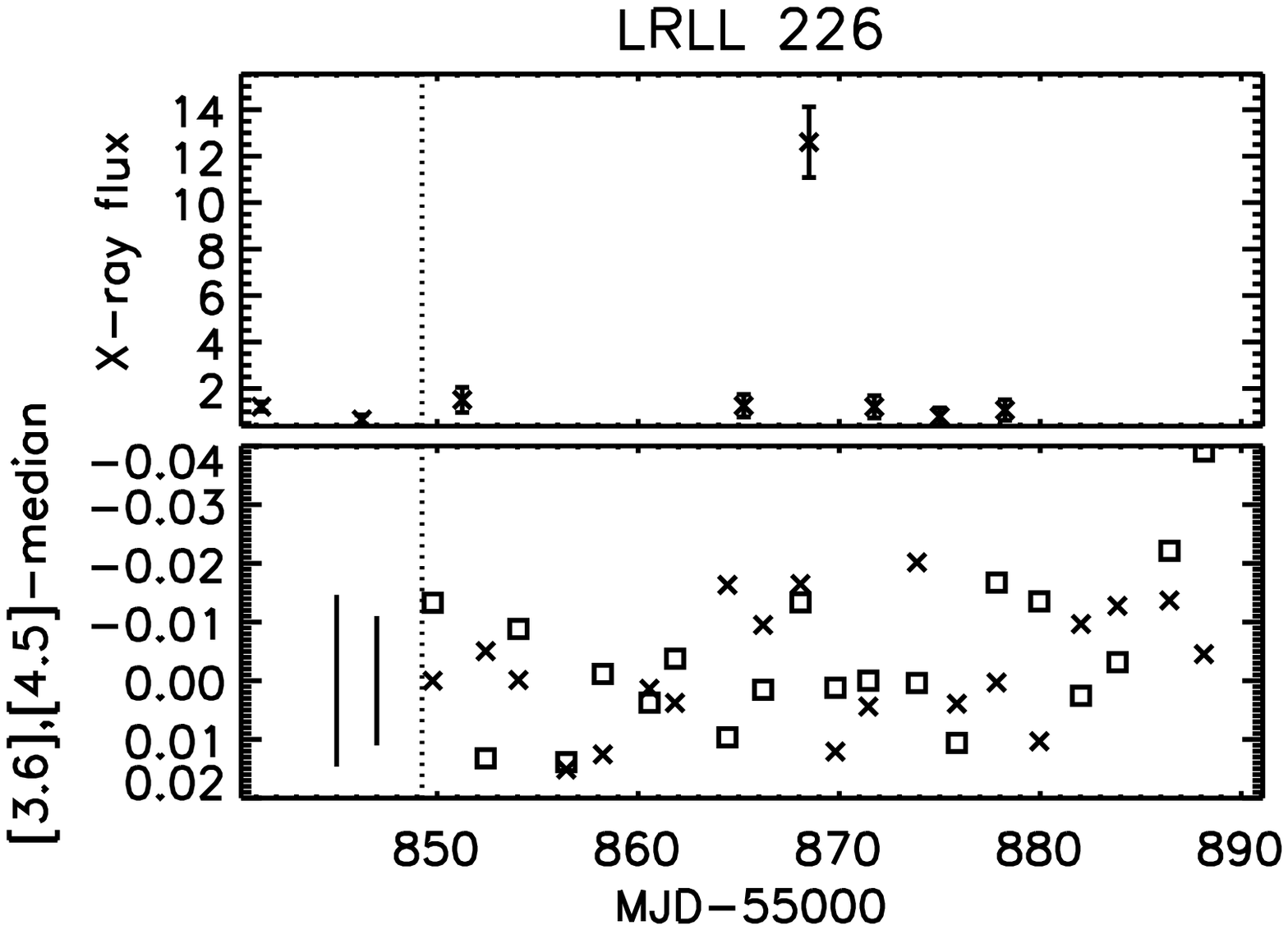}
\figsetgrpnote{X-ray and infrared light curves. The top panel show the X-ray light curve, in units of 1e-6 photons per cm$^2$ per sec. Points to the left of the vertical dashed line are derived from earlier Chandra observations of IC 348 \citep{pre02,ale12}. The infrared light curves are shown in the bottom panel. Error bars for [3.6] (squares) and [4.5] (crosses) photometry are shown as the left and right vertical lines.}
\figsetgrpend

\figsetgrpstart
\figsetgrpnum{3.37}
\figsetgrptitle{LRLL 1401 light curves}
\figsetplot{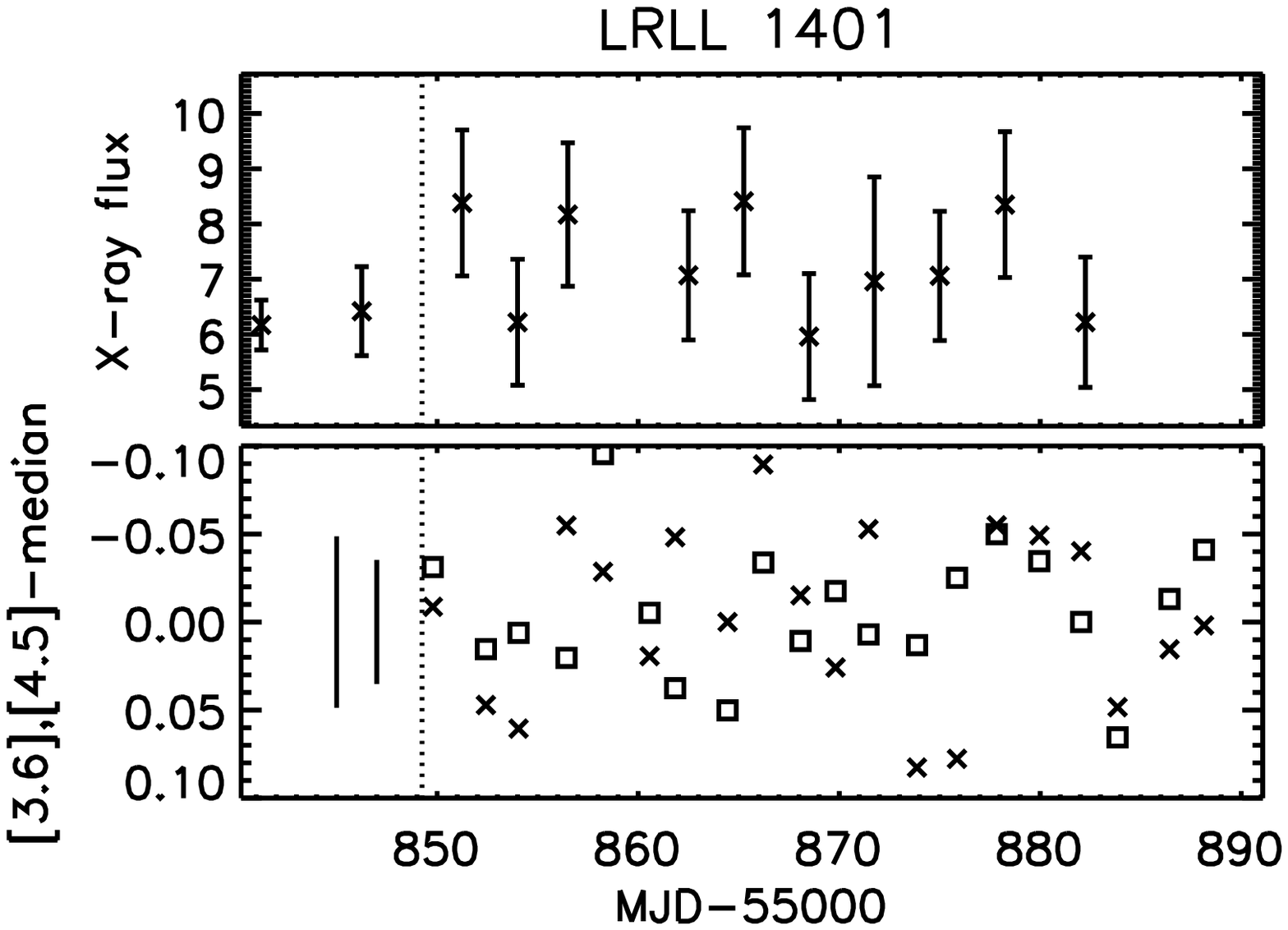}
\figsetgrpnote{X-ray and infrared light curves. The top panel show the X-ray light curve, in units of 1e-6 photons per cm$^2$ per sec. Points to the left of the vertical dashed line are derived from earlier Chandra observations of IC 348 \citep{pre02,ale12}. The infrared light curves are shown in the bottom panel. Error bars for [3.6] (squares) and [4.5] (crosses) photometry are shown as the left and right vertical lines.}
\figsetgrpend

\figsetgrpstart
\figsetgrpnum{3.38}
\figsetgrptitle{LRLL 1872 light curves}
\figsetplot{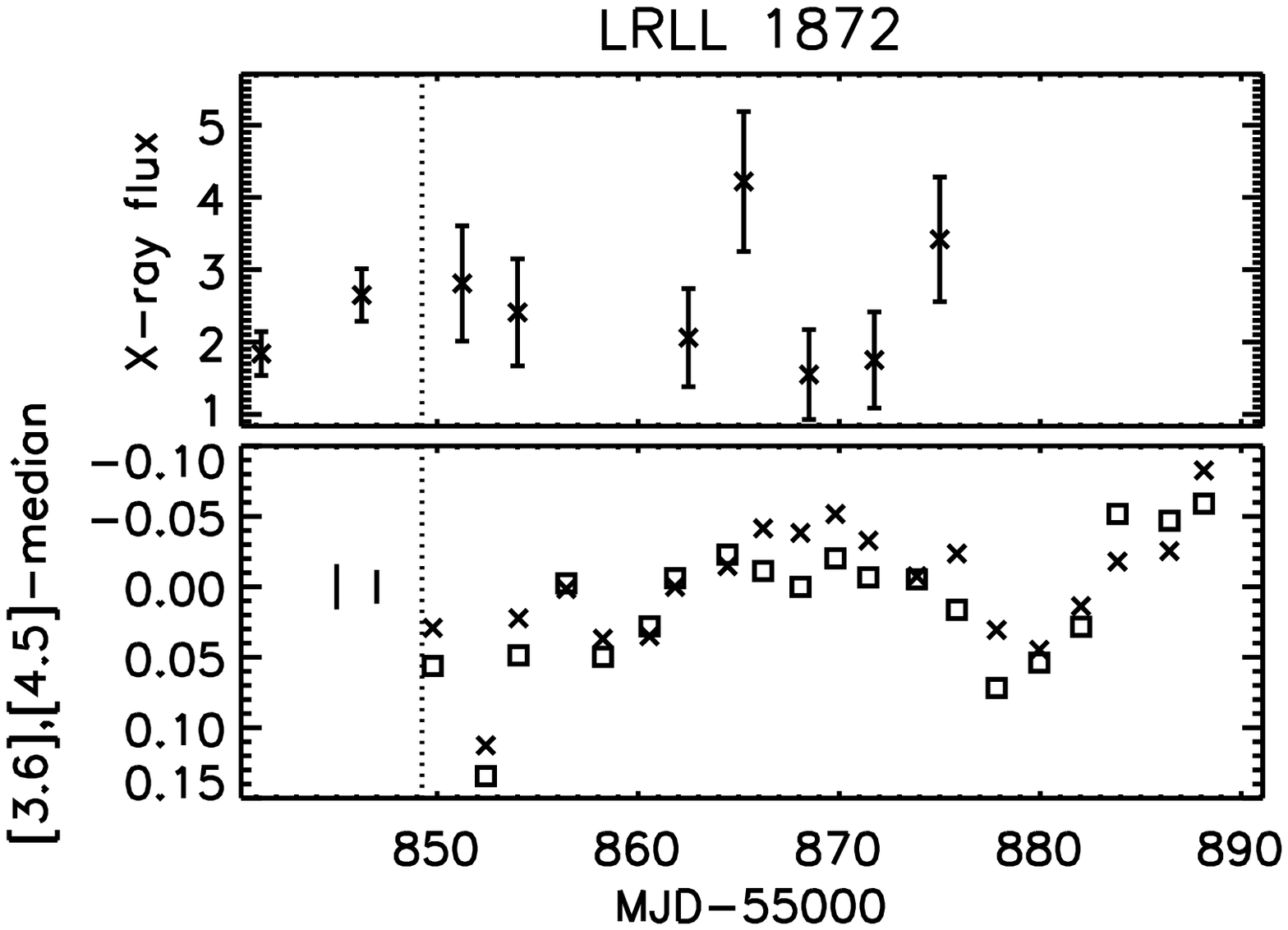}
\figsetgrpnote{X-ray and infrared light curves. The top panel show the X-ray light curve, in units of 1e-6 photons per cm$^2$ per sec. Points to the left of the vertical dashed line are derived from earlier Chandra observations of IC 348 \citep{pre02,ale12}. The infrared light curves are shown in the bottom panel. Error bars for [3.6] (squares) and [4.5] (crosses) photometry are shown as the left and right vertical lines.}
\figsetgrpend

\figsetgrpstart
\figsetgrpnum{3.39}
\figsetgrptitle{LRLL 9024 light curves}
\figsetplot{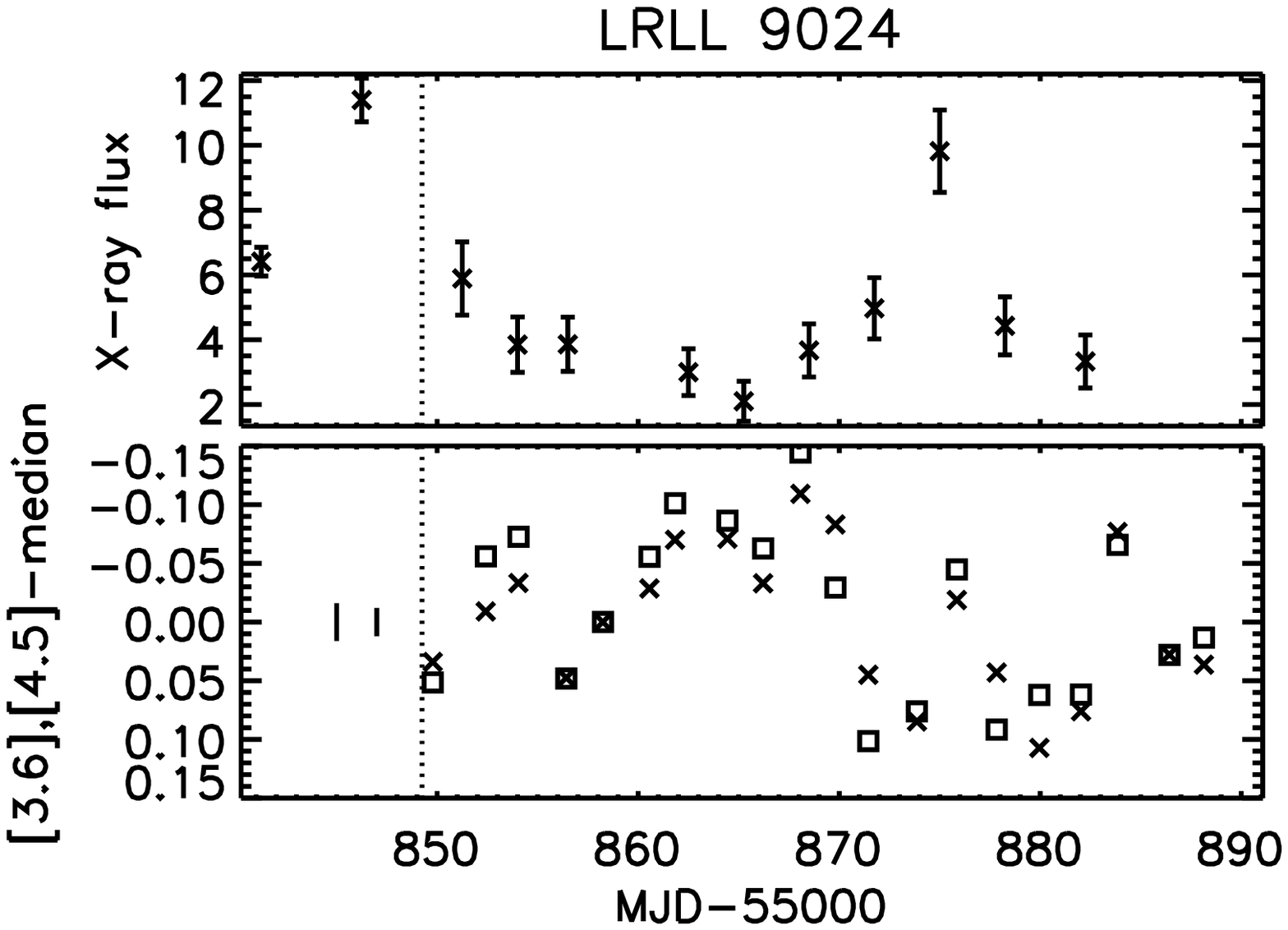}
\figsetgrpnote{X-ray and infrared light curves. The top panel show the X-ray light curve, in units of 1e-6 photons per cm$^2$ per sec. Points to the left of the vertical dashed line are derived from earlier Chandra observations of IC 348 \citep{pre02,ale12}. The infrared light curves are shown in the bottom panel. Error bars for [3.6] (squares) and [4.5] (crosses) photometry are shown as the left and right vertical lines.}
\figsetgrpend

\figsetend

\begin{figure}
\plotone{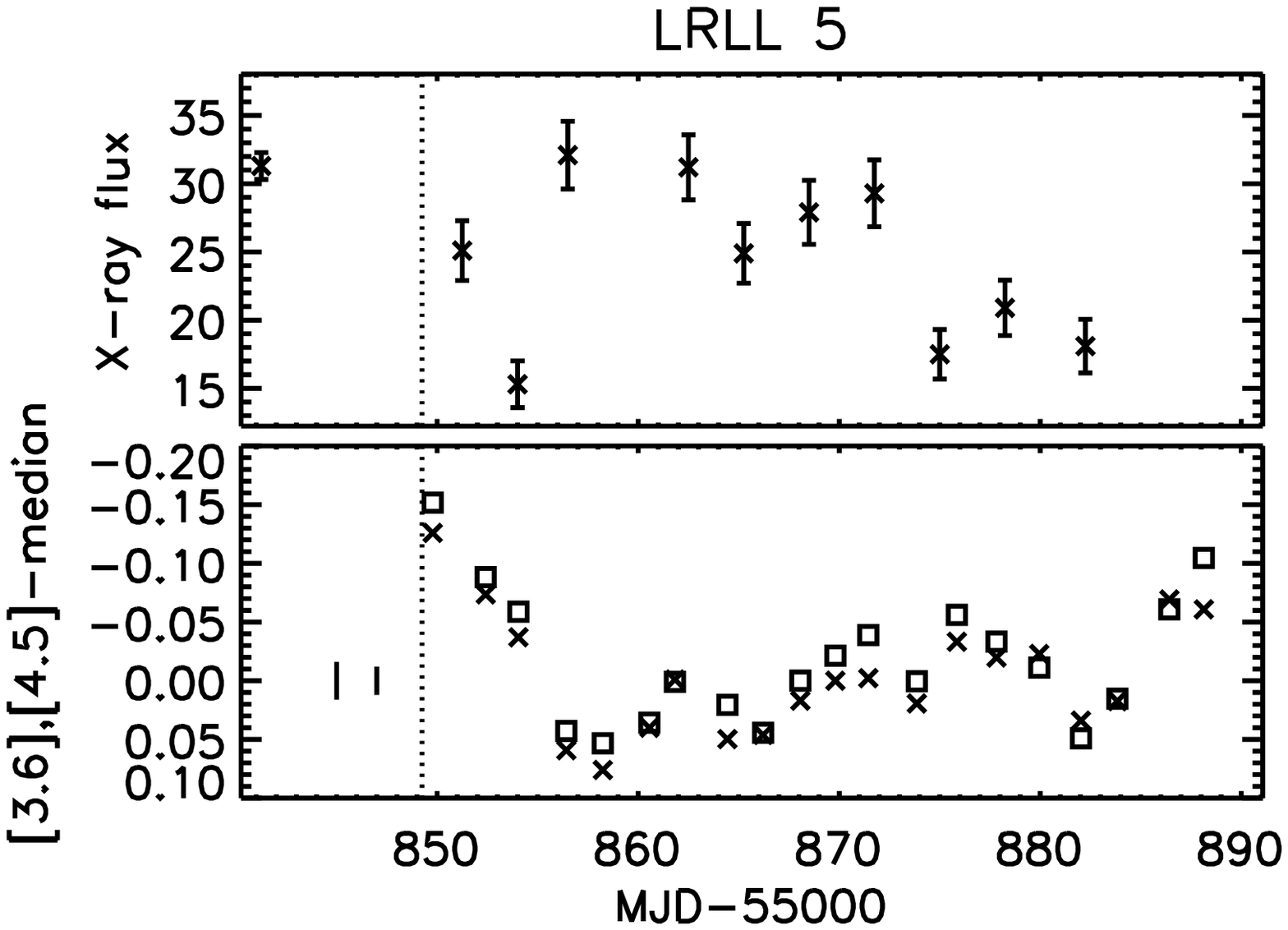}
\caption{X-ray and infrared light curves. The top panel show the X-ray light curve, in units of 1e-6 photons per cm$^2$ per sec. Points to the left of the vertical dashed line are derived from earlier Chandra observations of IC 348 \citep{pre02,ale12}. The infrared light curves are shown in the bottom panel. Error bars for [3.6] (squares) and [4.5] (crosses) photometry are shown as the left and right vertical lines.\label{all_lc}}
\end{figure}

\end{document}